\documentclass[twocolumn,fleqn]{aastex631}   
\usepackage{multirow}
\usepackage{float}
\usepackage{hyperref}
\usepackage{mathtools}
\usepackage{comment}
\usepackage{capt-of}

\begin{document}

\title{A Large-Scale Search for Photometrically Variable AGNs in Dwarf Galaxies Using the Young Supernova Experiment}

\author[0000-0002-9158-5408]{Alexander Messick}
\affiliation{Washington State University, Department of Physics and Astronomy, WA 99164, USA}

\author[0000-0003-4703-7276]{Vivienne Baldassare}
\affiliation{Washington State University, Department of Physics and Astronomy, WA 99164, USA}

\author[0000-0002-6230-0151]{David O. Jones}
\affiliation{Institute for Astronomy, University of Hawaii, 640 A'Ohoku Pl, Hilo, HI 96720, USA}

\author[0000-0002-4235-7337]{K. Decker French}
\affiliation{Department of Astronomy, University of Illinois at Urbana-Champaign, 1002 W. Green St., IL 61801, USA}

\author[0000-0002-6248-398X]{Sandra I. Raimundo}
\affiliation{DARK, Niels Bohr Institute, University of Copenhagen, Jagtvej 128, 2200 Copenhagen, Denmark}
\affiliation{Department of Physics and Astronomy, University of Southampton, Highfield, Southampton SO17 1BJ, UK}

\author[0000-0003-1714-7415]{Nicholas Earl}
\affiliation{Department of Astronomy, University of Illinois at Urbana-Champaign, 1002 W. Green St., IL 61801, USA}

\author[0000-0002-4449-9152]{Katie Auchettl}
\affiliation{Department of Astronomy and Astrophysics, University of California, Santa Cruz, CA 95064, USA}
\affiliation{School of Physics, The University of Melbourne, VIC 3010, Australia}

\author[0000-0003-4263-2228]{David A. Coulter}
\affiliation{Space Telescope Science Institute, Baltimore, MD 21218, USA}

\author[0000-0003-1059-9603]{Mark E. Huber}
\affiliation{Institute for Astronomy, University of Hawaii, 2680 Woodlawn Drive, Honolulu, HI 96822, USA}

\author[0000-0003-1535-4277]{Margaret E. Verrico}
\affiliation{Department of Astronomy, University of Illinois at Urbana-Champaign, 1002 W. Green St., IL 61801, USA}
\affiliation{Center for Astrophysical Surveys, National Center for Supercomputing Applications, Urbana, IL, 61801, USA}

\author[0000-0001-5486-2747]{Thomas de Boer}
\affiliation{Institute for Astronomy, University of Hawaii, 2680 Woodlawn Drive, Honolulu, HI 96822, USA}

\author[0000-0001-6965-7789]{Kenneth C. Chambers}
\affiliation{Institute for Astronomy, University of Hawaii, 2680 Woodlawn Drive, Honolulu, HI 96822, USA}

\author[0000-0003-1015-5367]{Hua Gao}
\affiliation{Institute for Astronomy, University of Hawaii, 2680 Woodlawn Drive, Honolulu, HI 96822, USA}

\author[0000-0002-7272-5129]{Chien-Cheng Lin}
\affiliation{Institute for Astronomy, University of Hawaii, 2680 Woodlawn Drive, Honolulu, HI 96822, USA}

\author[0000-0002-1341-0952]{Richard J. Wainscoat}
\affiliation{Institute for Astronomy, University of Hawaii, 2680 Woodlawn Drive, Honolulu, HI 96822, USA}

\begin{abstract}
We conduct an analysis of over 60,000 dwarf galaxies ($7\lesssim \log{M_*/M_\odot} \lesssim10$) in search of photometric variability indicative of active galactic nuclei (AGNs).
Using data from the Young Supernova Experiment (YSE), a time domain survey on the Pan-STARRS telescopes, we construct light curves for each galaxy in up to four bands (\textit{griz}) where available.
We select objects with AGN-like variability by fitting each light curve to a damped random walk (DRW) model.
After quality cuts and removing transient contaminants, we identify 1100 variability-selected AGN candidates (representing 2.4\% of the available sample).
We analyze their spectra to measure various emission lines and calculate black hole (BH) masses, finding general agreement with previously found mass scaling-relations and nine potential IMBH candidates.
Furthermore, we re-analyze the light curves for our candidates to calculate the dampening timescale $\tau_{DRW}$ associated with the DRW and see a similar correlation between this value and the BH mass.
Finally, we estimate the active fraction as a function of stellar mass and see evidence that active fraction increases with host mass.
\end{abstract}
\keywords{Active Galactic Nuclei, Dwarf Galaxies, Time Domain Astronomy, Photometric Variability}

\section{Introduction} \label{sec:intro}
The center of massive galaxies ($M_*\gtrsim10^{10}M_\odot$) are now understood to contain a supermassive black hole (SMBH) that can be roughly millions to tens of billions of times the mass of the Sun.
However, we do not yet understand how these black holes (BHs) grew to their incredible sizes.
Astronomers have narrowed down the formation of these massive compact objects to a few potential pathways summarized in \citet{2020ARA&A..58..257G}.
\begin{itemize}
    \item Gravitational Runaway: a series of mergers, accretion, and gravitational collapse events within dense stellar clusters \citep{1975Natur.256...23B, 1978MNRAS.185..847B, 1990ApJ...356..483Q, 1993ApJ...418..147L}
    \item Population III Stars: a theoretical population of stars in the early universe, where conditions would have allowed stars to become more massive prior to collapse \citep{1984ApJ...280..825B, 2001ApJ...551L..27M}
    \item Direct Collapse: the collapse of a massive gas cloud directly into a BH without undergoing all of the phases of stellar evolution \citep{1993MNRAS.263..168H, 1994ApJ...432...52L, 2004MNRAS.354..292K}
\end{itemize}
These models differ in a number of ways; for instance, population III stars and direct collapse are thought to only be possible at high redshift ($z\gtrsim10$), while gravitational runaway should be possible at any cosmic time.
The models predict different BH mass functions (the number of BHs as a function of BH mass), occupation fractions (the fraction of galaxies containing BHs as a function of host mass), and BH mass scaling relations, particularly for low-mass galaxies and BHs \citep{2008MNRAS.383.1079V, 2009MNRAS.400.1911V, 2010MNRAS.408.1139V}.
This low-mass regime currently remains unconstrained as such objects are simply more difficult to detect or study directly; the gravitational sphere of influence of a BH around $10^5\,M_\odot$ is only large enough to be resolved within about five megaparsecs and therefore largely unresolvable outside the Local Group.
Simultaneously, if galaxies in this low-mass regime harbor central massive BHs, they are expected to fall in the elusive intermediate mass range ($2\lesssim\log{M_{BH}/M_\odot}\lesssim6$).

As such, scientists have had to rely on alternative methods of detecting these BHs, usually via their interactions with surrounding matter.
These systems include Active Galactic Nuclei (AGNs), central massive BHs surrounded by a disk of accreting gas and dust \citep{1993ARA&A..31..473A}.
Due to observation effects and the complex geometry of an AGN, a ``zoo'' of AGN classes with different properties has arisen.
For some AGNs, the light can vary stochastically over time and across the electromagnetic spectrum \citep{doi:10.1146/annurev.astro.35.1.445, 2003AJ....125....1G}.
This photometric variability is well modeled by a Damped Random Walk (DRW) \citep{Schmidt_2010}, giving researchers a tool for AGN detection that has been used extensively over the past few decades. 
More recently, variability searches have been used to identify AGNs in low-mass galaxies \citep{2018ApJ...868..152B, 2020ApJ...896...10B, 2021MNRAS.504..543B}, a population that can be missed by other detection methods such as BPT diagnostics \citep{Baldwin_1981, 2006MNRAS.372..961K, 10.1111/j.1365-2966.2006.10812.x, Cann_2019}.
This is because [\ion{N}{2}], the numerator in the horizontal axis of the BPT diagram, is a robust indicator of metallicity and trends with galaxy mass, so low-mass or low-metallicity systems will shift towards the left on the diagram.
Simultaneously, low BH mass will also decrease the value of both axes since hardening of the spectral energy distribution would change the ionization structure of the system.

This method of AGN detection is particularly applicable to large-scale time domain surveys like the Palomar Transient Factory, Zwicky Transient Factory, and the upcoming Legacy Survey of Space and Time (LSST).
The Young Supernova Experiment (YSE) is another such survey \citep{2021ApJ...908..143J, 2022zndo...7278430C, 2023PASP..135f4501C}.
Using data from the Pan-STARRS telescope, it observes 1500 square degrees of sky with a three-day cadence in four bands, which are ideal conditions for finding supernovae (as intended) but also photometrically variable AGNs.
We compile a large list of low-mass galaxies within YSE fields with the goal of identifying low-mass AGNs.

This paper is organized as follows:
In Section \ref{sec:targets}, we describe the galaxies used in our study, their sources, their properties, and how they were derived.
In Section \ref{sec:pipeline}, we discuss the steps we took to collect, clean, and analyze the data from the YSE, as well as our AGN selection criteria.
Next, in Section \ref{sec:spec}, we download and analyze the spectra of the resulting AGN candidates, calculating BH mass where possible, as well as reanalyzing the variable light curves for DRW parameters.
Finally, we discuss the results of these analyses and compare them with previously found relations in Section \ref{sec:discussion}.
We also calculate the active fraction, conduct a BPT analysis on the data, compare our findings to known AGNs, and investigate the difference in our results across filters.

Herein, all quoted uncertainties are reported to 1$\sigma$ unless otherwise stated.

\section{Target Selection} \label{sec:targets}
\begin{figure*}
    \gridline{
        \fig{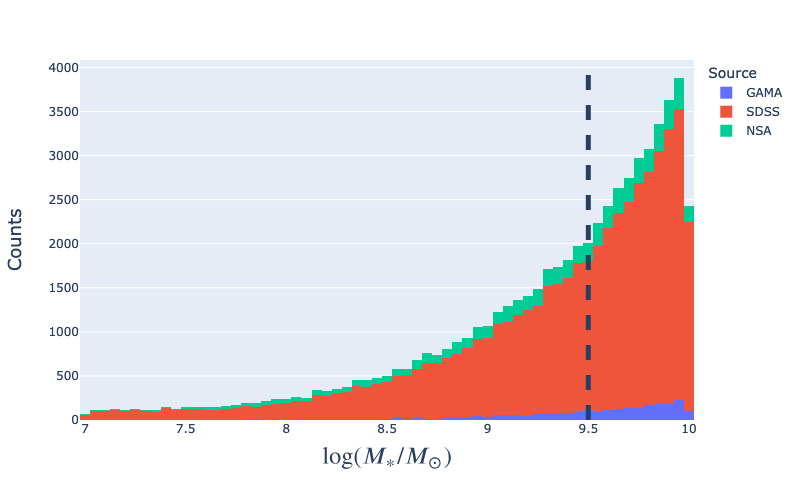}{0.5\textwidth}{}
        \fig{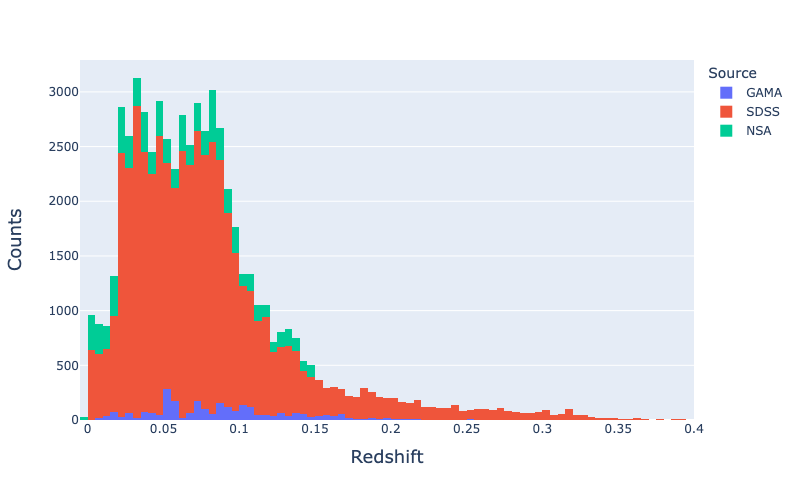}{0.5\textwidth}{}}
    \caption{Histograms of the stellar masses and redshifts of the sample of galaxies used in this paper. We include galaxies within the mass range $10^{9.5-10}\,M_\odot$, which in fact comprise half the sample. We also truncate the plot on the right at a redshift of 0.4 as there are only 98 objects past this value, which reach a redshift of up to 0.935.}
    \label{fig:hist}
\end{figure*}

All of the galaxies in our target sample are taken from the Galaxy And Mass Assembly (GAMA), the Sloan Digital Sky Survey (SDSS), or the NASA-Sloan Atlas (NSA).
We use the \textit{astroquery} Python package to select galaxies with masses in the desired mass range.
Although some definitions of dwarf galaxy cap off at $10^{9.5} M_\odot$, we use the extended range $7\leq\log{M_*/M_\odot}\leq10$.
This additional mass range comprises half of our final sample, as shown in Figure \ref{fig:hist}.

\subsection{GAMA} \label{subsec:GAMA}
The Galaxy And Mass Assembly is a spectroscopic survey carried out on the Anglo-Australian Telescope.
GAMA builds on previous surveys such as the SDSS and is designed to study astronomical structures from kpc to Mpc scales, with data stored in Data Management Units (DMUs).
All GAMA data was acquired via \textit{astroquery}, which is pulled from the Data Release 3 \citep{2010MNRAS.404...86B}.

We query IDs, redshift, and stellar masses (with errors) from the \textit{StellarMasses} DMU, which are estimated using Stellar Population Synthesis modeling \citep{2011MNRAS.418.1587T}.
We select objects whose stellar masses fall between $10^7$ and $10^{10} M_\odot$ (inclusive) and find the corresponding coordinate data in the \textit{SpecAll} DMU, matching with the catalog ID.

\subsection{SDSS} \label{subsec:SDSS}
The Sloan Digital Sky Survey is a large survey containing data for nearly a billion galaxies.
By default at the time of analysis, \textit{astroquery} uses data from Data Release 17, the final release of the SDSS-IV \citep{2022ApJS..259...35A}.
This release contains eight tables with stellar mass estimates for galaxies.

Four of these tables give masses calculated with the Granada Flexible Stellar Population Synthesis \citep{2009ApJ...699..486C}, which uses spectroscopic redshift, optical photometry, and SED modelling.
The flexibility of this model allows for fitting early formation-time scenarios or a wide range of formation-times, each with or without dust (yielding four tables).

Two tables contain stellar masses calculated with the Portsmouth method \citep{2013MNRAS.435.2764M}, corresponding to the passive and star-forming models.
Another table contains stellar mass estimates from \citet{2012MNRAS.421..314C}, which uses principal component analysis with stellar population synthesis models from \citet{2011MNRAS.418.2785M}.
The final table was SDSS's \textit{galSpecExtra}, whose stellar masses are calculated using MPA-JHU measurements, named after the Max Planck Institute for Astrophysics and the Johns Hopkins University who developed the technique based on work by \citet{2003MNRAS.341...33K, 2004MNRAS.351.1151B, 2004ApJ...613..898T}.
For multiple instances of the same spectrum ID, we select those with the best $\chi^2$ fit.
Then, for multiple instances of the same target ID, we select those with the best reduced $\chi^2$ fit.

\subsection{NSA} \label{subsec:NSA}
The NASA Sloan Atlas (NSA) is a catalog derived from the SDSS with additional data from the Galaxy Evolution Explorer.
We use both versions of the NSA, which we refer to as NSA v0\footnote{nsa\_v0\_1\_2.fits} and NSA v1\footnote{nsa\_v1\_0\_1.fits} for post-analysis comparison and target selection, respectively.
NSA v0 contains fewer galaxies, but more features including spectroscopic measurements, which is useful for post-analysis.
We use NSA v1 for our target selection, which was released with SDSS Data Release 13 \citep{2017ApJS..233...25A} and uses an improved background subtraction technique \citep{2011AJ....142...31B} relative to SDSS DR8 \citep{2011ApJS..193...29A}.
Relative to NSA v0, this catalog adds elliptical Petrosian aperture photometry derived from the r-band, including stellar mass estimates from a K-correction fit.
These stellar masses are given in $M_\odot/h^2$, so we use $h=0.7$ cosmology in our conversion.

\subsection{Combining Sources}
\begin{figure}[htb]
    \centering
    \includegraphics[width=0.48\textwidth]{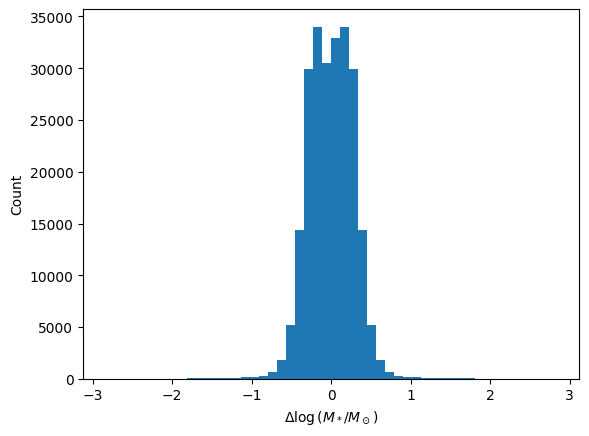}
    \caption{Plot of differences in masses reported for objects in multiple sources. We find 512 of the 237,092 overlapping objects with a difference of at least 1 (in logarithm space)}
    \label{fig:mass_diff}
\end{figure}

After collecting our galaxies from the individual sources, we first join them all together, then we cross-match the entire sample against itself.
We keep only the first instance of each unique object, which selected in the following order: NSA, GAMA, SDSS.
This order was selected to ensure that mass measurements were as accurate and uniform as possible.
This cross-matching is to avoid redundancies, especially since the NSA is a subset of the SDSS.
Although some objects appeared in multiple catalogs, we only query and compare objects below $10^{10}\,M_\odot$.
Therefore, if an object is calculated to have a stellar mass above this threshold in one source and below in another, only the latter would appear.
We also compare the masses reported for objects found in different sources, shown in Figure \ref{fig:mass_diff}, and find good agreement.
Altogether, this resulted in a target list of 60,468 galaxies with stellar masses between $10^7$ and $10^{10}\,M_\odot$.

\section{Light Curve Pipeline}\label{sec:pipeline}
We create light curves for each object using forced photometry.
We use a modified version of the YSE light curve analysis pipeline described in \citet{2021ApJ...908..143J}.

\subsection{Forced Photometry} \label{subsec:forced_phot}
For each target, we require the flux from the given coordinate observed at different points in time.
Because these images will necessarily have different resolutions (due to weather, atmospheric effects, etc.), simply calculating the resulting fluxes would lead to errors and artifacts, so the images must be degraded to a common resolution.
This is done by the Image Processing Pipeline (IPP) at the University of Hawaii's Institute for Astronomy \citep{2020ApJS..251....5M}, which provides difference and stack images.
To create the stack image, the data is first processed (i.e. detrended and warped), then the individual observations are added together \citep{Waters_2020}.
These stack images are used as templates alongside the individual observations which are scaled and convolved to a target point spread function (PSF).
The stack image is then subtracted from each reprocessed image, creating the difference images, from which the PSF flux is calculated.
The pipeline returns the difference image flux and error in microJanskys.
Since this flux is relative to a subtracted image, we request the stack images and measure the flux within a 2.5 arcsecond aperture, which is added back to the difference image flux so that they can be converted to magnitudes.
We selected this aperture because 2.5 arcseconds is roughly double the median FWHM for YSE observations.

\subsection{Light Curve Model Fitting} \label{subsec:qso_fit}
We utilize \textit{qso\_fit} \citep{2011AJ....141...93B} to analyze the light curves.
This software fits a DRW model to the given magnitudes and dates of a light curve, returning the best-fit model parameters and significances $\sigma_{vary}$, $\sigma_{QSO}$, and $\sigma_{notQSO}$ with corresponding $\chi^2_{vary}$, $\chi^2_{QSO}$, and $\chi^2_{notQSO}$.
These respectively represent the significance that the object is variable, that the source variability is well described by the DRW model, and that the source variability is better described as random.
The software uses these values to predict a class: if $\sigma_{vary}$ and $\sigma_{QSO}$ are both greater than 3, then the object is a ``QSO''. If $\sigma_{vary}$ and $\sigma_{notQSO}$ (but NOT $\sigma_{QSO}$) are greater than 3, then the object is classified as ``not\_qso''. In all other cases, the light curve is given the class ``ambiguous''.
We also apply an alternative criterion for initial classification, following along \citet{2018ApJ...868..152B}:
\begin{equation} \label{eq:crit}
    \sigma_{vary}>2,\; \sigma_{QSO}>2, \;\mathrm{and}\; \sigma_{QSO}>\sigma_{notQSO}
\end{equation}
If a light curve satisfies either criteria (i.e. has class ``QSO'' or satisfies the inequalities in Equation \ref{eq:crit}), then it is said to ``pass'' this test.

After running the fitting software, we also calculate the fractional variability \citep{2019Galax...7...62S} of each light curve when possible:
\begin{equation}
    F_{var}=\sqrt{\frac{S^2-\langle \sigma_{err}^2 \rangle}{\langle x \rangle ^2}}
\end{equation}
where $S^2$ is the variance of the data and $\langle \sigma_{err}^2 \rangle$ is the mean squared uncertainty.
Because of this formulation, any light curve for which $\langle \sigma_{err}^2 \rangle > S^2$ would result in an complex fractional variability, which we discard.

\subsubsection{Cleaning Process} \label{subsub:clean}
At the beginning of our analysis, we ``clean'' each band of each light curve.
We remove any data points that were flagged by the IPP, which can include artifacts such as difference spikes, ghosting, being off-chip, saturation, and general defects.
Afterward, we also remove any data points corresponding to a negative flux values, assuming them to be anomalous.
Finally, we apply a $5\sigma$ clipping to the data and convert the remaining flux values from microJanksys to magnitudes $\left(m=23.9-2.5\log{f}\right)$, as well as calculating their mean and standard deviations.
If there are more than 10 data points remaining in a (cleaned) band, it is ready for model fitting.
For reference, the mean number of data points after cleaning is 42.4, 52.0, 39.1, and 30.4 for the \textit{griz} bands, respectively.

\subsubsection{Boot-Strapping Uncertainty} \label{subsub:boot}
To get a distribution of the outputs, we bootstrap the uncertainties.
For each data point, we add or subtract from a random sampling of a normal distribution scaled by the reported errors.
We repeat this process 100 times for each band of each light curve, cleaning each curve after sampling.
We record the mean and standard deviation of the outputs, with the exception of the ``class'', for which we record the most common designation.
We also keep track of what fraction of iterations pass the initial classification described in Section \ref{subsec:qso_fit}.

\subsubsection{Additional Tests} \label{subsub:tests}
After cleaning and analyzing the light curves, we had many objects that were being classified as variable seemingly based on obvious individual outliers (even after sigma clipping).
To counteract this, we reanalyze every light curve.
We first calculate the mean magnitude, find the data point that lies farthest away from this mean, and remove it.
After removing this point, we run the \textit{qso\_fit} code and check again for AGN classification.
We repeat this process, except having removed the single data point with the largest uncertainty instead (although these two data points were the same for many light curves).
The results of these two additional runs are called the ``Point Tests''.
This test was necessary in order to remove many dubious light curves that would have otherwise been classified as exhibiting AGN-like variability.
Upon visual inspection, this test was successful in removing many, if not all, of these objects without discarding apparently ``real'' AGNs.

\begin{figure*}[htb]
    \centering
    \includegraphics[width=\textwidth]{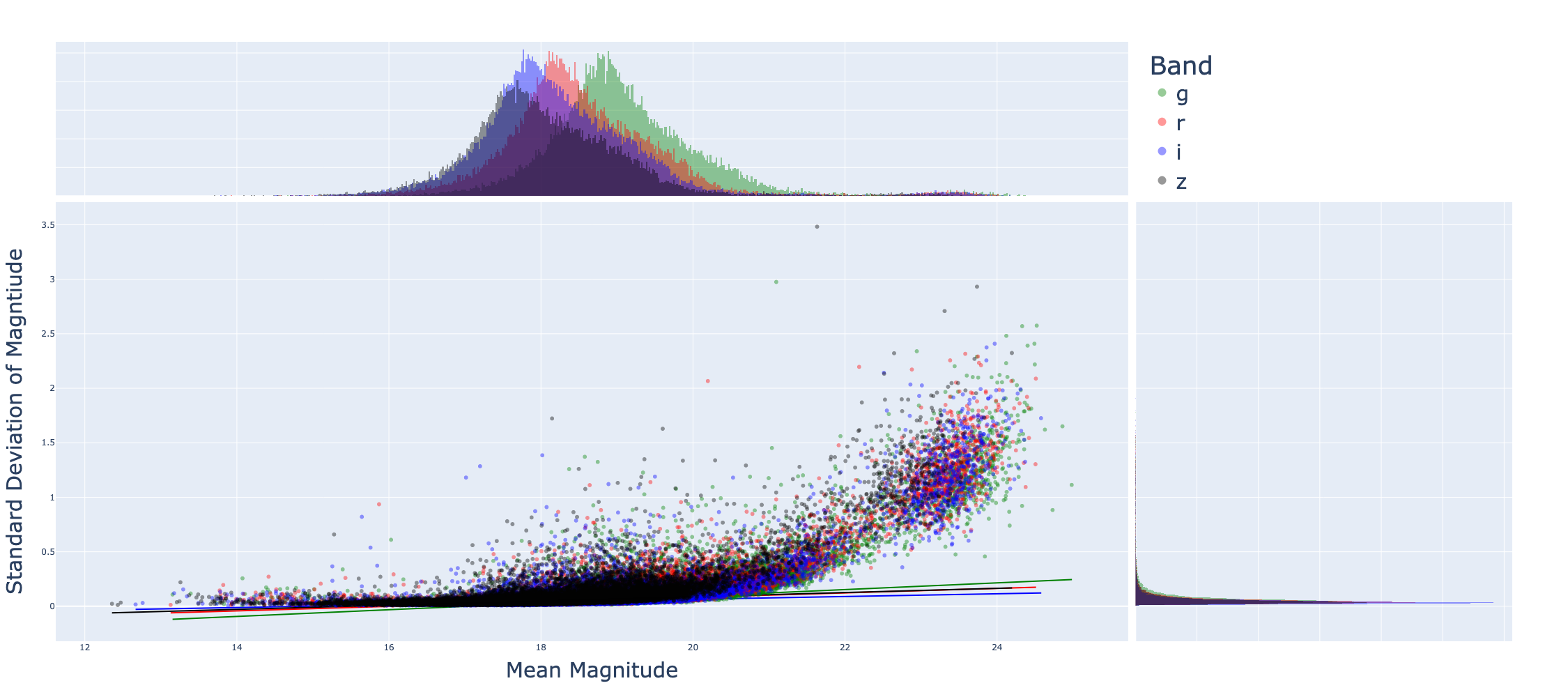}
    \caption{Plot of the standard deviation versus the mean magnitude of each band of each (clean) light curve with color corresponding to the band observed. We show the histogram for each axis to demonstrate the concentration of values with a low standard deviation, as well as the linear trend for each band. Observations that fall above their respective trend line are said to have excess variance.}
    \label{fig:excess}
\end{figure*}

For the ``clean'' light curves, we plot the standard deviation of the magnitude for each band of each object versus their mean magnitude, shown in Figure \ref{fig:excess}.
For each band, we perform a linear fit between the standard deviation and the mean; objects that fall above this linear fit are said to have excess variance.

\subsection{AGN Selection} \label{subsec:agn_select}
There are a number of selection criteria we apply to each band of each galaxy based on the results of the previous section.
\begin{enumerate}
    \item Boot-Strap: we select light curves that passed the AGN classification discussed in Section \ref{subsec:qso_fit} in at least half of their boot-strapped iterations
    \item Point Tests: we select curves that also passed the AGN classification for both point tests described in Section \ref{subsub:tests}
    \item Excess Variance: we select curves for which the standard deviation of the magnitude exceeds a linear fit relative to its mean magnitude
\end{enumerate}

\begin{figure}[htb]
    \centering
    \includegraphics[width=0.48\textwidth]{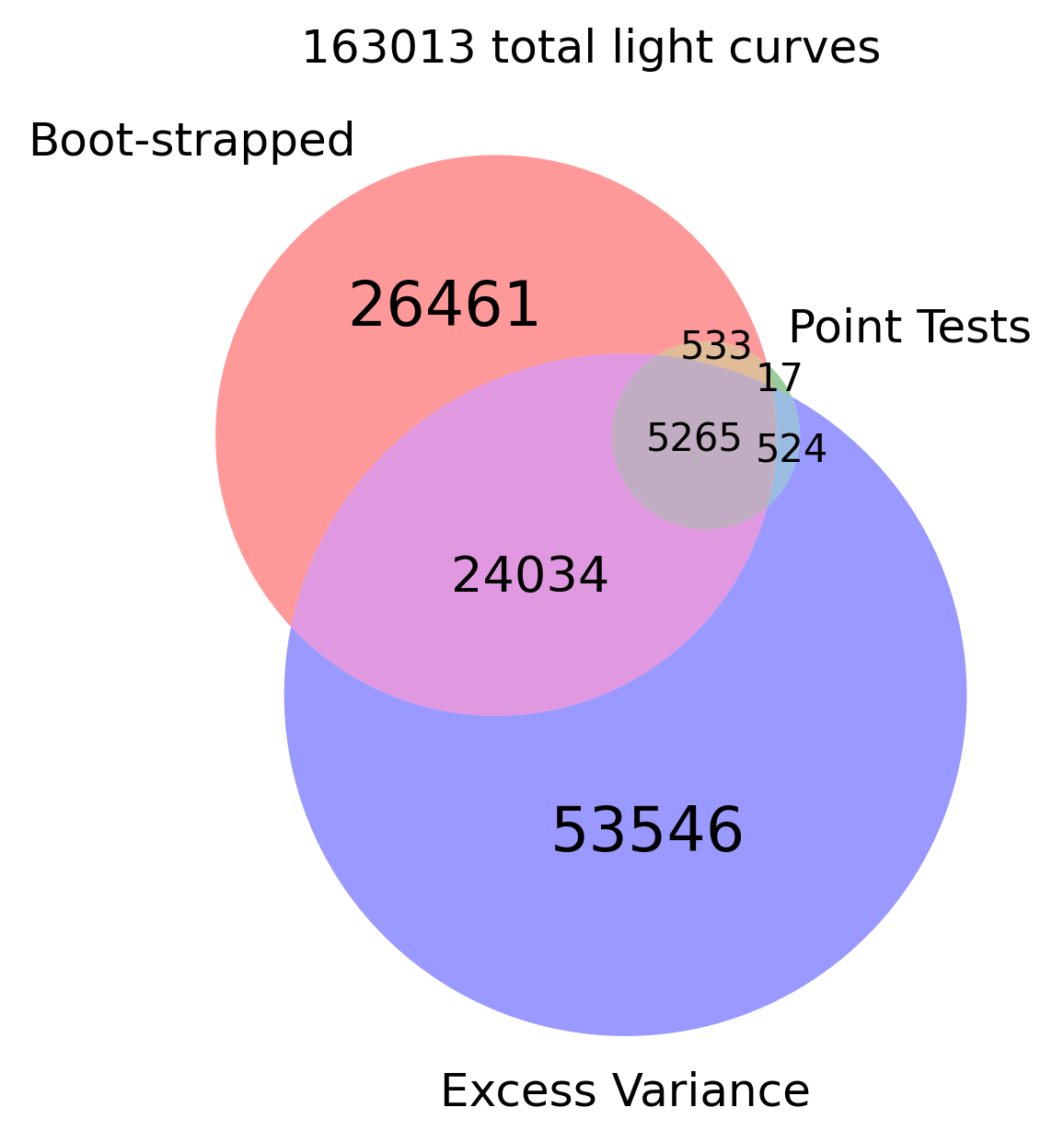}
    \caption{A Venn diagram of the sets of individual light curves that satisfy the AGN selection criteria listed in Section \ref{subsec:agn_select}. We found 5265 light curves that satisfy all criteria belonging to 3725 unique galaxies. After removing supernovae and other transient objects, 5070 light curves belonging to 3632 galaxies remained.}
    \label{fig:AGN_crit}
\end{figure}

\begin{figure*}
    \gridline{\fig{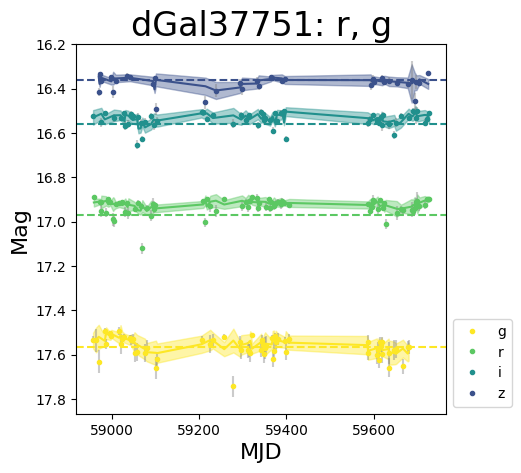}{0.3\textwidth}{Failed the Boot-Strap Condition; This object would have been variable in the r and g bands, but only 36\% and 49\% of the bootstrapped iterations were found to be variable for each band, respectively.}
        \fig{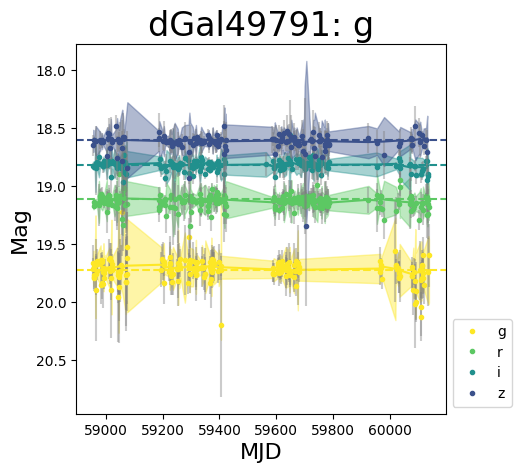}{0.3\textwidth}{Failed the Point Test; the g-band for this object is found to have AGN-like variability, but not when one excludes the farthest point from the mean: the dimmest observation (which is also the point with the largest uncertainty in this example).}
        \fig{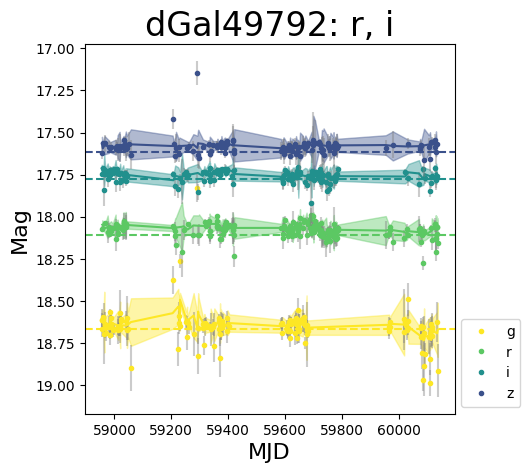}{0.3\textwidth}{Failed the Excess Variance Condition; the r and i bands for this object would have been classified as variable, but their variances across time are small relative to their mean magnitude (below a linear fit of all clean light curves for each band).}}
    \caption{Example light curves of objects that pass all but one of each criteria. In other words, each of these objects would have been included as an AGN candidate if not for a single one of these criteria. Above each plot is the identification number of each galaxy and the bands that would have otherwise passed AGN candidacy. For each band in each light curve, we show the host galaxy light (calculated from forced photometry on the stack image) as a dashed horizontal line.}
    \label{fig:crit_fails}
\end{figure*}

A summary of these three criteria is shown in Figure \ref{fig:AGN_crit}.
Examples of objects that failed a single of these additional criteria is shown in Figure \ref{fig:crit_fails}.

Our analysis flagged 5265 light curves as having AGN-like variability, which we then inspected for contamination of supernovae or other transient objects.
First, we inspected visually, finding and removing 37 objects with burst-like light curves from our list of potential AGNs.
We also cross-matched with the publicly available objects from the Transient Name Server \citep{2021AAS...23742305G}.
We found an additional 56 transients that were discovered during their respective light curves used in this analysis, so we discarded these.
We give an overview of the remaining objects with at least one light curve displaying AGN-like photometric variability in Table \ref{tab:var_gal}, including identifying information, positions, and stellar mass for each object.
In Table \ref{tab:phot_var}, we display some resulting properties for all of the individual variable light curves, such as fractional variability and the fit significances.

\begin{deluxetable*}{C c C C C C C c}[htb]
    \tablecolumns{8}
    \tablecaption{Objects Demonstrating AGN-like Variability \label{tab:var_gal}}
    \tablehead{\colhead{$\#_{gal}$} & \colhead{Source} & \colhead{ID} & \colhead{R.A.} & \colhead{Decl.} & \colhead{Redshift (z)} & \colhead{$\log{M_*/M_\odot}$} & \colhead{Bands}}
    \startdata
          2 & GAMA &  49986 & 223.4078300 & -0.6404700 & 0.075 & 9.18 & r \\
        117 & GAMA & 198708 & 140.6489200 & -0.6487200 & 0.075 & 9.58 & i \\
        259 & GAMA & 205155 & 140.6612500 & -0.3840200 & 0.055 & 9.73 & gri \\
        263 &  NSA &  57562 & 140.7075519 & -0.2745433 & 0.055 & 9.56 & ri \\
        270 &  NSA &  57554 & 140.7894632 & -0.4168406 & 0.056 & 9.66 & ri \\
        324 &  NSA &  56426 & 133.1824039 &  0.1007204 & 0.067 & 9.52 & z \\
        330 & GAMA & 209611 & 134.0787900 &  0.0132300 & 0.088 & 9.24 & r \\
        335 & GAMA & 209673 & 134.4074600 &  0.1944300 & 0.110 & 9.72 & r \\
        364 &  NSA &  56817 & 136.5000296 &  0.0090270 & 0.059 & 9.56 & iz \\
        382 & GAMA & 210346 & 137.8966700 &  0.0660900 & 0.054 & 9.51 & i \\
        386 &  NSA &  57169 & 138.3062241 &  0.1492347 & 0.054 & 9.15 & i \\
        391 & GAMA & 210471 & 138.4888800 &  0.0723100 & 0.093 & 9.84 & i \\
        410 & GAMA & 210786 & 139.6179200 &  0.1823400 & 0.092 & 9.76 & r \\
        481 & GAMA & 216062 & 136.3720000 &  0.5848900 & 0.071 & 9.62 & r \\
        518 &  NSA &  57228 & 138.5570596 &  0.4275729 & 0.053 & 9.38 & r \\
        \vdots & \vdots & \vdots & \vdots & \vdots & \vdots & \vdots & \vdots \\
        \enddata
    \tablecomments{Galaxy properties of the objects that demonstrated AGN-like variability in at least one band. We give the galaxy number assigned for this project, the source catalog and corresponding ID, the position (Right Ascension and Declination in units of degrees), redshift, and stellar mass. In the final column, we give the bands that were found to have AGN-like variability. If an object is variable in more than one band, it is included in our AGN candidates.}
\end{deluxetable*}

\begin{deluxetable*}{C c C C C C C C C}[htb]
    \tablecolumns{9}
    \tablecaption{Photometrically Variable Light Curves \label{tab:phot_var}}
    \tablehead{\colhead{$\#_{gal}$} & \colhead{Band} & \colhead{Magnitude} & \colhead{Frac. Var.} & \colhead{$\sigma_{vary}$} & \colhead{$\sigma_{QSO}$} & \colhead{$\sigma_{not\_QSO}$} & \colhead{$\log{\tau_{DRW}}$} & \colhead{$\sigma_{DRW}$}}
    \startdata
        2 & r & 19.6\pm0.3 & 0.170\pm0.188 & 5.39\pm1.41 & 3.93\pm1.17 & 2.00\pm0.58 & 2.04\pm0.70 & 0.367\pm0.122 \\
        117 & i & 18.8\pm0.2 & \ldots & 3.83\pm0.93 & 4.13\pm1.88 & 1.25\pm0.45 & \ldots & \ldots \\
        259 & i & 18.0\pm0.1 & \ldots & 7.54\pm0.90 & 2.95\pm1.44 & 1.69\pm0.41 & 1.82\pm0.72 & 0.224\pm0.072 \\
        259 & r & 18.4\pm0.1 & \ldots & 4.67\pm1.13 & 3.77\pm1.35 & 1.17\pm0.50 & \ldots & \ldots \\
        259 & g & 19.0\pm0.1 & 0.048\pm0.145 & 4.26\pm0.96 & 2.43\pm0.50 & 1.46\pm0.52 & \ldots & \ldots \\
        263 & i & 17.7\pm0.0 & 0.023\pm0.107 & 5.58\pm1.09 & 3.29\pm1.00 & 1.24\pm0.50 & \ldots & \ldots \\
        263 & r & 18.1\pm0.1 & \ldots & 5.90\pm1.04 & 3.79\pm0.90 & 1.79\pm0.51 & 1.40\pm1.05 & 0.160\pm0.050 \\
        270 & i & 17.7\pm0.0 & 0.019\pm0.124 & 6.49\pm1.03 & 3.66\pm2.46 & 0.84\pm0.44 & \ldots & \ldots \\
        270 & r & 18.1\pm0.1 & \ldots & 5.43\pm0.95 & 3.82\pm1.35 & 1.82\pm0.45 & \ldots & \ldots \\
        324 & z & 18.2\pm0.2 & \ldots & 4.09\pm1.19 & 3.62\pm1.87 & 1.88\pm0.59 & \ldots & \ldots \\
        330 & r & 19.4\pm0.3 & \ldots & 6.03\pm0.89 & 6.66\pm1.57 & 2.57\pm0.43 & 1.76\pm1.56 & 0.327\pm0.112 \\
        335 & r & 19.2\pm0.2 & \ldots & 4.28\pm0.86 & 4.54\pm1.88 & 2.39\pm0.43 & 1.45\pm2.05 & 0.232\pm0.078 \\
        364 & z & 17.4\pm0.2 & 0.064\pm0.199 & 3.36\pm0.92 & 5.40\pm2.08 & 1.13\pm0.43 & \ldots & \ldots \\
        364 & i & 17.5\pm0.1 & 0.043\pm0.107 & 4.93\pm0.96 & 2.96\pm0.65 & 1.45\pm0.41 & \ldots & \ldots \\
        382 & i & 18.5\pm0.1 & \ldots & 4.44\pm1.01 & 3.12\pm0.80 & 1.18\pm0.49 & \ldots & \ldots \\
        \vdots & \vdots & \vdots & \vdots & \vdots & \vdots & \vdots & \vdots & \vdots \\
        \enddata
    \tablecomments{Properties of light curves that demonstrated AGN-like photometric variability. We give the galaxy number, filter, mean magnitude, and fractional variability. We also show the DRW fit significances from \textit{qso\_fit} $\sigma_{vary}$, $\sigma_{QSO}$, and $\sigma_{not\_QSO}$ and the dampening timescale $\log{\tau_{DRW}}$ and length scale $\sigma_{DRW}$ from \textit{taufit} for light curves for which the timescale was less than one fifth of the baseline.}
\end{deluxetable*}

\begin{figure*}
    \gridline{\fig{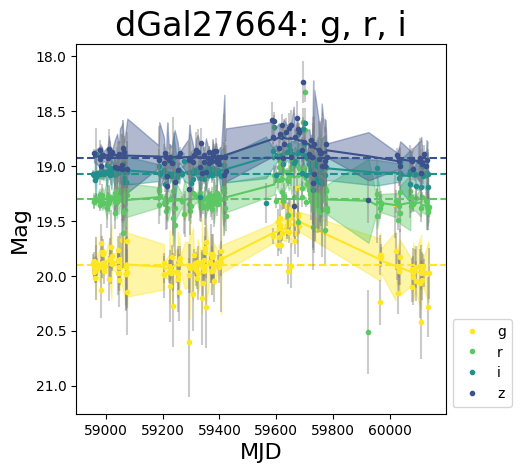}{0.3\textwidth}{}
        \fig{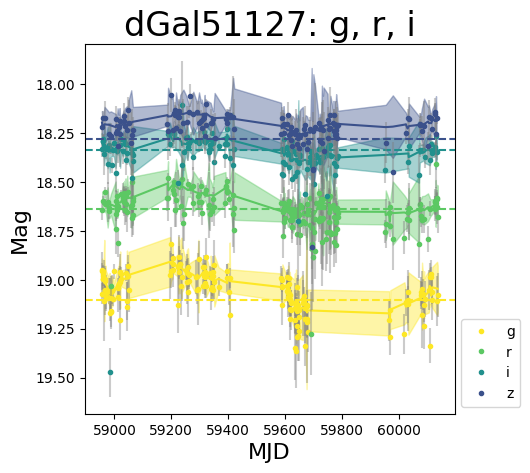}{0.3\textwidth}{}
        \fig{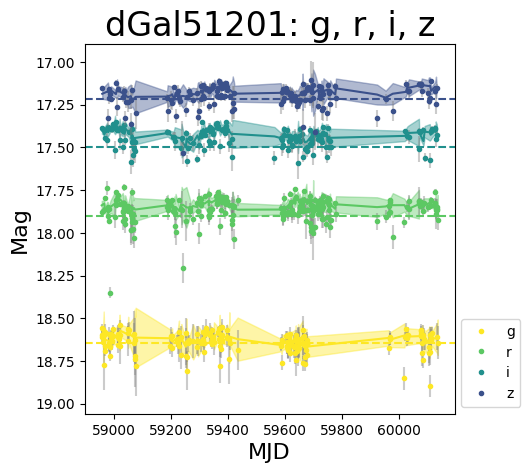}{0.3\textwidth}{}}
    \caption{Example light curves some of our AGN candidates (i.e. objects in which we found AGN-like variability in more than one band). Above each plot is the object ID number, a list of which bands were observed to have AGN-like variability, and the stellar mass of the host galaxy. For each band, we show the host galaxy light (calculated from forced photometry on the stack image) as a dashed horizontal line.}
    \label{fig:agn_cand}
\end{figure*}

\begin{figure}[htb]
    \centering
    \includegraphics[width=0.48\textwidth]{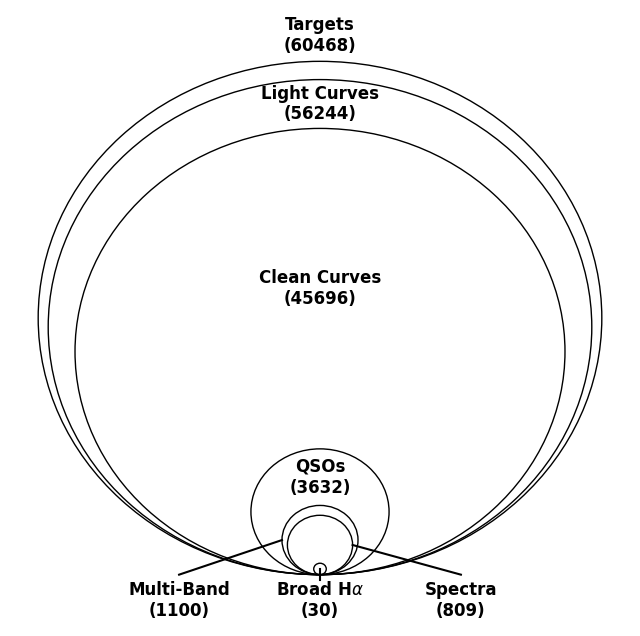}
    \caption{A summary of the successive subsets of galaxies analyzed in our experiment. Of the 60,468 objects in the initial sample, we were able to make light curves for 56,244. Only light curves for 45,696 had enough data points to be analyzed after cleaning. Our analysis resulted in 3632 objects with AGN-like variability in at least one band with 1100 variable in multiple bands. We collected spectra for 809 of the 1100 multi-band variable AGN candidates and found 30 with broad H$\alpha$ lines. In total, the number of multi-band AGNs accounts for just over 2.4 percent of the analyzed light curves.}
    \label{fig:sets}
\end{figure}

Any remaining galaxies with light curves that satisfy all of these criteria in multiple bands were selected as AGN candidates and had their spectra analyzed for emission features.
In total, we identified 5070 light curves that were selected by all our criteria corresponding to 3632 unique galaxies.
Of these, 1100 galaxies exhibited AGN-like variability in multiple bands, giving us our final list of AGN candidates.
Example light curves for our AGN candidates are shown in Figure \ref{fig:agn_cand}, and a breakdown of the successive subsets of galaxies is shown in Figure \ref{fig:sets}.

\section{Analysis of Variable AGN Candidates and Their Host Galaxies} \label{sec:spec}
We spectroscopically follow up our AGN candidates, all the objects whose light curves satisfy our AGN selection criteria in multiple bands.
For objects from the GAMA survey, we download their spectra utilizing GAMA's Single Object Viewer.
If there is more than one spectrum available for a single object, we download the spectrum with the ``IS\_BEST'' flag.
For objects from the NSA and/or SDSS, we downloaded their spectra using the astroquery package in Python, which downloads data from the SDSS Data Release 17 \citep{2022ApJS..259...35A}.

We were able to download spectra for 809 of our 1100 AGN candidates.
Thirty of the missing spectra were from the NSA, while the remainder were from the SDSS. 

\subsection{Spectral Line Fitting} \label{subsec:line_spec}
Various features of a galaxy's emission spectrum can be associated with properties of the AGN or central BH, such as BH mass \citep{2013ApJ...775..116R} or emission lines for BPT analysis.
To measure these, we use the software \textit{PyQSOFit} \citep{2018ascl.soft09008G, 2019ApJS..241...34S} version 2.0.0 to analyze the spectra of our AGN candidates.
This program takes an input spectrum and redshift and fits various components.
First, it decomposes the host galaxy and quasar components, using principal component analysis if necessary.
If either component is excessively negative, this decomposition is not applied.
Then, the continuum is fit using line-free windows, including a power-law, polynomial, and Balmer and \ion{Fe}{2} continua.
Finally, the program fits each line complex using broad and narrow Gaussian profiles.

We fit each object three times: once without any broad components, once with one broad H$\alpha$ component, once with two broad components.
These broad components are Gaussian distributions with widths corresponding to at least 500 km/s
and were fit within a window from 6450 to 6800 \AA.
If the inclusion of the broad components improved the outputted $\chi^2_{H\alpha}$ by at least twenty percent, then we kept the number of broad components corresponding to the best fit.
If more than one broad component was used, PyQSOFit gave a total effective total value for the output parameters.
After visually inspecting the the resulting fits, we were left with 30 objects with broad H$\alpha$ emission.

\subsection{Black Hole Mass Estimation} \label{subsec:BH_mass}
\begin{figure*}[htb]
    \centering
    \includegraphics[width=\textwidth]{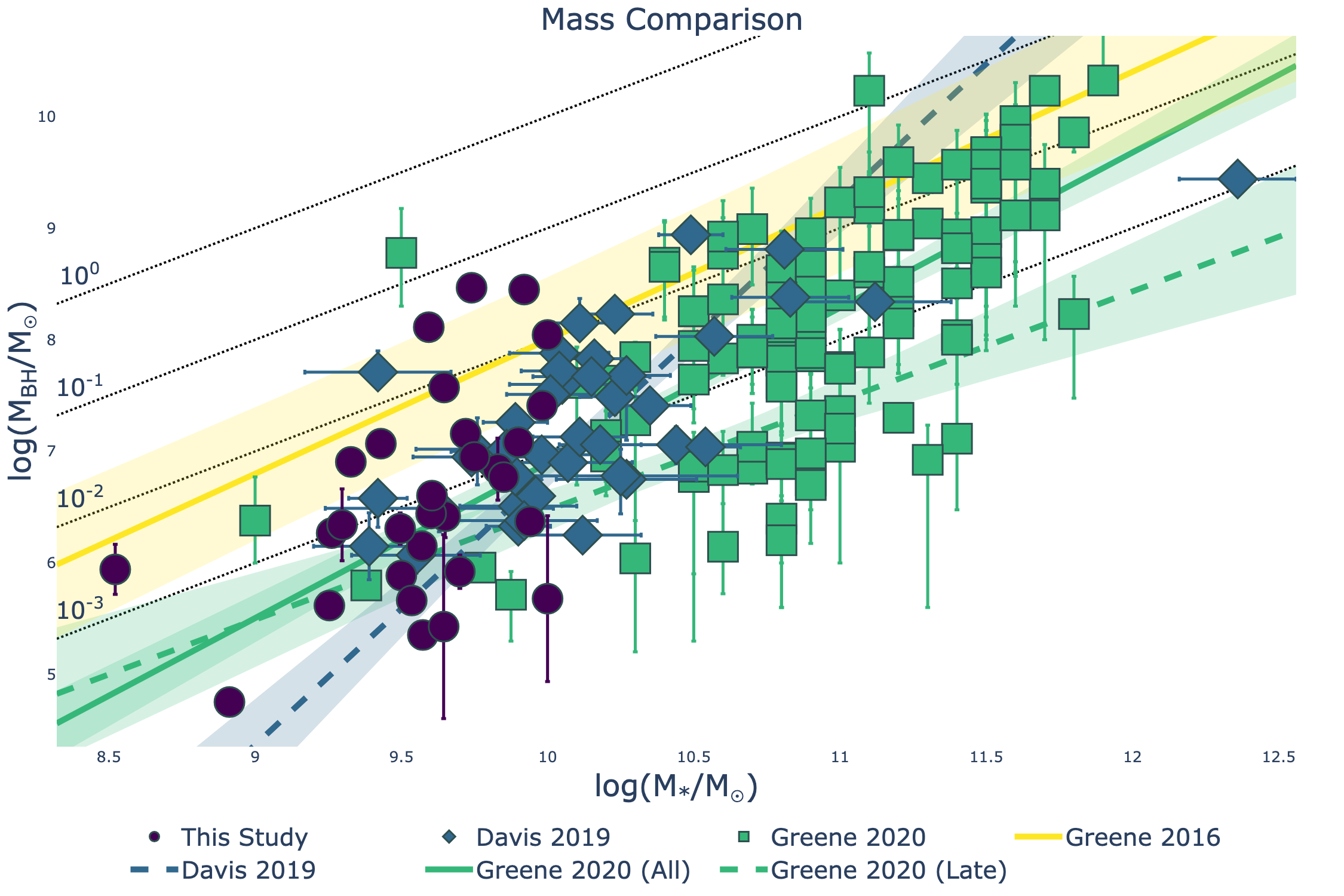}
    \caption{A comparison of BH mass calculated from the broad $\mathrm{H\alpha}$ emission versus the host galaxy stellar mass. The color and shape of each point correspond to the source of the data. For comparison, we show mass scaling relations from \citet{2016ApJ...826L..32G}, \citet{2019ApJ...873...85D}, and \citet{2020ARA&A..58..257G} (all galaxies (no limits) and late galaxies (no limits)). We show dashed lines to indicate fits that were performed on a sample of spiral or late type galaxies specifically. We include black dotted lines showing the proportion of the BH mass to host stellar mass ($M_{BH}/M_*$), labeled on the left.}
    \label{fig:mass_comp}
\end{figure*}

Understanding the mass of the central BH could help us gain insights on the nature of scaling relations in the low-mass range, as well as the potential for finding intermediate-mass BHs (IMBHs).
To calculate the BH mass, we use the broad component of the H$\alpha$ line where available with the following formula from \citet{2013ApJ...775..116R}:
\begin{equation}
    \begin{array}{c}
         \log{\left(\frac{M_{BH}}{M_\odot}\right)} = 0.47\log{\left(\frac{L_{H\alpha}}{10^{42} erg/s}\right)}\\
         + 2.06\log{\left(\frac{FWHM_{H\alpha}}{10^3 km/s}\right) + 6.57}
    \end{array}
\end{equation}
Assuming flat $\Lambda$CDM cosmology with $h=0.7$, we use the \textit{astropy} package to convert redshifts to distances so that we can calculate the luminosities from the resulting flux values.
We give the resulting values in Table \ref{tab:BPT} and plot the resulting BH masses against the host galaxy stellar mass and dampening timescale $\tau_{DRW}$ in Figures \ref{fig:mass_comp} and \ref{fig:mass_tau}, respectively.
We find BH masses ranging between $10^{4.76\pm0.01}$ and $10^{8.47\pm0.01}\,M_\odot$

\subsection{Dampening Timescale} \label{sub:timescale}
With the relatively expensive cost of spectroscopic follow-up, astronomers have begun to look for features from photometry or the time domain to measure AGN or BH properties, such as BH mass.
One such method involves correlating the properties of the DRW model with BH mass \citep{2021Sci...373..789B}.
We refit the light curves (in every band) of all objects with AGN-like in at least one band with the software \textit{taufit} \citep{2021Sci...373..789B}.
This program models the light curve as a Gaussian process with the ability to fit for both the length scale $\sigma_{DRW}$ and timescale $\tau_{DRW}$ associated with a damped random walk.
A fit is deemed valid only if the returned timescale is less than 20 percent of the total baseline, or $t_{base}/\tau_{DRW} \geq 5$ \citep{2017A&A...597A.128K}.
We give the results of the good fits for light curves with AGN-like variability in Table \ref{tab:phot_var}.
Overall, we found good fits for 2170 light curves corresponding to 976 of our 1100 AGN candidates.

\section{Discussion} \label{sec:discussion}
\subsection{Active Fraction}
\begin{figure*}[htb]
    \centering
    \includegraphics[width=\textwidth]{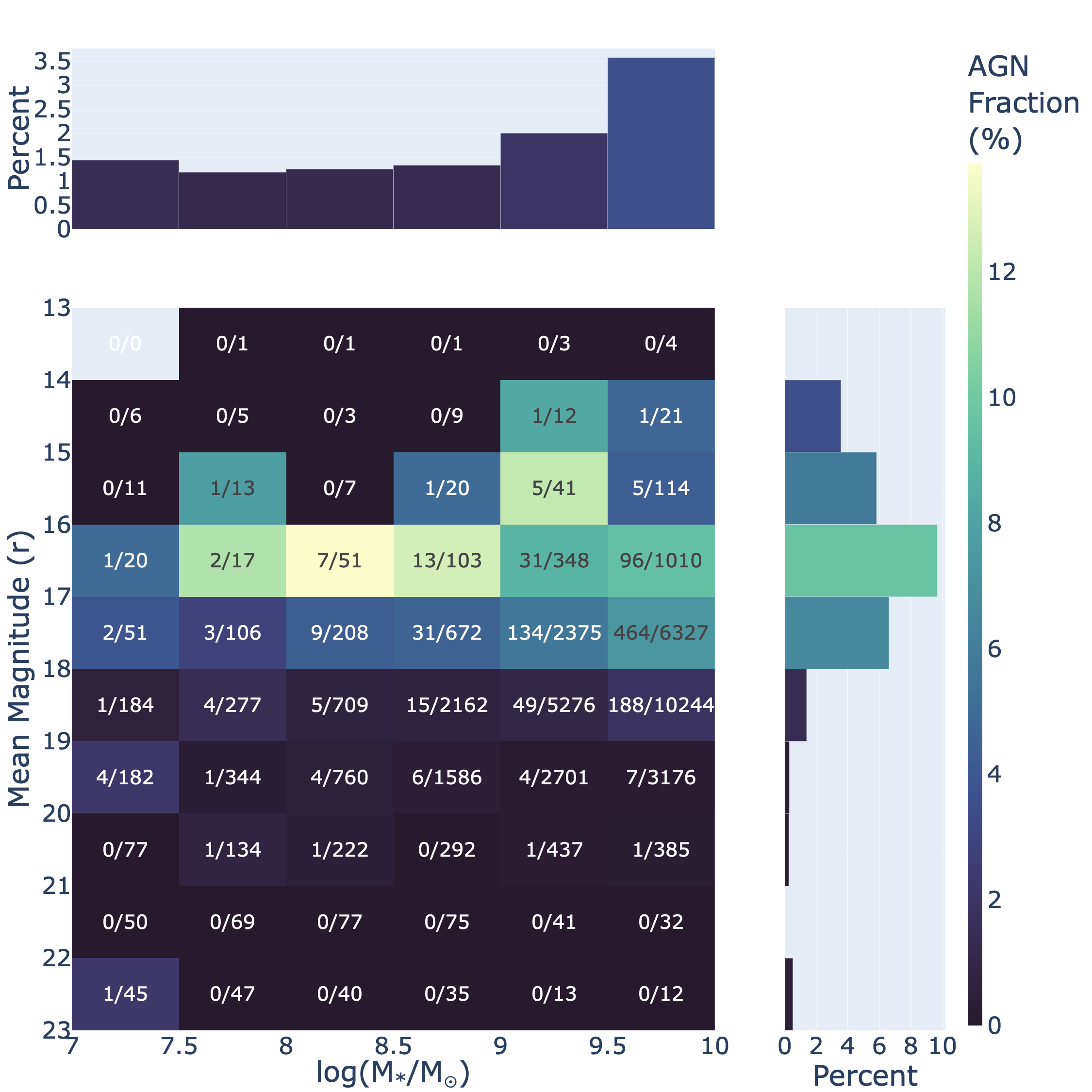}
    \caption{A heat map of the active fraction binned by host mass and mean r-magnitude with corresponding histograms above and to the right, respectively. We exclude data with a mean magnitude above 23, leaving out 650 target galaxies, none of which displayed AGN-like variability. The histogram above shows the active fraction (i.e. the percent of each mass bin that contains AGNs) prior to magnitude unbiasing, which increases with stellar mass, though perhaps nonlinearly. Similarly, the histogram in the right subplot shows the na\"ive active fraction as a function of r-magnitude.}
    \label{fig:corner}
\end{figure*}

The active fraction describes the portion of galaxies hosting an AGN, either in a given mass range or overall.
While this number depends heavily on the population sampled and the detection method(s) used, multiwavelength active fractions in dwarf galaxies typically range from 5\% to about 30\% \citep{2021ApJ...920..134P}.
This number is important to constrain because of it is fundamental to understanding the formation of IMBHs and SMBHs.
We calculate the active fraction $\mathcal{A}(M_*)$ as a function of host stellar mass by binning our previous results.
As shown in Figure \ref{fig:corner}, binning the galaxies by mass gives different magnitude distributions within each bin where smaller galaxies are dimmer on average.
As such, simply binning the galaxies by mass would introduce a magnitude bias since the active fraction would be calculated from increasingly dimmer populations.
This na{\"i}ve active fraction is shown in the top panel of Figure \ref{fig:corner}.
To then control for magnitude, we sample from distributions with similar brightness properties, following a procedure similar to that in \citet{2020ApJ...896...10B}
First, we split the (clean) galaxy sample into mass bins 0.5 dex in size.
Then, for the heaviest bin, we create a histogram of the resulting r-band magnitudes and fit a Gaussian curve, recording the center $\mu$ and standard deviation $\sigma$ of this fit.

\begin{figure*}
    \gridline{
        \fig{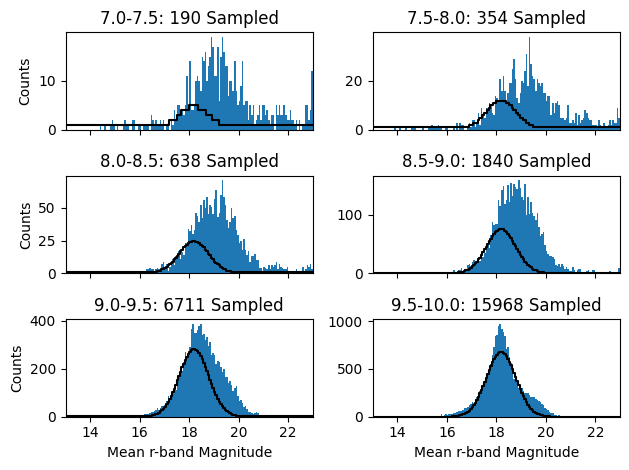}{0.5\textwidth}{(a) Magnitude Distributions Binned by Stellar Mass}
        \fig{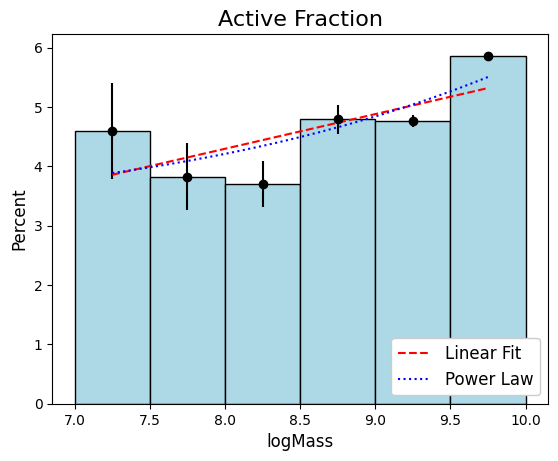}{0.5\textwidth}{(b) Magnitude-unbiased Active Fraction Binned by Stellar Mass}}
    \caption{Histograms of the r-band magnitude for the mass-binned galaxies (blue), a fitted Gaussian from which the data is sampled (black), and the resulting active fraction (right). Each subplot on the left is labeled by the logarithm of the host stellar mass range (i.e. "7.0-7.5" refers to a mass range of $7.0\leq\log{(M_*/M_\odot)}<7.5$, etc.) and the corresponding number of objects sampled in each iteration of the bootstrapping process. Note that the fitted Gaussians all share the same center and standard deviation across plots. On the right, we show a inverse-variance weighted linear fit of the data points (red dashed line), as well as a power law from \citet{2021ApJ...920..134P} that we fit to our data (blue dotted line).}
    \label{fig:AGN_frac}
\end{figure*}

We then fit a Gaussian to the magnitude histogram of the remaining mass bins, keeping $\mu$ and $\sigma$ fixed to the values determined in the heaviest bin.
We adjust each amplitude such that at least 80\% of the counts within the full-width-half-maximum (FWHM) region of each Gaussian were less than or equal to their corresponding histogram counts.
We chose to only focus on the FWHM region since the wings outside had many empty values.
The cutoff value of 80\% was chosen to maximize the overall number of galaxies sampled in each bin while allowing for some boxes to be undersampled. 
This method of Gaussian fitting ensures populations across mass bins have similar magnitude properties with a standardized amount of variety, and is shown in Figure \ref{fig:AGN_frac}, alongside the results of this calculation.

We sample from the magnitude histogram to match the fitted Gaussian and count the resulting number of active and inactive galaxies to calculate the active fraction.
We repeat this resampling 1000 times to get a distribution of the active fraction within each mass bin, from which we calculate get uncertainties.
We see active fractions in each bin between three and six percent, with an unsteady upward trend with stellar mass.
We fit the active fraction to a line as well as a power law of the form $\mathcal{A}\sim\log{(M_*/M_\odot)}^{4.5}$ from \citet{2021ApJ...920..134P}.
For both cases, the fitted was weighted by the inverse variance of the data.
The best fits (given as a percentage) were:

\begin{eqnarray}
    \mathcal{A}(\%) = \alpha_{lin}\log{\left(M_*/ 10^9M_\odot\right) + \beta_{lin}} \\
    \mathcal{A}(\%) = \alpha_{Pac}[\log{\left(M_*/M_\odot\right)]^{4.5} + \beta_{Pac}}
\end{eqnarray}
where $\alpha_{lin} = 0.6\pm0.3$, $\beta_{lin} = 4.9\pm0.1$, $\alpha_{Pac} = (7.7\pm2.7)\times10^{-5}$, and $\beta_{Pac} = 3.3\pm0.6$.

\subsection{Black Hole Mass and Scaling Relations}\label{subsec:scaling}
We use broad H$\alpha$ emission from our AGN candidates to calculate the mass from their central BHs, given in Table \ref{tab:BPT}.
Depending upon the definition, between one and ten of our objects had BH masses in the intermediate-mass range ($2\lesssim\log{M_{BH}/M_\odot}\lesssim6$).
We investigate these further in Section \ref{subsec:cases}, as well as galaxy 38344 which was overmassive relative to previous scaling relations.

We plot the BH masses against the host stellar masses in Figure \ref{fig:mass_comp} and compare our results with mass scaling relations previously found in \citet{2016ApJ...826L..32G}, \citet{2019ApJ...873...85D}, and \citet{2020ARA&A..58..257G} (all galaxies and late galaxies, in both cases without upper limit estimates for BH masses).
We also plot data from the latter two papers (and sources therein) to investigate whether the slopes of these relations remain consistent in the higher mass range.

As with the previous data, there was a large amount of scatter between host and BH masses.
However, we still observed a general upward trend, with the majority of data fitting between the relations from \citet{2020ARA&A..58..257G} (late) and \citet{2016ApJ...826L..32G}.
We quantify the goodness of these fits on our data and the total of all three samples in Table \ref{tab:fit_stats}.
The scaling relation from \citet{2020ARA&A..58..257G} for all galaxies is the best fit for both groups, having the smallest mean squared error and largest coefficient of determination ($R^2$) in both cases.
The data from \citet{2019ApJ...873...85D} also fits within these trends.
Although their data in isolation found a steeper relation, we do not see evidence of this relation continuing below $M_*\sim10^{9.5}M_\odot$ or above $M_*\sim10^{11}M_\odot$.
Instead, the sum of data over this extended mass range appears fairly consistent in slope.
This consistency in the slope towards the low-mass range could be indicative of a history of gravitational runaway events since the other models predict a shallower or even flat relation \citep{2020ARA&A..58..257G}.

\begin{deluxetable}{c c c c}[htb]
    \tablecolumns{4}
    \tablecaption{Goodness of Linear Fits\label{tab:fit_stats}}
    \tablehead{\colhead{Data} & \colhead{Line} & \colhead{MSE} & \colhead{$R^2$}}
    \startdata
        \multirow{4}{*}{\rotatebox[origin=c]{90}{This Study}}
        & Greene 2016        & 1.476 & -0.777 \\
        & Davis 2019         & 1.384 & -0.667 \\
        & Greene 2020 (All)  & 0.719 & 0.134 \\
        & Greene 2020 (Late) & 0.938 & -0.129 \\
        \hline
        \multirow{4}{*}{\rotatebox[origin=c]{90}{All Data}}
        & Greene 2016        & 1.435 & -0.089 \\
        & Davis 2019         & 1.507 & -0.144 \\
        & Greene 2020 (All)  & 0.592 & 0.551 \\
        & Greene 2020 (Late) & 1.114 & 0.154 \\
    \enddata
    \tablecomments{Comparison of the goodness of fit for each scaling relation on our data and the total data. For each, we report the mean squared error and the coefficient of determination ($R^2$).}
\end{deluxetable}

\subsection{Timescale Relations}\label{subsec:timescale}
\begin{figure}[htb]
    \centering
    \includegraphics[width=0.48\textwidth]{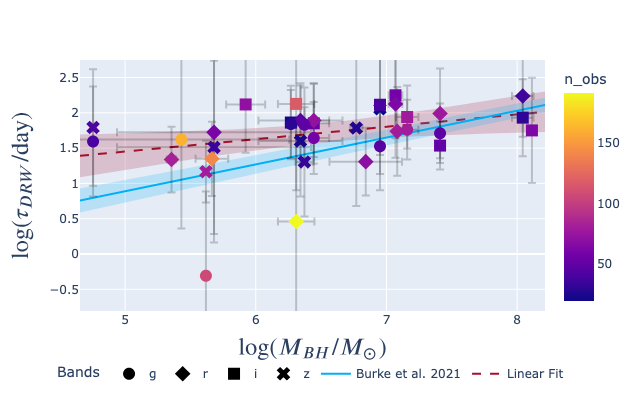}
    \caption{A plot of the DRW dampening timescale $\tau_{DRW}$ versus the BH mass. The shape of each point corresponds to the band analyzed, while the color corresponds to the number observations in each light curve. We show the relation found in \citet{2021Sci...373..789B} (solid blue) and a linear fit to our data using all available bands (dotted red); we choose not to show the individual band fits since they are relatively weak.}
    \label{fig:mass_tau}
\end{figure}

To investigate the relation between DRW dampening timescale and BH mass, we create a logarithmic plot of $\tau_{DRW}$ versus $\mathrm{M_{BH}}$ (calculated from broad $H\alpha$ emission).
Although we only had a small number of data points, we fit our results to a linear equation of the form:
\begin{equation}
    \log{\frac{\tau_{DRW}}{day}} = \alpha\log{\left(\frac{M_{BH}}{10^{\gamma}M_\odot}\right)}+ \beta.
\end{equation}
While \citet{2021Sci...373..789B} uses $\gamma=8$, we chose $\gamma=6.5$ as it minimized uncertainty for the intercept (without affecting the slope).
First, we fit a line to all data from all bands, then per individual band.
We once again bootstrap the uncertainties from the data to get errors for the slope and intercept.
The overall linear relation is plotted in Figure \ref{fig:mass_tau} and the line parameters are shown in Table \ref{tab:tau_fit}.
Although our overall linear fit shows a shallower slope than \citet{2021Sci...373..789B}, even within 1$\sigma$, it is more consistent with the slope of 0.21 previously reported in \citet{MacLeod_2010}.
We also have larger uncertainties for both the slope and intercept, likely as the result of the small number of data points available (37 data points belonging to 21 galaxies, compared to the 67 data from 67 galaxies used by \citet{2021Sci...373..789B}).

\begin{deluxetable}{c C R C}[htb]
    \tablecolumns{4}
    \tablecaption{Linear Relation Parameters\label{tab:tau_fit}}
    \tablehead{\colhead{Band} & \colhead{$\mathrm{N_{obs}}$} & \colhead{Slope ($\alpha$)} & \colhead{Intercept ($\beta$)}}
    \startdata
        Total & 37 &  0.18\pm0.15 & 1.71\pm0.14  \\
        g     &  9 &  0.23\pm0.32 & 1.54\pm0.21 \\
        r     & 11 &  0.29\pm0.31 & 1.62\pm0.21 \\
        i     & 10 & -0.10\pm0.26 & 1.98\pm0.27 \\
        z     &  7 &  0.14\pm0.46 & 1.66\pm0.42 \\
        Burke & 67 &  0.38\pm0.04 & 2.03\pm0.10 \\
    \enddata
    \tablecomments{Comparison of the linear fit parameters for relations of the form $\log{\frac{\tau_{DRW}}{day}} = \alpha\log{\left(\frac{M_{BH}}{10^{\gamma}M_\odot}\right)}+ \beta$. We include fits for the total data as well as per band and the fit found in \citet{2021Sci...373..789B}. Note that for our linear fits, we use $\gamma=6.5$, while the previous relation used $\gamma=8$. }
\end{deluxetable}

\subsection{BPT Diagnostic}\label{subsec:BPT}
\begin{figure*}
    \gridline{
        \fig{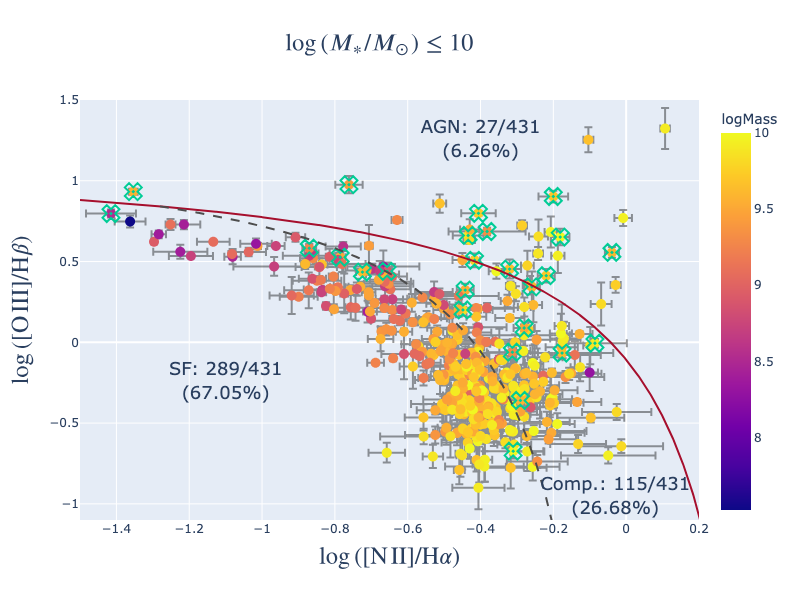}{0.5\textwidth}{(a)}
        \fig{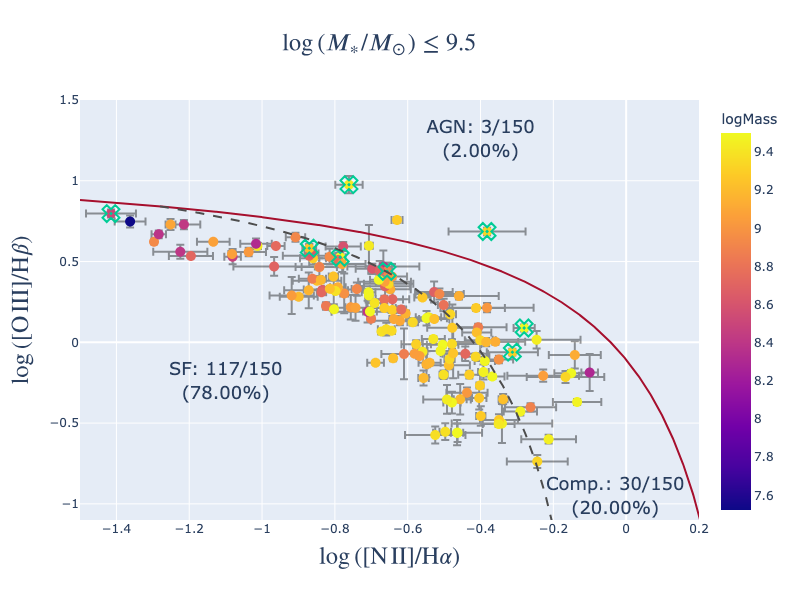}{0.5\textwidth}{(b)}}
    \caption{BPT diagrams of our AGN candidates using the fluxes from PyQSOFit.
    Each data point is shown with 1$\sigma$ uncertainties along both axes and is colored by the logarithm of its host galaxy's stellar mass.
    Plot (a) shows the entire sample of AGNs for which $\mathrm{SNR}\geq3$ for all four emission lines, while plot (b) shows the subset of those galaxies with stellar mass $M_*\leq10^{9.5}\,M_\odot$. In both plots, we highlight objects for which we detected broad H$\alpha$ emission with open green `x's.}
    \label{fig:BPT}
\end{figure*}

\begin{deluxetable*}{C C C C C c C C C}[htb]
    \tablecolumns{9}
    \tablecaption{Spectral Properties of AGN Candidates\label{tab:BPT}}
    \tablehead{\colhead{$\#_{gal}$} & \colhead{$F_{[N\,II]}$} & \colhead{$F_{H\beta,n}$} & \colhead{$F_{[O\,III]}$} & \colhead{$F_{H\alpha,n}$} & \colhead{BPT Class} & \colhead{Min. SNR} & \colhead{$L_{H\alpha,br}$} & \colhead{$\log{\frac{M_{BH}}{M_\odot}}$}}
    \startdata
        259 & 0.449\pm0.105 & 0.143\pm0.205 & 0.0117\pm1.0790 & 0.0301\pm0.1062 & AGN & 0.0108 & \ldots & \ldots \\
        263 & 11.4\pm11.8 & 9.81\pm11.20 & 5.68\pm8.13 & 1.04\pm1.14 & AGN & 0.698 & \ldots & \ldots \\
        270 & 5.77\pm4.78 & 0.157\pm4.302 & 11.4\pm10.4 & 0.291\pm1.444 & AGN & 0.0365 & \ldots & \ldots \\
        364 & 28.2\pm4.5 & 39.9\pm14.7 & 17.8\pm8.6 & 10.1\pm2.7 & AGN & 2.08 & \ldots & \ldots \\
        891 & 7.74\pm3.09 & 71.1\pm5.9 & 25.4\pm7.4 & 6.80\pm2.37 & Comp. & 2.50 & \ldots & \ldots \\
        1508 & 74.0\pm3.5 & 93.5\pm12.6 & 109\pm12 & 301\pm6 & SF & 7.43 & \ldots & \ldots \\
        2212 & 126\pm3 & 82.3\pm4.1 & 76.6\pm4.2 & 472\pm6 & SF & 18.4 & \ldots & \ldots \\
        2379 & \ldots & 12.9\pm1.2 & 1.32\pm0.43 & \ldots & \ldots & \ldots & \ldots & \ldots \\
        3113 & 41.3\pm7.0 & 0.433\pm0.402 & 12.3\pm9.6 & 11.5\pm4.6 & AGN & 1.08 & \ldots & \ldots \\
        3120 & 2.20\pm2.89 & 0.0584\pm0.1447 & 19.1\pm7.1 & 0.254\pm0.286 & AGN & 0.404 & \ldots & \ldots \\
        3155 & 157\pm5 & 61.4\pm5.8 & 23.5\pm4.5 & 379\pm11 & SF & 5.19 & \ldots & \ldots \\
        3170 & 221\pm5 & 154\pm7 & 182\pm7 & 664\pm15 & SF & 23.1 & \ldots & \ldots \\
        3187 & 35.9\pm3.3 & 3.97\pm4.83 & 9.85\pm9.59 & 60.2\pm4.7 & AGN & 0.822 & \ldots & \ldots \\
        3190 & 0.335\pm0.684 & 24.7\pm16.3 & 8.91\pm4.85 & 0.0372\pm0.1188 & AGN & 0.313 & \ldots & \ldots \\
        3215 & 26.0\pm3.9 & 0.0882\pm0.2442 & 12.1\pm6.7 & 31.1\pm4.7 & AGN & 0.361 & \ldots & \ldots \\
        \vdots & \vdots & \vdots & \vdots & \vdots & \vdots & \vdots & \vdots & \vdots \\
    \enddata
    \tablecomments{Table containing the fluxes necessary for BPT analysis (reported in units of $10^{-17}erg/s/cm^2$) as well as the results of the diagnostic and the minimum SNR of the fluxes (we suggest only using data for which this value is at least 3).}
\end{deluxetable*}
The BPT diagram \citep{Baldwin_1981} is a logarithmic plot of lines flux ratios with empirical cuts that can characterize the source of high energy emission as coming from star formation, AGN activity, or signatures of both \citep{2003MNRAS.346.1055K, 2006MNRAS.372..961K}.
From the fluxes measured with \textit{PyQSOFit}, we use the BPT diagnostic to characterize the emission from our AGN candidates, displayed in Table \ref{tab:BPT} (alongside the results of the BH mass calculation).
To ensure we use only good fits for spectroscopic fluxes, we select objects that have a signal-to-noise ratio of at least 3 for all four emission lines used in the BPT diagnostic ([\ion{O}{3}], H$\beta$, [\ion{N}{2}], and H$\alpha$), resulting in 431 galaxies of the 809 AGN candidates with available spectra, shown in Figure \ref{fig:BPT}.
Of these, 289 (67.1\%) fall in the star-forming region, 115 (26.7\%) in the Composite region, and 27 (6.3\%) are classified as AGNs, despite all objects demonstrating AGN-like photometric variability in multiple bands.
This was expected since the BPT diagnostic is already known to miss low-mass galaxies since their data points on the plot tend down and to the left compared to more massive AGNs \citep{10.1111/j.1365-2966.2006.10812.x, Cann_2019}.
Of particular note is the distribution of galaxies with Broad H$\alpha$ emission Lines (BL); of the 30 BL AGNs in our sample, 25 had strong enough emission lines for BPT analysis.
Of those 25, 5 are star-forming, 9 are composite, and 11 are AGNs.

Selecting for only dwarf galaxies ($\log(M_*/M_\odot)\leq9.5$) leaves 150 galaxies: 117 star-forming (78.0\%), 30 composite (20.0\%) , and 3 AGNs (2.0\%).
We will examine these three low-mass BPT AGNs more in Section \ref{subsec:cases}.
If we once again focus on (low-mass) BL galaxies, 3 are star-forming, 3 are composite, and 2 are in the AGN region of the diagram.

\subsection{Comparison to Known Dwarf AGN}\label{subsec:known}
\begin{figure}[htb]
    \centering
    \includegraphics[width=0.48\textwidth]{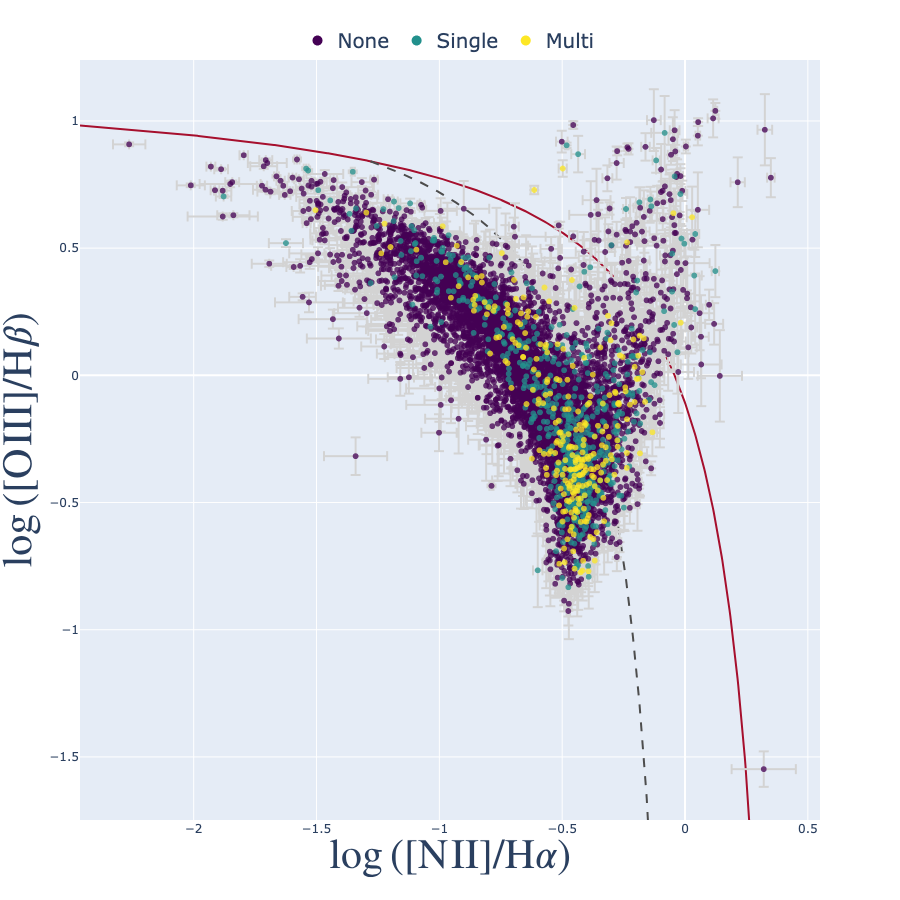}
    \caption{A BPT diagram using data from the NSA v0 for objects in our study that had sufficiently strong enough lines for this analysis. The color of each point corresponds to whether we found AGN-like variability in zero (purple), one (teal), or multiple bands (yellow).}
    \label{fig:NSA_BPT}
\end{figure}

We sought to compare our AGN candidates against previously found AGNs, both overall and in the low-mass regime.
We not only compared the list of objects directly, looking for any of our candidates that had been previously identified as AGNs, but we also compared the properties of the different sources for AGNs.
To make this comparison, we used two sources:
\begin{enumerate}
    \item The Active Dwarf Galaxy Database \citep[ADGD]{2024ApJ...971...68W}: a complete catalog of the known bona fide AGNs in dwarf galaxies to date
    \item The NSA v0
\end{enumerate}
From the NSA, we performed a BPT analysis on any galaxy that had a SNR of at least 3 for all emission lines involved.
We cross-matched these objects with our variability analysis results within two arcseconds and found 8107 objects in common.
The majority of these sources (6547) were in the SF region of the NSA BPT diagram and had no AGN-like variability, as seen in Figure \ref{fig:NSA_BPT} and Table \ref{tab:NSA_BPT}.
There were 216 objects in the SF region that demonstrated AGN-like variability in multiple bands, while 135 objects in the AGN region demonstrated no AGN-like variability.
Only 6 of our AGN candidates (which exhibited AGN-like variability in multiple bands) were in the AGN region of the BPT diagram, one of which had a host stellar mass less than $10^{9.5}\,M_\odot$.
We will examine these six galaxies in Section \ref{subsec:cases}.
Overall, we found 272 objects with strong emission lines in the NSA that exhibited AGN-like variability in multiple bands.

\begin{deluxetable}{c c | C C C}[htb]
    \tablecolumns{5}
    \tablecaption{NSA BPT Results for Objects in Our Study \label{tab:NSA_BPT}}
    \tablehead{& & \multicolumn{3}{c}{NSA BPT Results} \\
    \colhead{Mass} & Variability & \colhead{Star-Forming} & \colhead{Composite} & \colhead{AGN}}
    \startdata
        \multirow{3}{*}{$\leq10$}
        & None     & 6547 & 549 & 135 \\
        & Single   & 485  & 95  & 24  \\
        & Multiple & 216  & 50  & 6   \\
        \hline
        \multirow{3}{*}{$\leq9.5$}
        & None     & 5231 & 131 & 21 \\
        & Single   & 301  & 15  & 2  \\
        & Multiple & 111  & 4   & 1  \\
    \enddata
    \tablecomments{A tabular representation of the BPT diagram shown in Figure \ref{fig:NSA_BPT}, which plots objects from our study that had strong emission lines in the NSA v0. The rows indicate how many objects were found to be variable in no bands, one band, or multiple bands (our AGN candidates; note that only 272 of our 1100 AGN candidates had corresponding data in the NSA v0). The columns indicate how many objects were found in the Star-forming, Composite, and AGN regions of the BPT diagram, respectively. The top half of the table displays the data corresponding to our entire sample, while the bottom half displays the numbers corresponding to galaxies with $\log{(M_*/M_\odot)}\leq9.5$.}
\end{deluxetable}

Next, we compared our results to the ADGD, once again cross-matching within two arcseconds.
There were 98 objects in common: 8 with AGN-like variability in multiple bands, 12 with AGN-like variability in one band, and 78 with no AGN-like variability.
Interestingly, 22 of those 78 were originally selected for the variability in \citet{2020ApJ...896...10B} and 1 from \citet{2022ApJ...933...37W}.
Many of the other galaxies were selected using BPT diagnostics or mid-infrared color cuts, though four showed X-ray emission, five showed broad line emission, and eleven showed \ion{He}{2} or coronal line emission.

There were 55 galaxies in this study that were also NSA BPT galaxies and in the ADGD.
Two of these were BPT AGNs: one with AGN-like variability in multiple bands, and one with none.
Excluding the 6 NSA BPT AGNs and the 8 AGNs in the ADGD (with 1 object in both), our analysis resulted in 1087 new AGN candidates out of our 1100.

\begin{figure*}
    \gridline{
        \fig{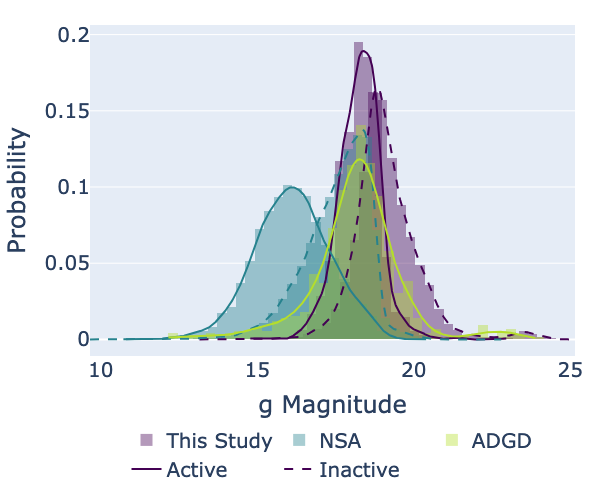}{0.32\textwidth}{}
        \fig{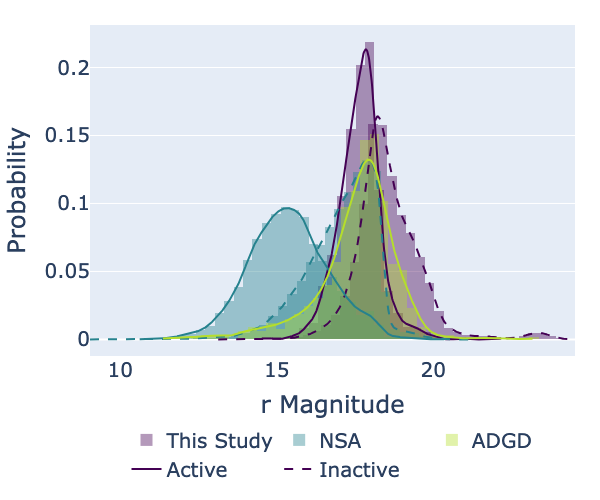}{0.32\textwidth}{}
        \fig{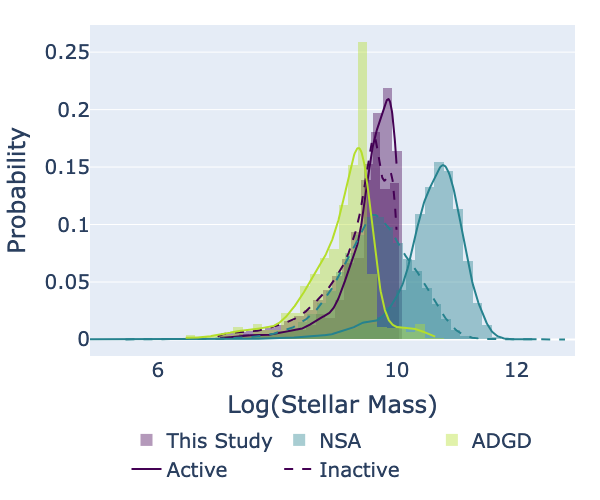}{0.32\textwidth}{}}
    \gridline{
        \fig{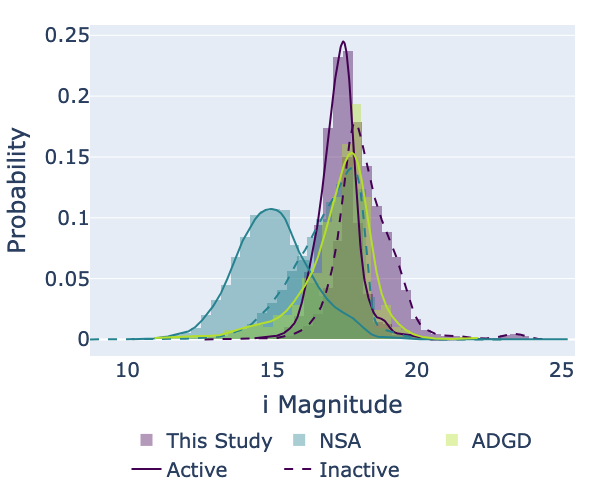}{0.32\textwidth}{}
        \fig{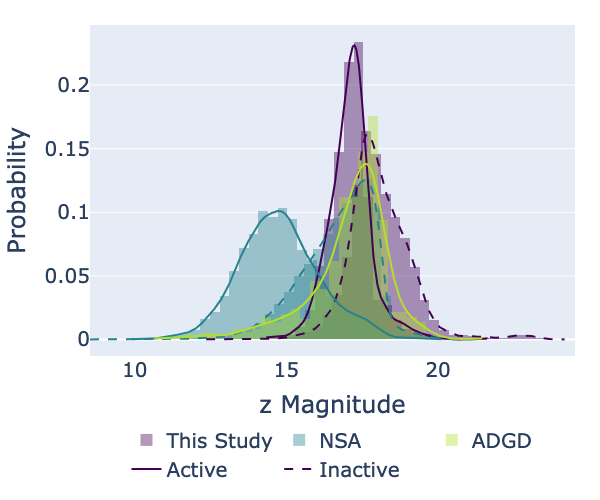}{0.32\textwidth}{}
        \fig{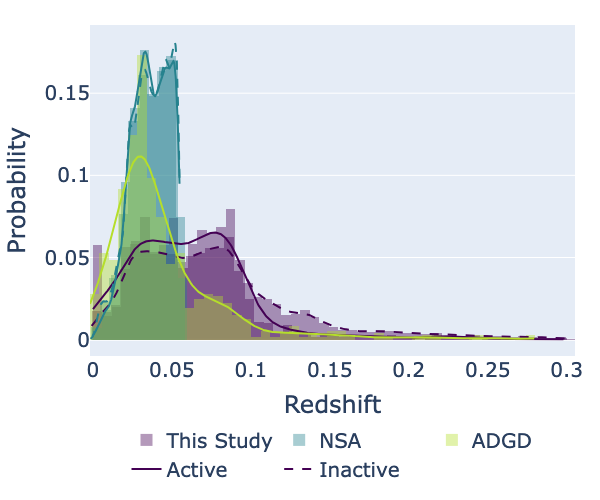}{0.32\textwidth}{}}
    \caption{Comparison of relative distributions of galaxy properties: mean magnitudes, stellar mass, and redshift. Galaxies from different AGN sources are indicated by color and separated by whether they are active (solid line) or not (dashed line). As a reminder, all of the objects in the ADGD are active by definition.}
    \label{fig:agn_diff}
\end{figure*}

\begin{figure}[htb]
    \centering
    \includegraphics[width=0.48\textwidth]{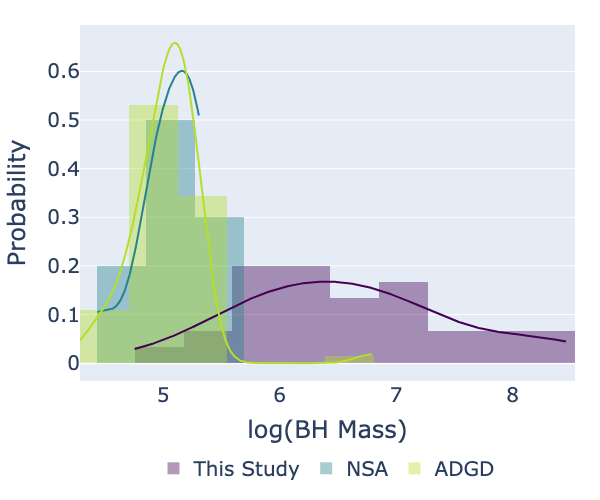}
    \caption{Relative distributions of the BH masses colored by their source. It is worth noting the small number of BH mass calculations available for active galaxies from  each source: 30 from this study, 10 from the NSA, and 64 from the ADGD.}
    \label{fig:agn_diff_MBH}
\end{figure}

\begin{figure*}[htb]
    \centering
    \includegraphics[width=\textwidth]{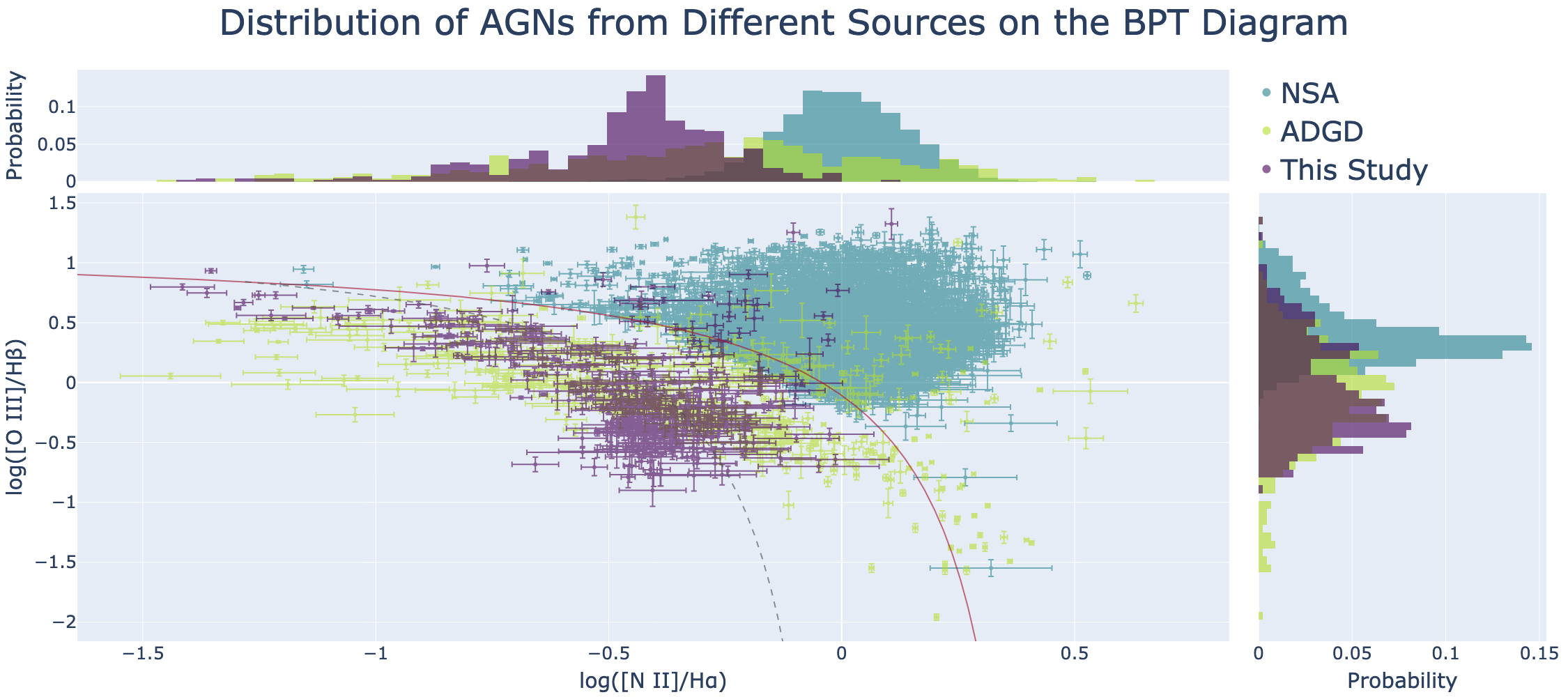}
    \caption{A comparison of the distribution of AGNs on the BPT diagram using data from three sources, as well as their distributions along each axis.}
    \label{fig:agn_diff_BPT}
\end{figure*}

After examining the overlapping membership of these sets, we then compare the properties of the galaxies from each AGN source, shown in Figures \ref{fig:agn_diff}, \ref{fig:agn_diff_MBH}, and \ref{fig:agn_diff_BPT}.
In Figure \ref{fig:agn_diff}, we also compare the properties of active versus inactive galaxies from each data set.
Across all bands, the galaxies from our study are dimmer on average than those from the NSA.
In fact, the active and inactive galaxies from our study and galaxies from the ADGD all had relatively similar distributions in magnitude to the inactive galaxies from the NSA.
This isn't surprising since we and \citet{2024ApJ...971...68W} specifically probe lower mass galaxies, more similar to the inactive NSA sample.
For each band shown, the population of active galaxies has a brighter distribution than the inactive galaxies from the same source.
However, the difference between the mean magnitude of active and inactive bands is larger for the NSA galaxies than those from our study by about a factor of two for every band.
For stellar host mass, there is not much difference between the active and inactive populations from our study, while active galaxies in the NSA are nearly an order of magnitude more massive than their inactive counterparts on average.
Additionally, the distributions for redshift seem to be consistent within each source, regardless of AGN status.

In Figure \ref{fig:agn_diff_MBH}, we compare the BH masses from the active galaxies from each source.
Although there are no objects with mass estimates from our analysis and one of the other sources we can use for direct comparison, the BH masses from this paper have a heavier and wider distribution than the NSA and ADGD.

Finally, in Figure \ref{fig:agn_diff_BPT}, we plot the three AGN data sets on the BPT diagram, as well as histograms for each axis.
As a reminder, BPT diagnostics were used to determine AGN candidacy for the NSA sample, so they are all in the AGN region by definition.
The majority (94\%) of AGN candidates from our study land in the star-forming and composite regions.
This is also true of the ADGD, but with a larger portion in the AGN region (nearly one fifth).
Along the x-axis, AGN candidates from our study have a distinctly lower distribution than those from the NSA, while objects in the ADGD have a much broader spread that encompasses the range of both other sets.
On the y-axis, the AGNs from our study and the ADGD have similar distributions, while the NSA AGNs have higher values on average.

\subsection{Case Studies and IMBH Candidates}\label{subsec:cases}
\begin{figure*}[htb]
    \gridline{
        \fig{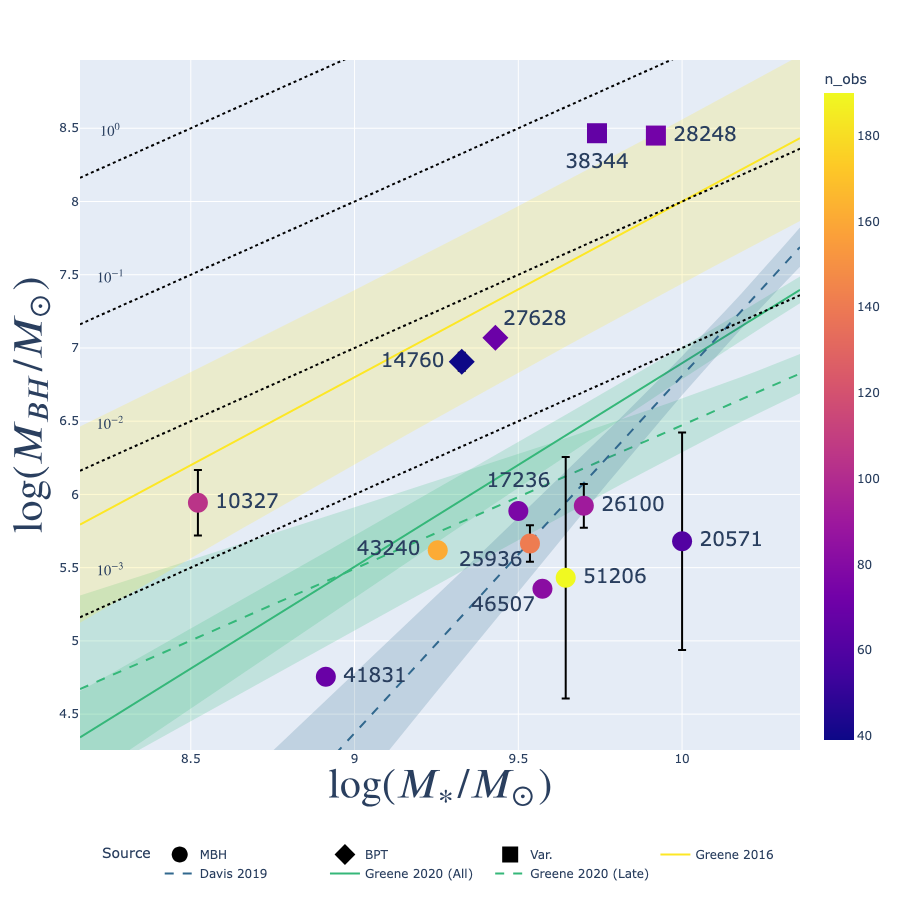}{0.5\textwidth}{(a) Comparison of the BH mass calculated from the broad H$\alpha$ emission versus the host galaxy stellar mass. In this plot, BPT AGNs are depicted with circles, AGN candidates with notable BH mass are diamonds, and AGN candidates with notable variability properties are shown as squares.}
        \fig{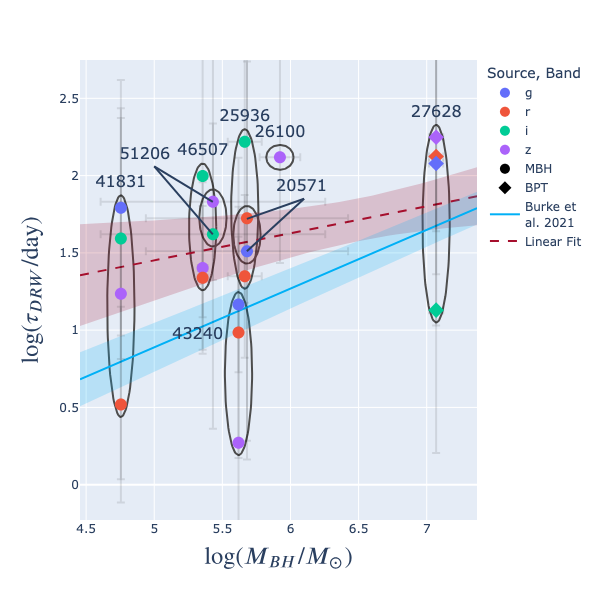}{0.5\textwidth}{(b) Comparison of the DRW dampening timescale $\tau_{DRW}$ versus the BH mass calculated from broad H$\alpha$ emission. In this plot, AGN candidates with notable BH mass are depicted with circles and BPT AGNs are shown as diamonds, while colors indicate bands.}}
    \caption{Recreations of Figures \ref{fig:mass_comp} and \ref{fig:mass_tau}, respectively, focusing on the individually selected case studies. In both plots, we show previously found relations. For plot (a), we also show lines of constant proportionality as dotted lines. In plot (b), we show our fitted line and group data from the same objects with vertical ellipses with different bands indicated by their color. All galaxies are labeled by their assigned ID numbers with marker shape indicating the reason for inclusion in this section.}
    \label{fig:case_mass}
\end{figure*}

\begin{figure}[htb]
    \centering
    \includegraphics[width=0.48\textwidth]{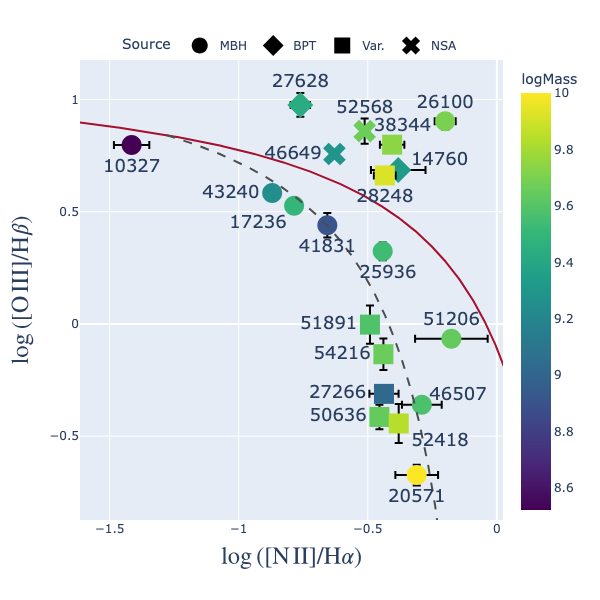}
    \caption{A recreation of Figure \ref{fig:BPT} focusing on our case studies. All galaxies are labeled by their assigned ID numbers and colored by their stellar mass. BPT AGNs are depicted with circles, AGN candidates with notable BH mass are rhombuses, AGN candidates with notable variability properties are squares, and BPT AGNs from the NSA are shown are `x's.}
    \label{fig:case_BPT}
\end{figure}

Here, we examine the properties of particularly notable objects from our results.
We plot the mass scaling and timescale relations for our case studies in Figure \ref{fig:case_mass}, as well as a BPT diagram in Figure \ref{fig:case_BPT}.
In each of these plots, all data are labeled by their ID numbers and the marker shapes correspond to the notable property that lead to their inclusion in this section.
Note that objects 38344 and 46649 were selected as case studies for multiple reasons, though only one is shown for each.
In Figure \ref{fig:case_mass}(b), we used color to indicate the calculations made using different bands to demonstrate that large scatter is present in every individual band.

Light curves for all objects and the spectra for the 9 IMBH candidates are shown in Appendix \ref{sec:app}.

\subsubsection{Notable Black Hole Masses}
First, we examine objects whose BH masses were notable.
We found nine potential IMBH candidates and one object with a relatively overmassive BH: 10327, 17236, 20571, 25936, 26100, 41831, 43240, 46507, 51206, and 38344 respectively.
Object 38344 had a BH mass of $10^{8.5}$, nearly 10\% of its host's stellar mass, placing it above all the previously found mass relations \citep{2016ApJ...826L..32G, 2019ApJ...873...85D, 2020ARA&A..58..257G}.
The remaining objects all have BH masses $M_{BH}\leq10^6M_\odot$, as low as $M_{BH}=10^{4.76\pm0.01}M_\odot$ for galaxy 41831.

As shown Figure \ref{fig:case_mass}(a), nearly all these BH masses lie near or below the scaling relation for late type galaxies from \citet{2020ARA&A..58..257G} with two exceptions.
While galaxy 10327 has a BH mass of $10^{5.94\pm0.22}M_\odot$, its host's low stellar mass of $10^{8.52}M_\odot$ places it above the both relations from \citet{2020ARA&A..58..257G}, but within 1$\sigma$ of the relation found in \citet{2016ApJ...826L..32G}.
Galaxy 38344 was notable for being overmassive with $M_{BH}=10^{5.94\pm0.22}M_\odot$ and $M_*=10^{8.52}M_\odot$, lying more than 1$\sigma$ above all relations.

Seven of these ten objects had a long enough baseline to calculate the dampening timescale, shown in Figure \ref{fig:case_mass}(b).
Nearly all of the measurements for $\tau_{DRW}$ lie well above the relation found in \citet{2021Sci...373..789B}, with r-band data from galaxy 41831 falling below.
For object 43240, calculations made from the r and z bands align well with the relation from \citet{2021Sci...373..789B}, but the $\tau_{DRW}$ derived from the i band is nearly an order of magnitude below.
With the exception of 43240, these data seem to support the flatter slope found in our paper.

The spectra of all these galaxies showed the necessary emission lines for BPT analysis.
Of these ten objects, two landed in the AGN region (26100 and 38344) and  four fell in the composite region (25936, 41831, 46507, and 51206).
The remaining four galaxies (10327, 17236, 20571, 43240) were in the star-forming region, though they all lie near the empirical cut-off.

We re-examined the spectral fits used to calculate the BH masses.
If we narrow down our objects to those for which $\chi_{\nu,H\alpha}^2\leq2$, then that leaves galaxies 17236, 20571, 25936, 26100, and 46507, whose BH and host masses are shown in Table \ref{tab:imbh}.
It is worth noting that for two of these (20571 and 26100), their BH masses lie within 1$\sigma$ of the $10^6M_\odot$ cutoff, potentially placing them outside the intermediate-mass range for BHs.
Interestingly, the hosts of all five of the reduced IMBH candidates have stellar masses between $10^{9.5}$ and $10^{10}M_\odot$, in the upper portion of our low-mass range for galaxies.
This could be related to the relatively weak spectra of the smaller galaxies.

\begin{deluxetable}{C c C c}[htb]
    \tablecolumns{4}
    \tablecaption{Intermediate-Mass Black Hole Candidates\label{tab:imbh}}
    \tablehead{\colhead{Object ID} & \colhead{$\log{M_*/M_\odot}$} & \colhead{$\log{M_{BH}/M_\odot}$} & \colhead{$\chi_{\nu,H\alpha}^2$}}
    \startdata
        17236^\dag &  9.50 & 5.89\pm0.02 & 1.98 \\
        20571 & 10.00 & 5.68\pm0.74 & 1.48 \\
        25936^\dag &  9.54 & 5.66\pm0.12 & 0.98 \\
        26100 &  9.70 & 5.92\pm0.15 & 1.18 \\
        46507^\dag &  9.57 & 5.36\pm0.04 & 1.35 \\
    \enddata
    \tablecomments{Host stellar mass and BH mass for the six objects in our study that had strong broad H$\alpha$ emission ($\chi_{\nu,H\alpha}^2\leq2$) corresponding to nasses in the intermediate-mass range ($M_{BH}\leq10^6M_\odot$). Objects marked with a dagger ($\dagger$) are more than 1$\sigma$ below the upper limit of $10^6M_\odot$.}
\end{deluxetable}

\subsubsection{Notable Variability Properties}
Next, we look at the objects with notable variability properties.
Of the 39 galaxies that showed AGN-like variability in all four of the \textit{griz} bands, we picked 10 with exceptionally high QSO significance ($\sigma_{QSO}\gtrsim5$) to examine: 27266, 43871, 50636, 51205, 51217, 51254, 51472, 51891, 52418, and 53781.
We also picked 4 objects with high $\sigma_{vary}$: 28248, 38344, 54216, and 56594 (maximum $\sigma_{vary}\sim10-120$).

Of these, 28248 and 38344 exhibited broad H$\alpha$ emission, with BH masses $10^{8.45\pm0.02}$ and $10^{8.47\pm0.01}M_\odot$, respectively.
These high BH masses land above all previous scaling relations discussed in this paper, though 28248 is within 1$\sigma$ of the relation from \citet{2016ApJ...826L..32G}.
Neither of these objects had long enough baselines to include their measurements for $\tau_{DRW}$.

Seven objects had sufficiently strong emission lines for BPT analysis.
Objects 28248 and 38344 fall in the AGN region of the diagram, the same two galaxies with broad H$\alpha$ emission.
The remaining five galaxies (27266, 50636, 51891, 52418, 54216) were all in the star-forming region.
None of these objects were found in the ADGD, but 27266 and 51217 had corresponding NSA data that placed them in the star-forming region (just as in our analysis).

\subsubsection{Notable BPT Diagnostics}
Our BPT analysis categorized the emission of only three dwarf galaxies as being AGN-dominated: 14760, 27628, 46649.
Object 46649 is both an NSA BPT AGN (corresponding to object 124249 in the NSA v0) and found in the ADGD, the only AGN candidate in our study for which this is true.

All three galaxies showed broad H$\alpha$ emission which corresponded to BH masses between $\sim10^{6.9}-10^{8.5} M_\odot$.
As stated previously, object 38344 lies above all previously found scaling relations used in this study.
The remaining two galaxies were just below the relation from \citet{2016ApJ...826L..32G}, but within 1$\sigma$ uncertainty.
Only object 27628 had a long enough baseline to accurately measure $\tau_{DRW}$.
Just as with many of our observations, the different measurements were higher than expected from \citet{2021Sci...373..789B} and our fitted line in all but the g-band, which was below both relations.

Finally, we focus on the six objects that showed AGN-like variability in multiple bands and were categorized as AGNs by the NSA BPT analysis: 35246, 40161, 41451, 46649, 50828, 52568.
We were able to download spectra corresponding to objects 46649, 50828, 52568, but did not find broad H$\alpha$ emission in any.
Despite their NSA BPT classification, only objects 46649 and 52568 had strong enough emission lines for our BPT analysis; both objects landed in the AGN region of our diagram as well.

\subsection{Differences in Bands}
\begin{deluxetable}{c C C C C C C}[htb]
    \tablecolumns{7}
    \tablecaption{Photometric Variability Across Bands\label{tab:agn_bands}}
    \tablehead{\colhead{Band} & \colhead{Curves} & \colhead{Variable} & \colhead{AGNs} & \colhead{Active} & \colhead{Rec.}}
    \startdata
        g & 41143 & 814  & 1095 & 424 & 0.39 \\
        r & 41777 & 1804 & 1100 & 853 & 0.78 \\
        i & 43677 & 1517 & 1099 & 738 & 0.67 \\
        z & 35981 & 935  & 1097 & 523 & 0.47 \\
    \enddata
    \tablecomments{A comparison of the variability and AGN candidacy per band. For each band, we give the total number of (analyzable) light curves, curves with AGN-like variability, the number of available light curves corresponding to our AGN candidates, and active curves, as well as their recall. Here, we define ``active'' as a light curve that demonstrates AGN-like variability (in that band) AND corresponds to one of our AGN candidates.}
\end{deluxetable}

We explore the differences in our results across bands, first in their variability, then their derived properties (e.g. luminosity, BH mass).
We compare the variability results for the curves available in each band, shown in Table \ref{tab:agn_bands}.
There are a few reasons why an object could have AGN-like variability in one band but not another: variability amplitudes tend to be higher in bluer bands \citep{Kimura_2020, 2004ApJ...601..692V}, but there is also a trade off with S/N since observations in bluer bands tend to be dimmer.
However, galaxies with more extinction might be more inclined to be observed as variable in redder filters.
We can define a ``recall'' value for each band by dividing the number of AGN candidates with AGN-like variability (in that band) by the number of AGN candidates with available data (in that band).
The recall is a measure of the completeness of a classification ($R=1$ means that all available AGN candidates showed AGN-like variability in that band).
The r-band is able to achieve a recall of nearly 0.8, meaning nearly 4 out of every 5 AGN candidates were ``found''.
This comes, however, at the cost of potential overprediction since 1804 objects had AGN-like variability in the r-band, the majority of which are not variable in other bands, and so are not included in our AGN candidates.
This seems to be the trend overall as every band finds a number of galaxies with AGN-like variability that is approximately double the number that corresponds to our final AGN candidates.
It appears, then, that there is no single band that can accurately and discriminantly classify our AGN candidates without including a large number of potential ``contaminants''.

\begin{figure}[htb]
    \centering
    \includegraphics[width=0.48\textwidth]{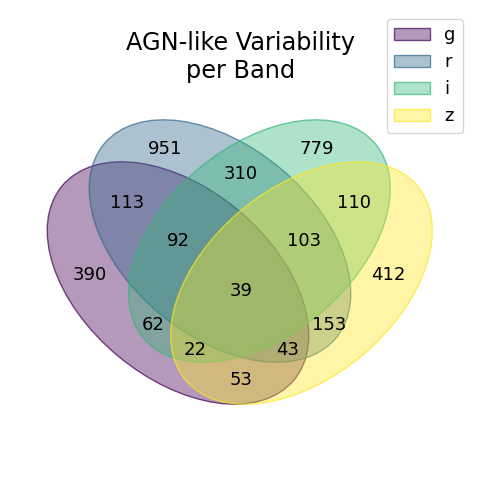}
    \caption{A breakdown of the photometric variability per band in our analysis. The number within each region is the number of galaxies with AGN-like variability in the corresponding subset of bands. For instance, only 39 galaxies had this variability in all four bands. By definition, our AGN candidates are any objects that were variable in multiple bands.}
    \label{fig:band_venn}
\end{figure}

This prompts the question: is there a subset of the bands that can select AGN candidates?
We break down the photometric variability of the bands of our AGN candidates in Figure \ref{fig:band_venn}.
Examining each pair of bands, if we still select objects that were variable in both bands, then the pair of bands that capture the most AGNs is `r' and `i', which select 544 AGN candidates out of our total 1100.
Similarly, if we examine triplets of bands and select objects which were variable in at least two, then the most effective triplet would be `r', `i', and `z' together, capturing 872 of our 1100 AGN candidates.

\begin{figure*}
    \gridline{
        \fig{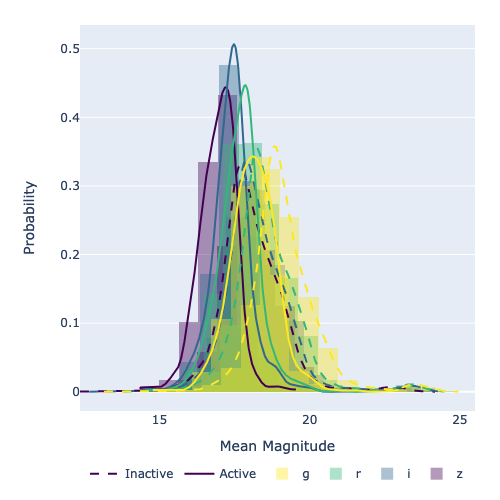}{0.32\textwidth}{}
        \fig{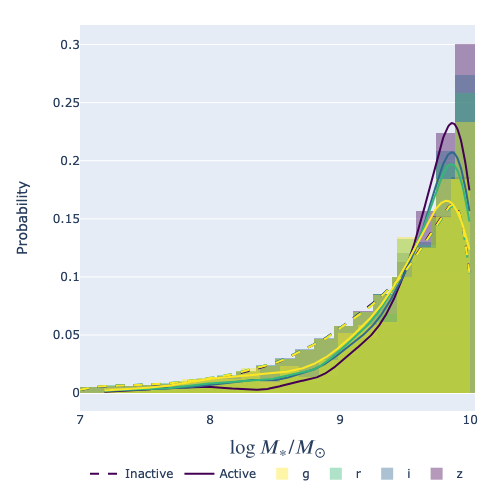}{0.32\textwidth}{}
        \fig{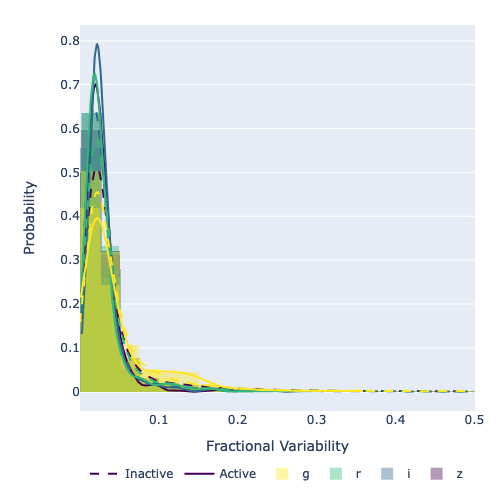}{0.32\textwidth}{}}
    \gridline{
        \fig{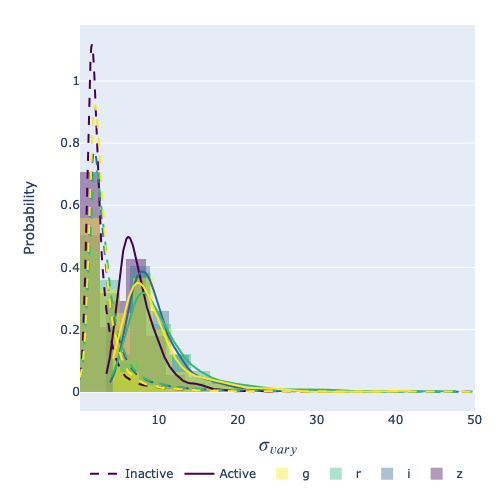}{0.32\textwidth}{}
        \fig{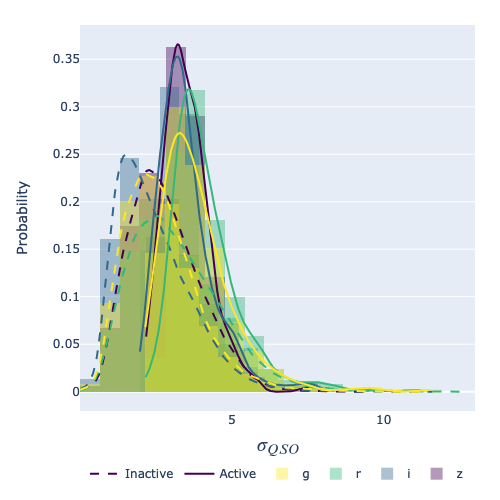}{0.32\textwidth}{}
        \fig{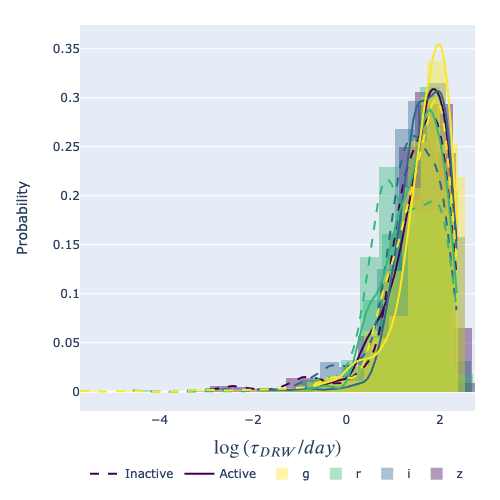}{0.32\textwidth}{}}
    \caption{Comparison of output parameters with histograms and kernel density estimate plots. Colors correspond to different bands, while active and inactive galaxy populations are shown with solid and dashed lines, respectively. An ``active'' light curve refers to one that is photometrically variable and corresponds to one of our AGN candidates, while a curve is ``inactive'' if neither is true.}
    \label{fig:band_diff}
\end{figure*}

Next, we compare the mean magnitude, stellar mass, and variability parameters of each band for active and inactive galaxies in Figure \ref{fig:band_diff}.
Some of these distributions look superficially similar, so we employ the Friedman test for any band-dependent variables (all but stellar mass).
This method can test the hypothesis that multiple observations of the same objects under different circumstances (e.g. observation bands) are sampled from the same distribution.
If the returned p-value does not exceed a given confidence level (typically 1-5\%), then the null hypothesis can be rejected meaning that the different observations of the same object are sampled from different distributions.
Based on our data, we can reject this null hypothesis for mean magnitude, $\mathrm{\sigma_{vary}}$, and $\mathrm{\sigma_{QSO}}$.
This is to be expected for mean magnitude since multiple factors (e.g. dust, reddening, intrinsic SED, etc.) contribute to the difference in brightness between bands.
All three of these features gave a p-value less than $10^{-16}$ while fractional variability and $\log{\tau_{DRW}}$ had values of 0.011 and 0.22, respectively.
For fractional variability, this indicates that the hypothesis can still be rejected within a reasonable confidence.
On the other hand, the results for $\log{\tau_{DRW}}$ indicate that there is not a statistically significant difference between observations across bands.

\section{Conclusion} \label{sec:conclusion}
 We compiled a list of 60,468 low-mass galaxies ($7\leq\log{(M_*/M_\odot)}\leq10$) from the Sloan Digital Sky Survey, NASA Sloan Atlas, and Galaxy and Mass Assembly within the observation fields for the Young Supernova Experiment.
 Using YSE data, we created light curves for 56,244 (93.0\%) of our targets, 45,696 (81.2\%) of which had enough data points to be analyzed for photometric variability.
 3632 objects were found to have at least one light curve with AGN-like variability, and 1100 were variable in multiple bands (7.9\% and 2.4\% of analyzable galaxies, respectively).
 We present these 1100 objects as our final list of AGN candidates.
 \begin{itemize}
    \item Spectra were available for 809 of the 1100 AGN candidates, which we used to calculate the fluxes necessary for BPT analysis and the BH masses via broad H$\alpha$ emission.
    \item We found broad H$\alpha$ emission in 30 of our candidates, corresponding to BH masses between $10^{4.76\pm0.01}$ and $10^{8.47\pm0.01}\,M_\odot$.
    \item Our calculated BH masses generally agree with previously found mass scaling relations from \citet{2020ARA&A..58..257G}; this consistency in the slope of the relation from the high-mass to the low-mass range could be indicative of a history of gravitational runaway events.
    \item We found a flatter relation between BH mass and the dampening timescale associated with a damped random walk than from \citet{2021Sci...373..789B}: $\log{\frac{\tau_{DRW}}{day}} = (0.18\pm0.15)\log{\left(\frac{M_{BH}}{10^{6.5}M_\odot}\right)}+ 1.71\pm0.14$.
    \item We found nine galaxies whose calculated central BH masses are within the intermediate-mass range ($2\lesssim\log{M_{BH}/M_\odot}\lesssim6$). Three of these are strong candidates for IMBHs, having broad H$\alpha$ fits with $\chi_{\nu,H\alpha}^2\leq2$ and masses more than 1$\sigma$ below the upper limit of this range.
    \item 431 of the 1100 AGN candidates had strong enough emission lines for BPT diagnostics, which characterized 27 objects as being dominated by AGN radiation with 115 objects in the composite region (only 3 and 30 of which, respectively, belong to dwarf galaxies with $\log{(M_*/M_\odot)}\leq9.5$). The BPT diagnostic is known to miss some AGNs in low-mass and low-metallicity galaxies, which could potentially explain why it only captures 2.5\% of the AGNs detected by photometric variability.
    \item We compared our candidates with a database of previously discovered dwarf AGNs and find 8 objects in common. We also compare our candidates with BPT diagnostics using NSA data, finding 6 candidates in the AGN region. 1,087 of our AGN candidates are new (i.e. not corresponding to AGNs in either source catalog).
    \item We estimated the active fraction as a function of host stellar mass by sampling from the AGN candidates binned by both mass and r-band magnitude (for a magnitude-unbiased calculation). We find evidence that these two values are related in this low-mass regime, as active fraction increases with stellar mass.
 \end{itemize}

\section{Acknowledgements} \label{sec:acknowledgements}

This material is based upon work supported by the National Science Foundation under Grant Number AAG 2206165.
Alexander Messick thanks the LSST-DA Data Science Fellowship Program, which is funded by LSST-DA, the Brinson Foundation, and the Moore Foundation; their participation in the program has benefited this work.
We also thank the reviewers for their helpful suggestions to improve this manuscript.

D.O.J. acknowledges support from HST grants HST-GO-17128.028 and HST-GO-16269.012, awarded by the Space Telescope Science Institute (STScI), which is operated by the Association of Universities for Research in Astronomy, Inc., for NASA, under contract NAS5-26555. D.O.J. also acknowledges support from NSF grant AST-2407632 and NASA grant 80NSSC24M0023.
K.D.F. acknowledges support from NSF grant AST–2206164.
M.E.V. acknowledges support from NSF grant AST–2206164, CAPS, and ISGC.
Parts of this research were supported by the Australian Research Council Discovery Early Career Researcher Award (DECRA) through project number DE230101069.
N. E. acknowledges support from NSF grant AST-2206164 and the Center for Astrophysical Surveys Graduate Fellowship.

Pan-STARRS is a project of the Institute for Astronomy of the University of Hawaii, and is supported by the NASA SSO Near Earth Observation Program under grants 80NSSC18K0971, NNX14AM74G, NNX12AR65G, NNX13AQ47G, NNX08AR22G, 80NSSC21K1572  and by the State of Hawaii.

The Young Supernova Experiment (YSE) and its research infrastructure is supported by the European Research Council under the European Union's Horizon 2020 research and innovation programme (ERC Grant Agreement 101002652, PI K.\ Mandel), the Heising-Simons Foundation (2018-0913, PI R.\ Foley; 2018-0911, PI R.\ Margutti), NASA (NNG17PX03C, PI R.\ Foley), NSF (AST--1720756, AST--1815935, PI R.\ Foley; AST--1909796, AST-1944985, PI R.\ Margutti), the David \& Lucille Packard Foundation (PI R.\ Foley), VILLUM FONDEN (project 16599, PI J.\ Hjorth), and the Center for AstroPhysical Surveys (CAPS) at the National Center for Supercomputing Applications (NCSA) and the University of Illinois Urbana-Champaign.

YSE-PZ was developed by the UC Santa Cruz Transients Team with support from The UCSC team is supported in part by NASA grants NNG17PX03C, 80NSSC19K1386, and 80NSSC20K0953; NSF grants AST-1518052, AST-1815935, and AST-1911206; the Gordon \& Betty Moore Foundation; the Heising-Simons Foundation; a fellowship from the David and Lucile Packard Foundation to R. J. Foley; Gordon and Betty Moore Foundation postdoctoral fellowships and a NASA Einstein fellowship, as administered through the NASA Hubble Fellowship program and grant HST-HF2-51462.001, to D. O. Jones; and a National Science Foundation Graduate Research Fellowship, administered through grant No. DGE-1339067, to D. A. Coulter. 

Funding for SDSS-III has been provided by the Alfred P. Sloan Foundation, the Participating Institutions, the National Science Foundation, and the U.S. Department of Energy Office of Science. The SDSS-III web site is http://www.sdss3.org/. SDSS-III is managed by the Astrophysical Research Consortium for the Participating Institutions of the SDSS-III Collaboration including the University of Arizona, the Brazilian Participation Group, Brookhaven National Laboratory, Carnegie Mellon University, University of Florida, the French Participation Group, the German Participation Group, Harvard University, the Instituto de Astrofisica de Canarias, the Michigan State/Notre Dame/JINA Participation Group, Johns Hopkins University, Lawrence Berkeley National Laboratory, Max Planck Institute for Astrophysics, Max Planck Institute for Extraterrestrial Physics, New Mexico State University, New York University, Ohio State University, Pennsylvania State University, University of Portsmouth, Princeton University, the Spanish Participation Group, University of Tokyo, University of Utah, Vanderbilt University, University of Virginia, University of Washington, and Yale University. 

GAMA is a joint European-Australasian project based around a spectroscopic campaign using the Anglo-Australian Telescope. The GAMA input catalogue is based on data taken from the Sloan Digital Sky Survey and the UKIRT Infrared Deep Sky Survey. Complementary imaging of the GAMA regions is being obtained by a number of independent survey programmes including GALEX MIS, VST KiDS, VISTA VIKING, WISE, Herschel-ATLAS, GMRT and ASKAP providing UV to radio coverage. GAMA is funded by the STFC (UK), the ARC (Australia), the AAO, and the participating institutions. The GAMA website is http://www.gama-survey.org/ .

\appendix
\section{Case Study Plots}\label{sec:app}
To better understand the individual objects discussed in Section \ref{subsec:cases}, we show the light curves and spectra used to calculate their properties.

\begin{figure*}
        \centering
        \includegraphics[width=.30\textwidth]{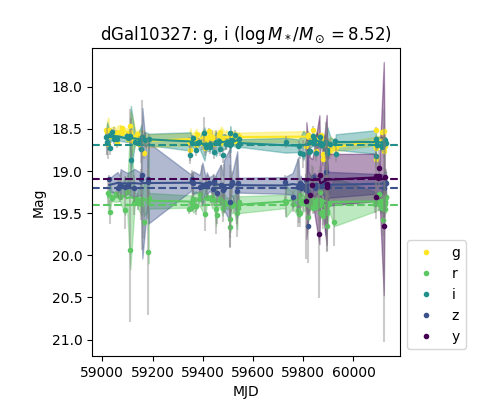} 
        \includegraphics[width=.30\textwidth]{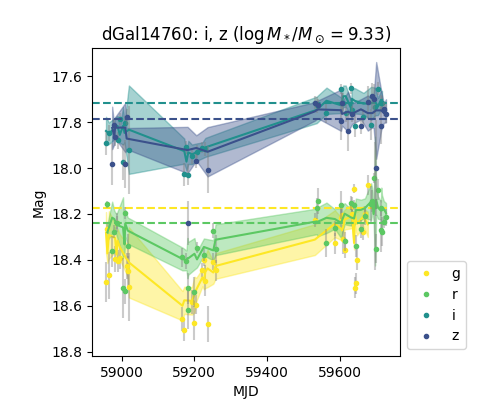} 
        \includegraphics[width=.30\textwidth]{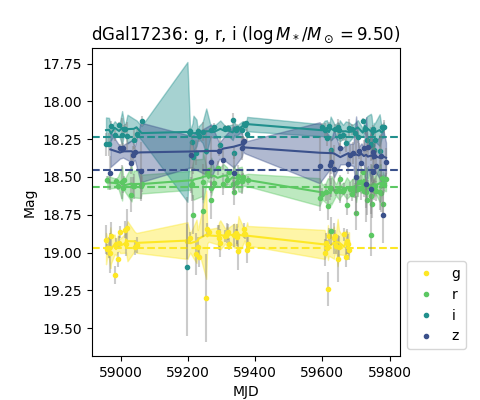} 
        
        \includegraphics[width=.30\textwidth]{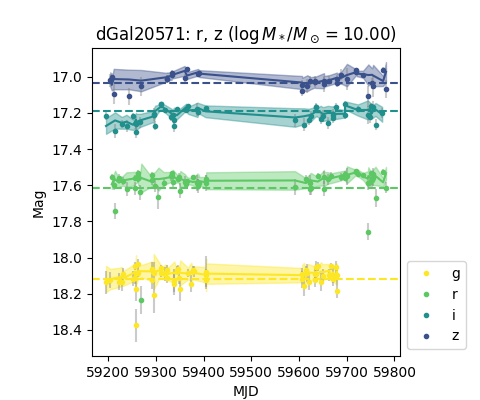}
        \includegraphics[width=.30\textwidth]{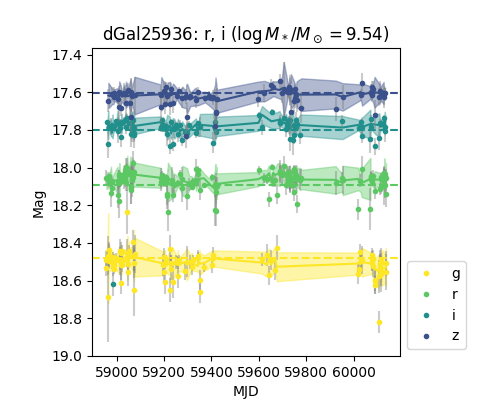} 
        \includegraphics[width=.30\textwidth]{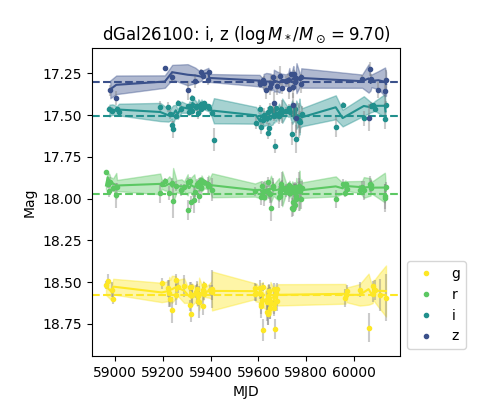}
        
        \includegraphics[width=.30\textwidth]{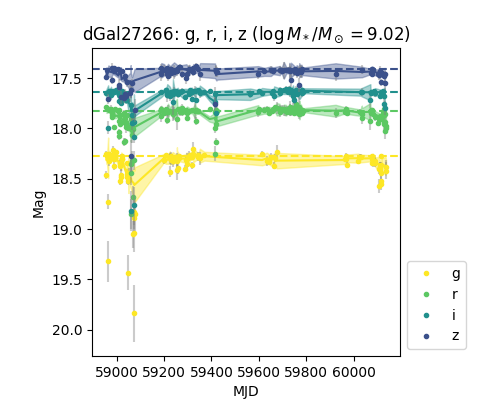} 
        \includegraphics[width=.30\textwidth]{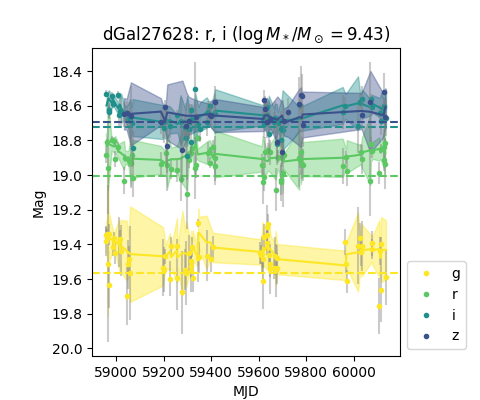}
        \includegraphics[width=.30\textwidth]{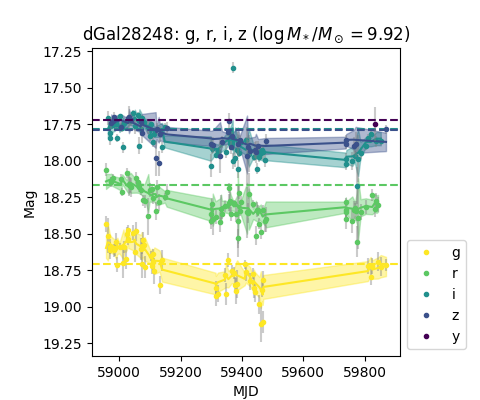}
        
        \includegraphics[width=.30\textwidth]{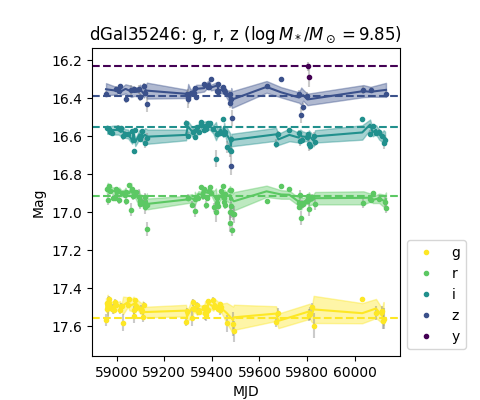}
        \includegraphics[width=.30\textwidth]{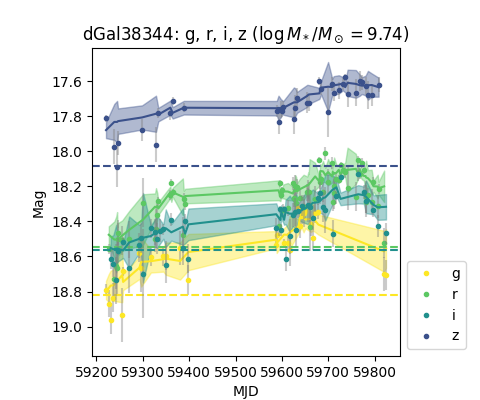}
        \includegraphics[width=.30\textwidth]{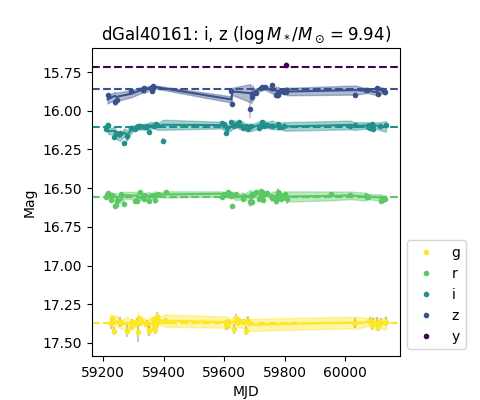} 
        \caption{Light curves for all of the case study objects mentioned in Section \ref{subsec:cases}. Above each plot is the object ID number, a list of which bands were observed to have AGN-like variability, and the stellar mass of the host galaxy. For each band, we show the host galaxy light (calculated from forced photometry on the stack image) as a dashed horizontal line. (Continued in Figures \ref{fig:app_curves2} and \ref{fig:app_curves3}).}
        \label{fig:app_curves1}
\end{figure*}

\begin{figure*}
        \centering
        \includegraphics[width=.30\textwidth]{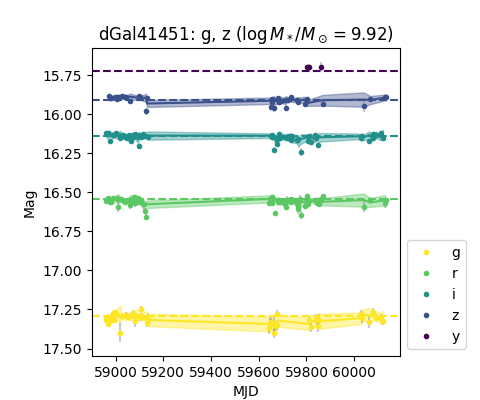}
        \includegraphics[width=.30\textwidth]{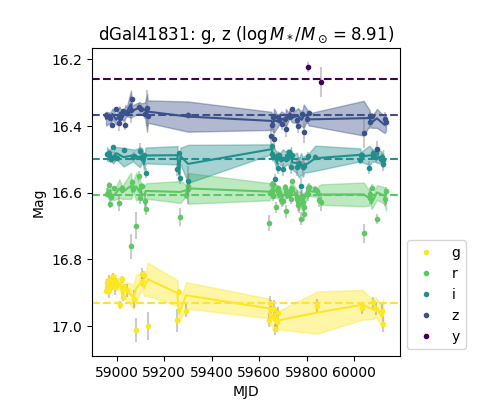}
        \includegraphics[width=.30\textwidth]{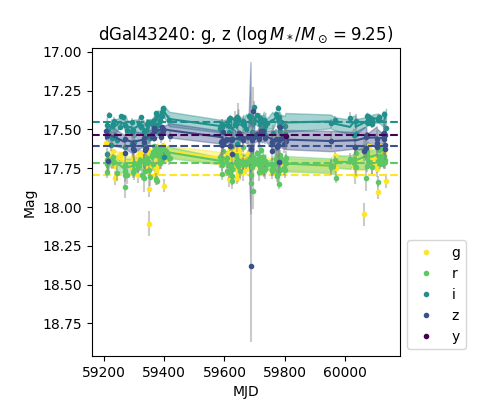}
        
        \includegraphics[width=.30\textwidth]{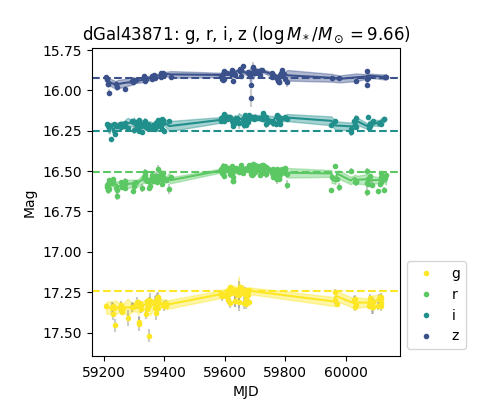} 
        \includegraphics[width=.30\textwidth]{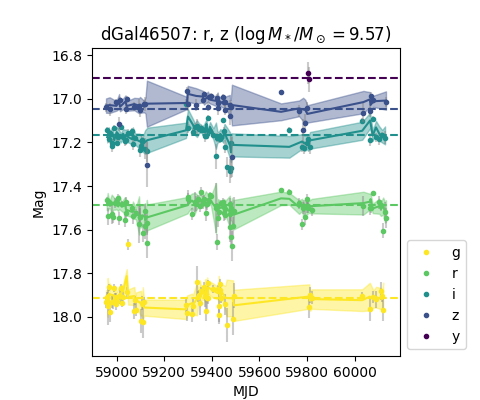}
        \includegraphics[width=.30\textwidth]{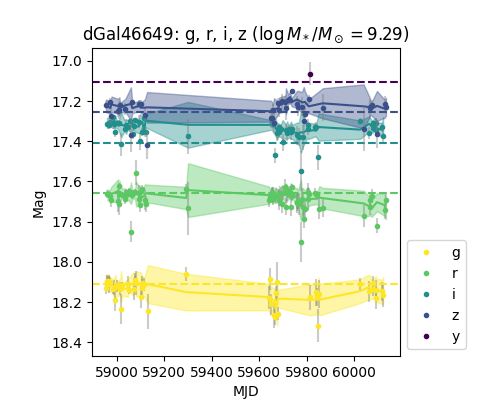}
        
        \includegraphics[width=.30\textwidth]{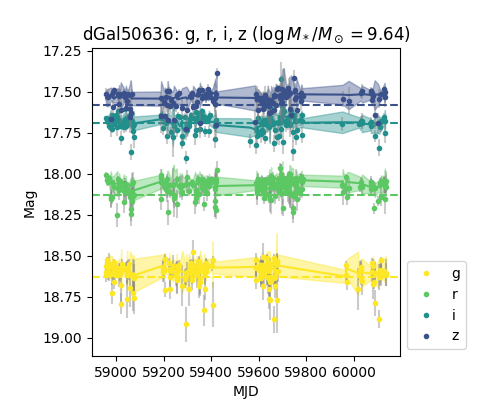}
        \includegraphics[width=.30\textwidth]{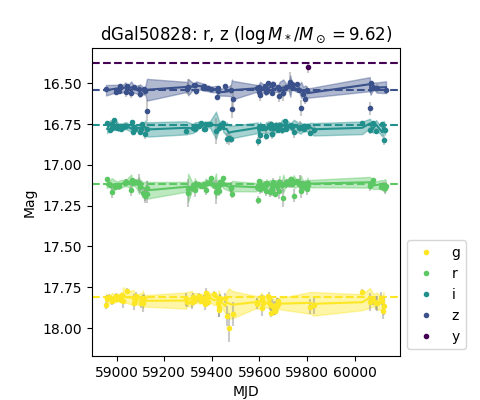}
        \includegraphics[width=.30\textwidth]{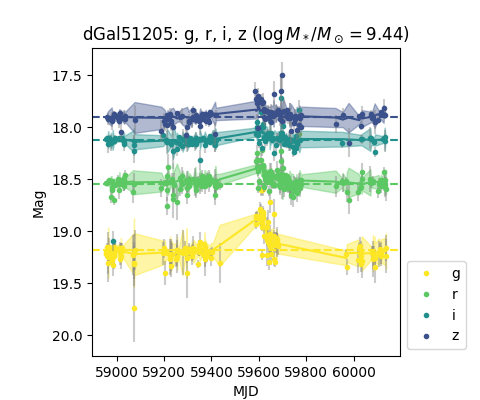}
        
        \includegraphics[width=.30\textwidth]{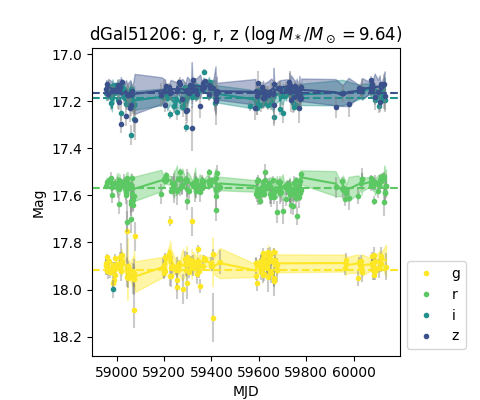}
        \includegraphics[width=.30\textwidth]{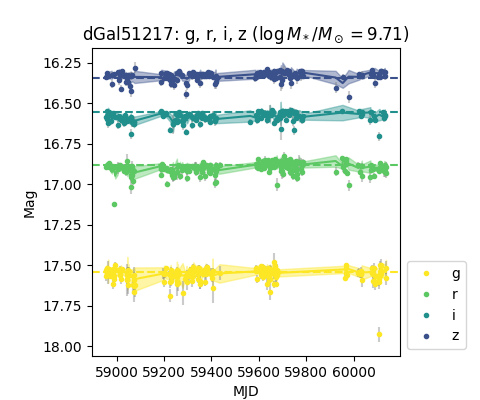}
        \includegraphics[width=.30\textwidth]{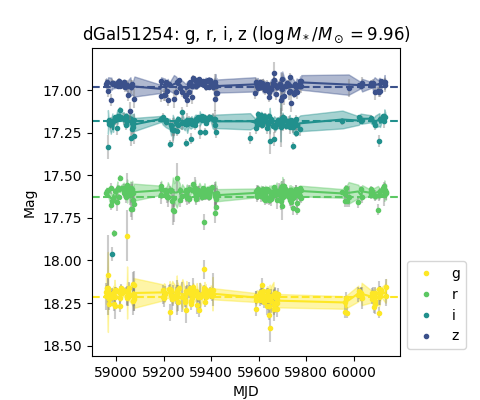} 
        \caption{Light curves for all of the case study objects mentioned in Section \ref{subsec:cases}. (Continuation of Figure \ref{fig:app_curves1}).}
        \label{fig:app_curves2}
\end{figure*}

\begin{figure*}
        \centering
        \includegraphics[width=.30\textwidth]{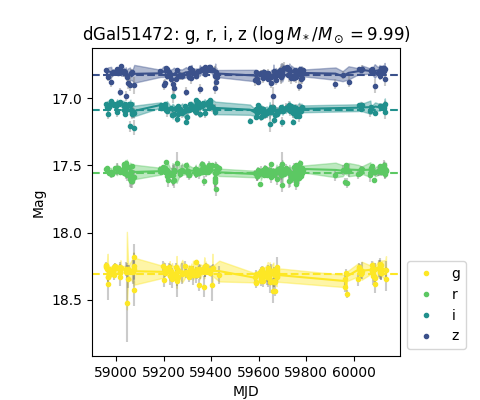}
        \includegraphics[width=.30\textwidth]{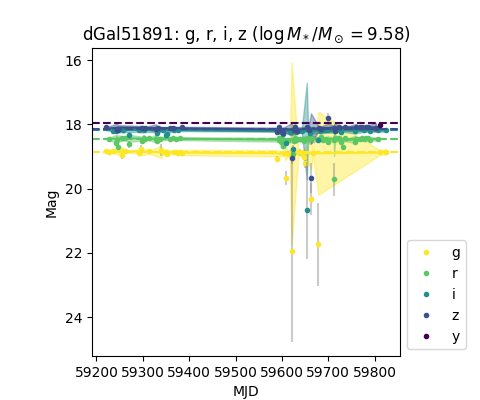}
        \includegraphics[width=.30\textwidth]{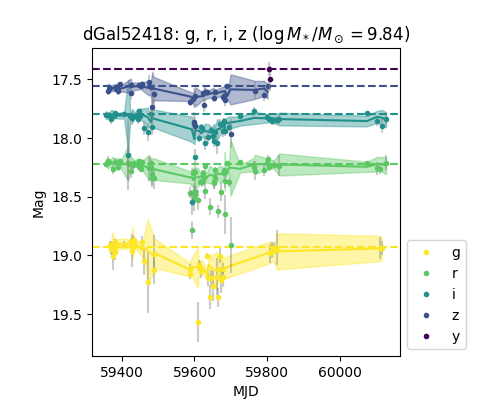}
        
        \includegraphics[width=.30\textwidth]{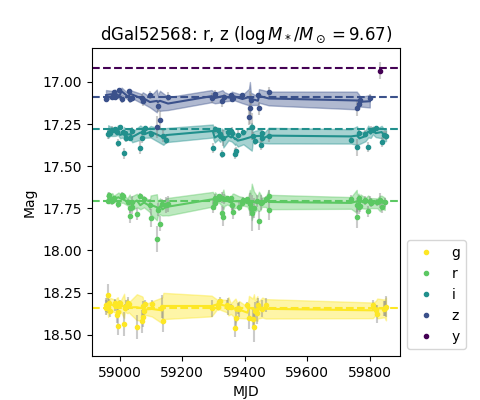} 
        \includegraphics[width=.30\textwidth]{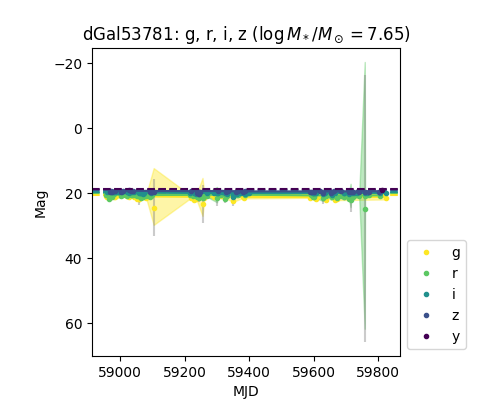}
        
        \includegraphics[width=.30\textwidth]{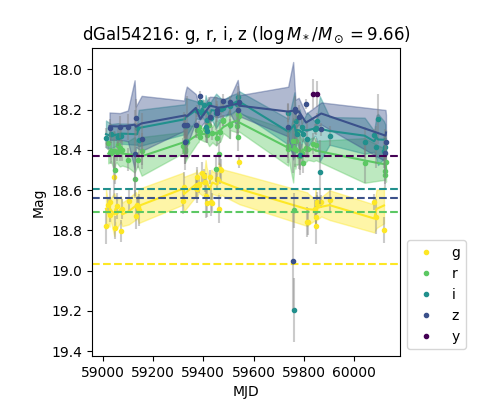}
        \includegraphics[width=.30\textwidth]{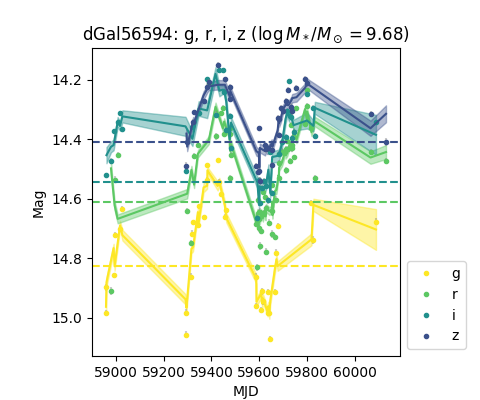}
        \caption{Light curves for all of the case study objects mentioned in Section \ref{subsec:cases}. (Continuation of Figure \ref{fig:app_curves1}).}
        \label{fig:app_curves3}
\end{figure*}

We give the individual light curves for all the case studies in Figures \ref{fig:app_curves1}-\ref{fig:app_curves3}, each labeled by their ID number, which bands demonstrated AGN-like variability, and their host galaxy's stellar mass.
More information about these is available in the downloadable Tables \ref{tab:var_gal}, \ref{tab:phot_var}, and \ref{tab:BPT}.

\begin{figure*}
    \centering
        \includegraphics[width=\textwidth]{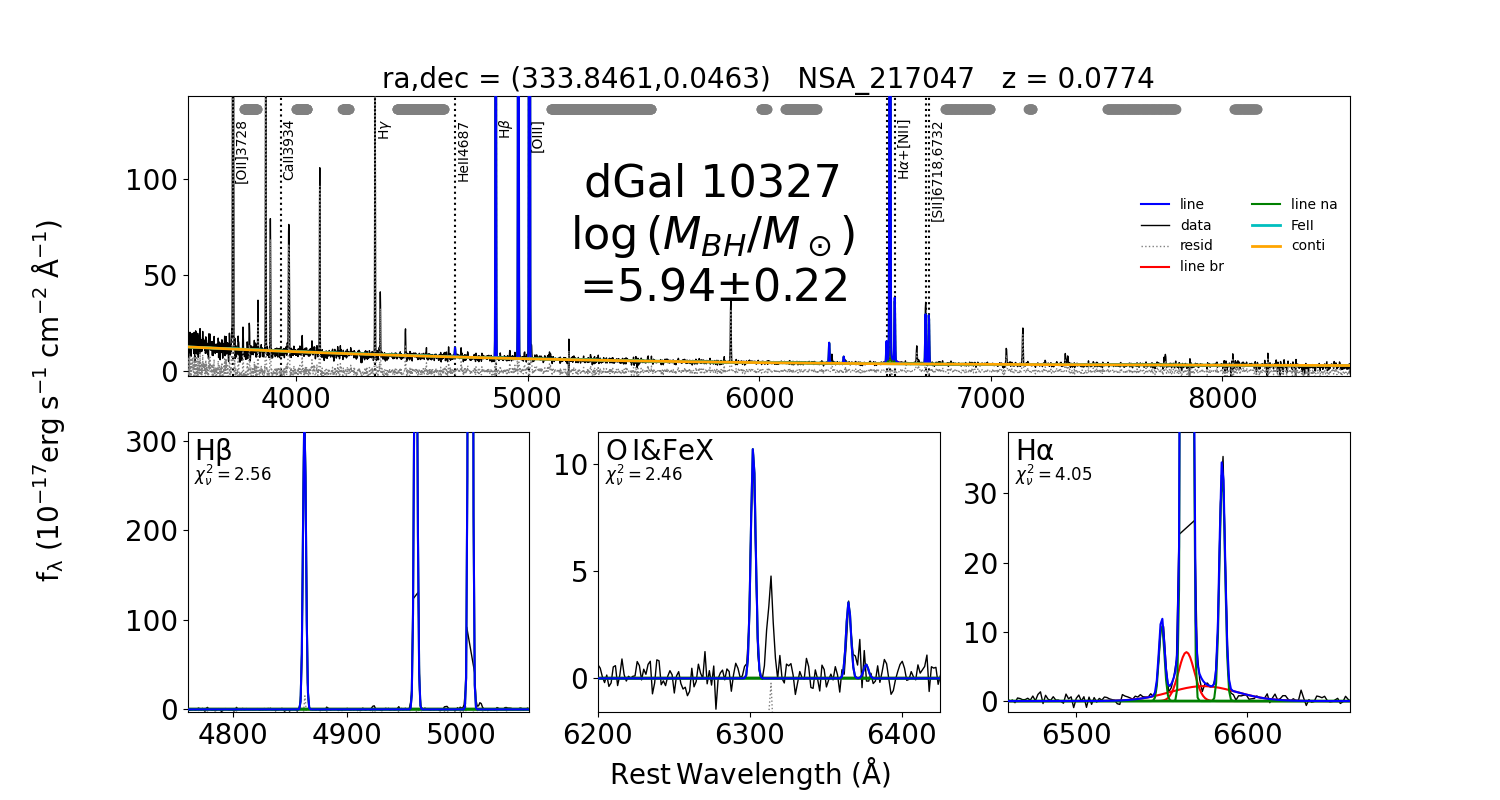} 
        
        \includegraphics[width=\textwidth]{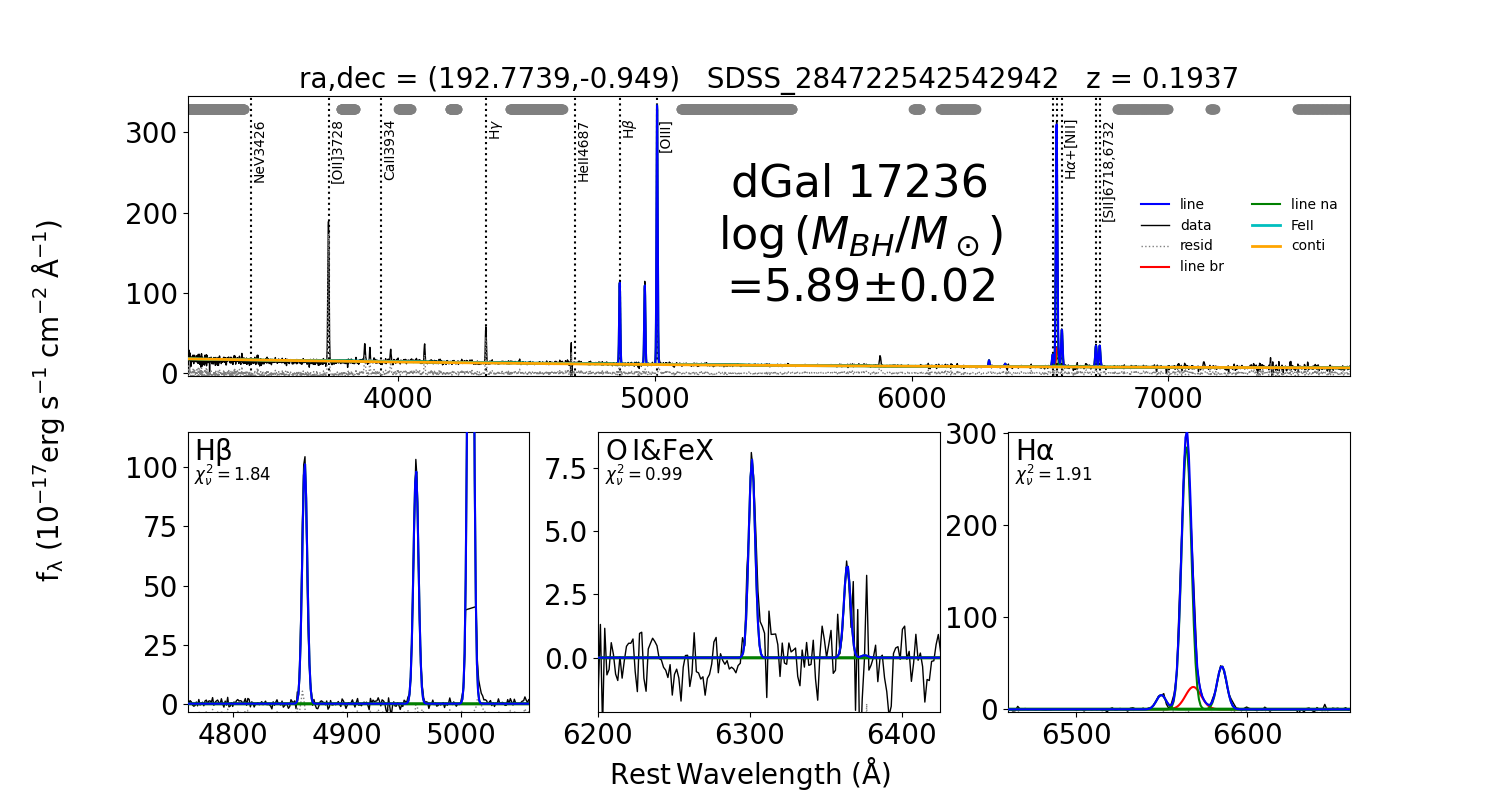}
        \caption{Spectra for the nine IMBH candidates ($M_{BH}\leq10^6M_\odot$), beginning with galaxies number 10327 and 17236, respectively. Each plot is comprised of four subplots. First, the total fitted spectrum consisting of multiple continuum and emission line components, on which we have added the calculated BH mass. We show the fitted line (blue), data (black), residuals (dotted), narrow (green) and broad (red) emission lines, and the \ion{Fe}{2} (cyan) and polynomial continua (orange). The three additional subplots show the continuum-subtracted H$\beta$, \ion{O}{1} \& \ion{Fe}{10}, and  H$\alpha$ complexes in greater detail. (Continued in Figures \ref{fig:BH_spectra2}-\ref{fig:BH_spectra5}).}
        \label{fig:BH_spectra1}
\end{figure*}

\begin{figure*}
    \centering
        \includegraphics[width=\textwidth]{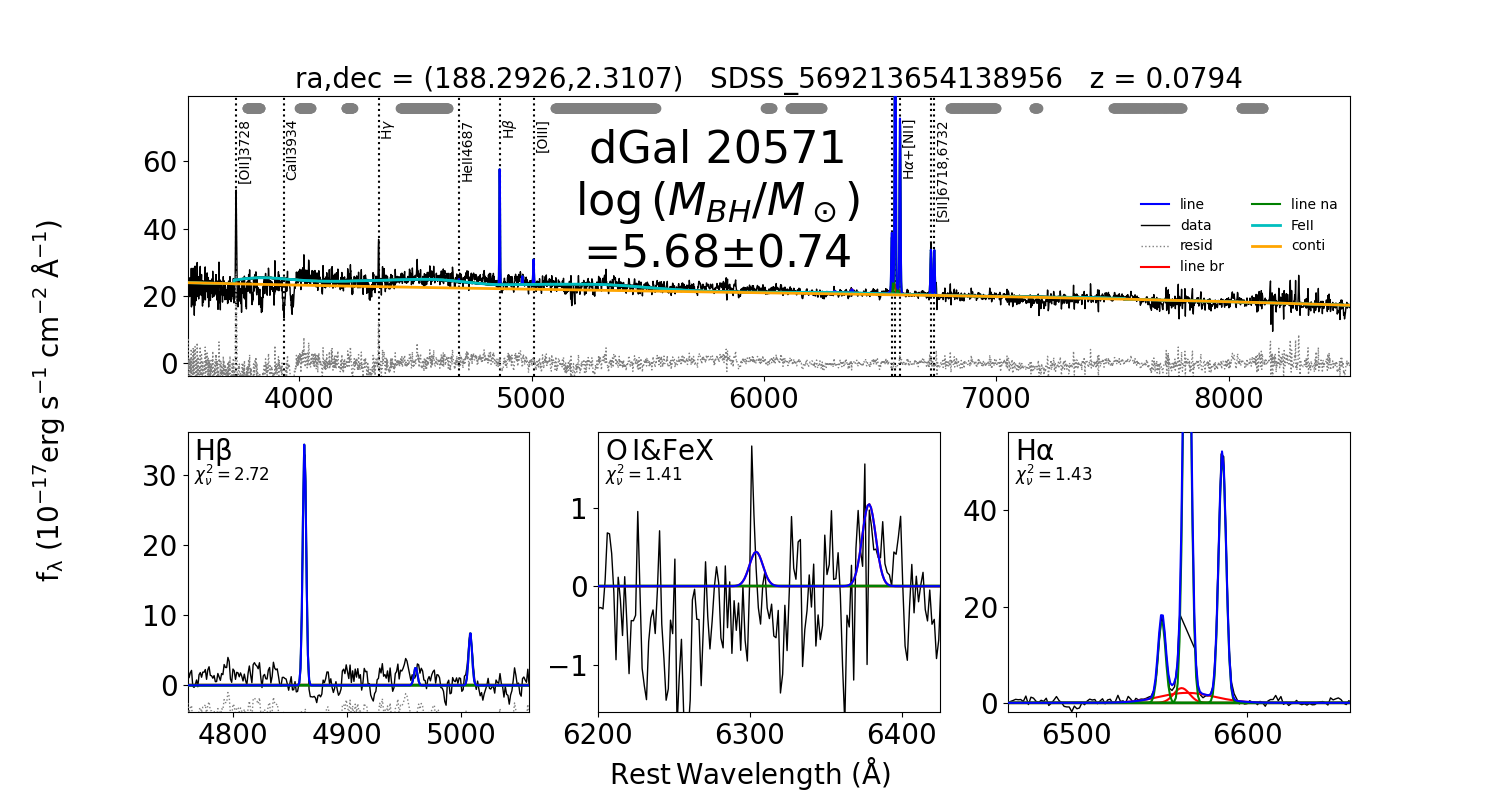}
        \includegraphics[width=\textwidth]{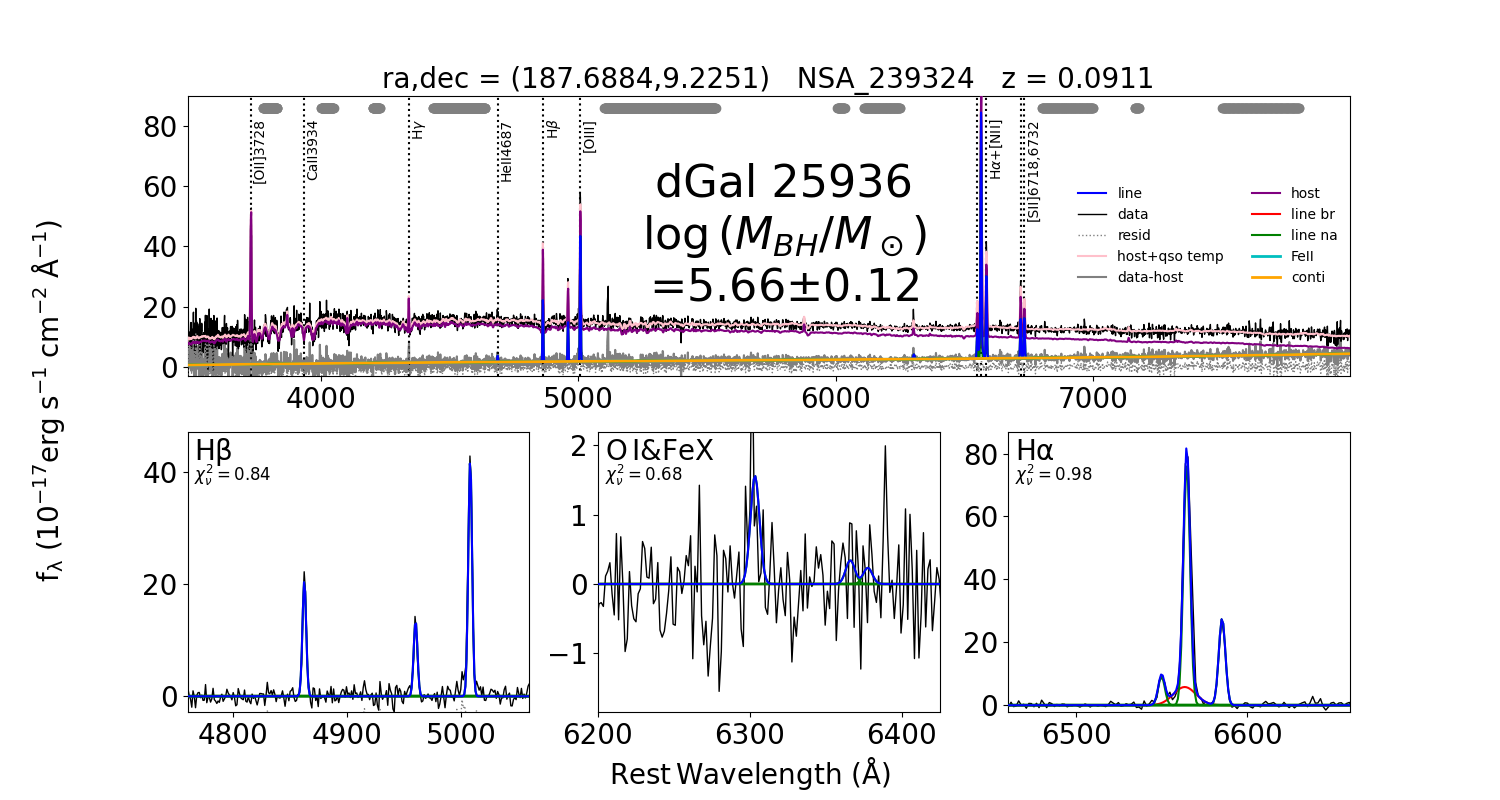} 
        \caption{Spectra for two of the nine IMBH candidates, galaxies number 20571 and 25936, respectively. For galaxy 25936, we were able to decompose the data using host and qso templates, which is shown in pink with the host-subtracted data in grey. (Continuation of Figure \ref{fig:BH_spectra1}).}
        \label{fig:BH_spectra2}
\end{figure*}

\begin{figure*}
    \centering
        \includegraphics[width=\textwidth]{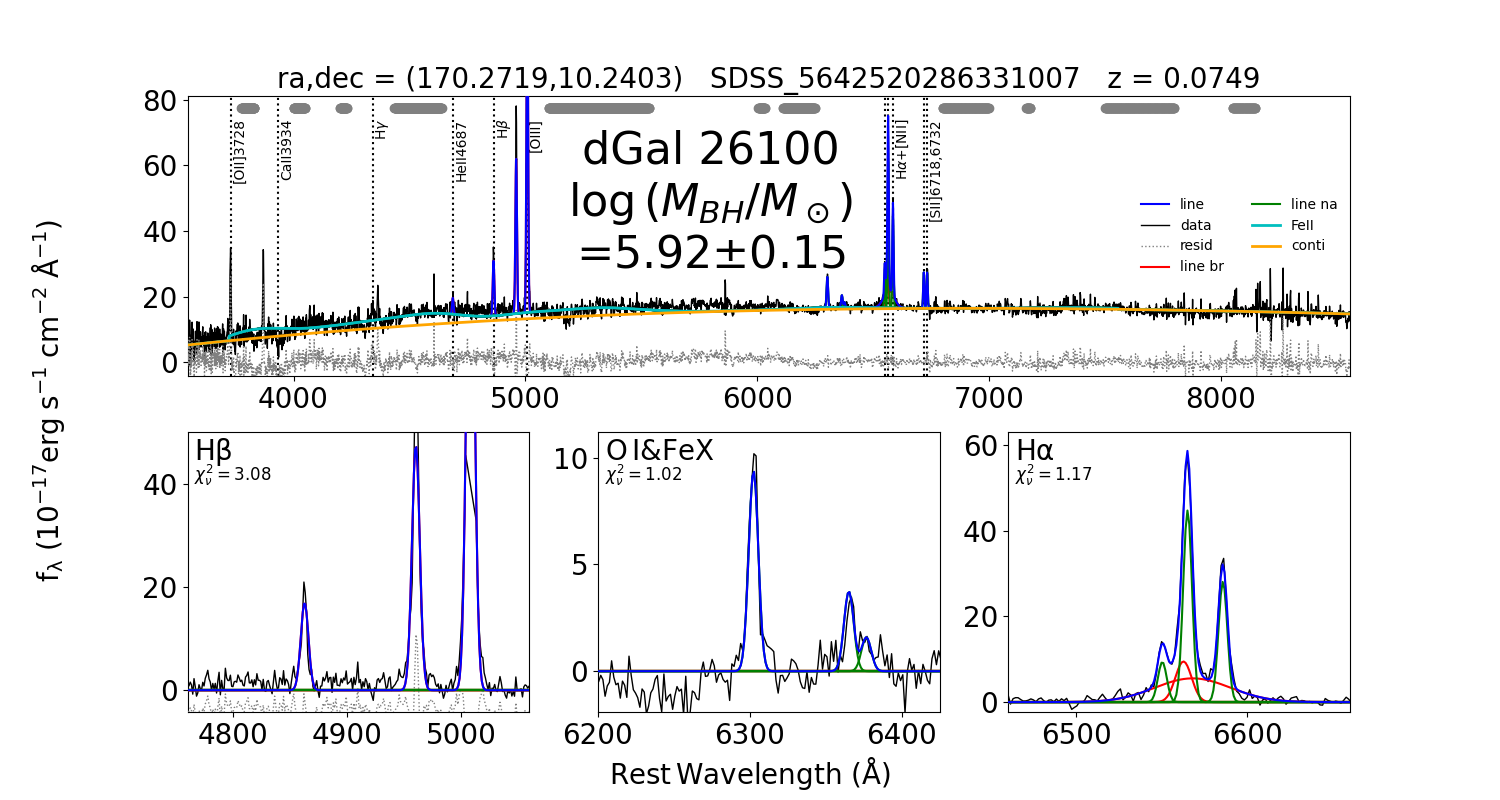} 
        \includegraphics[width=\textwidth]{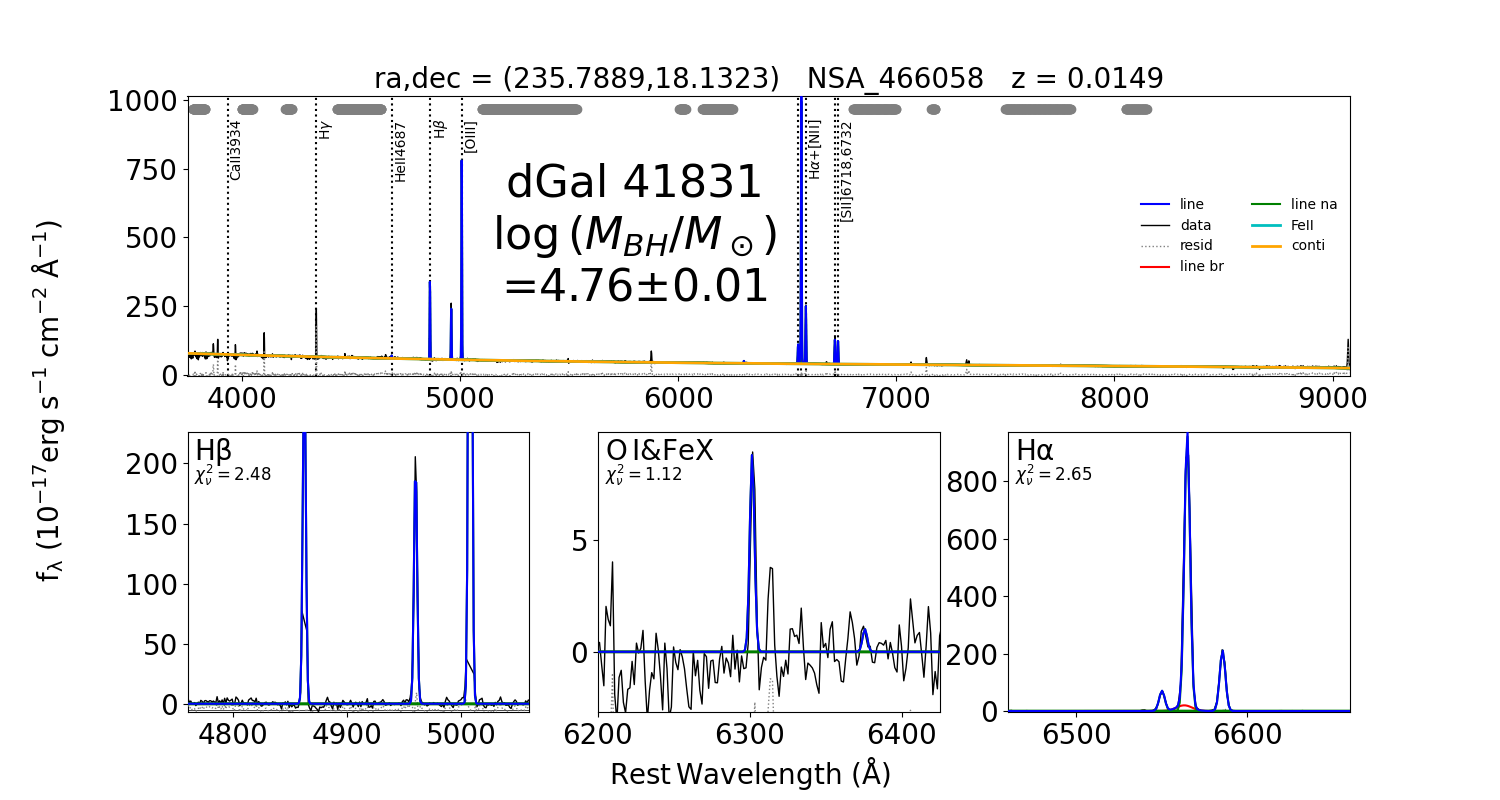} 
        \caption{Spectra for two of the nine IMBH candidates, galaxies number 26100 and 41831, respectively. (Continuation of Figure \ref{fig:BH_spectra1}).}
        \label{fig:BH_spectra3}
\end{figure*}

\begin{figure*}
    \centering
        \includegraphics[width=\textwidth]{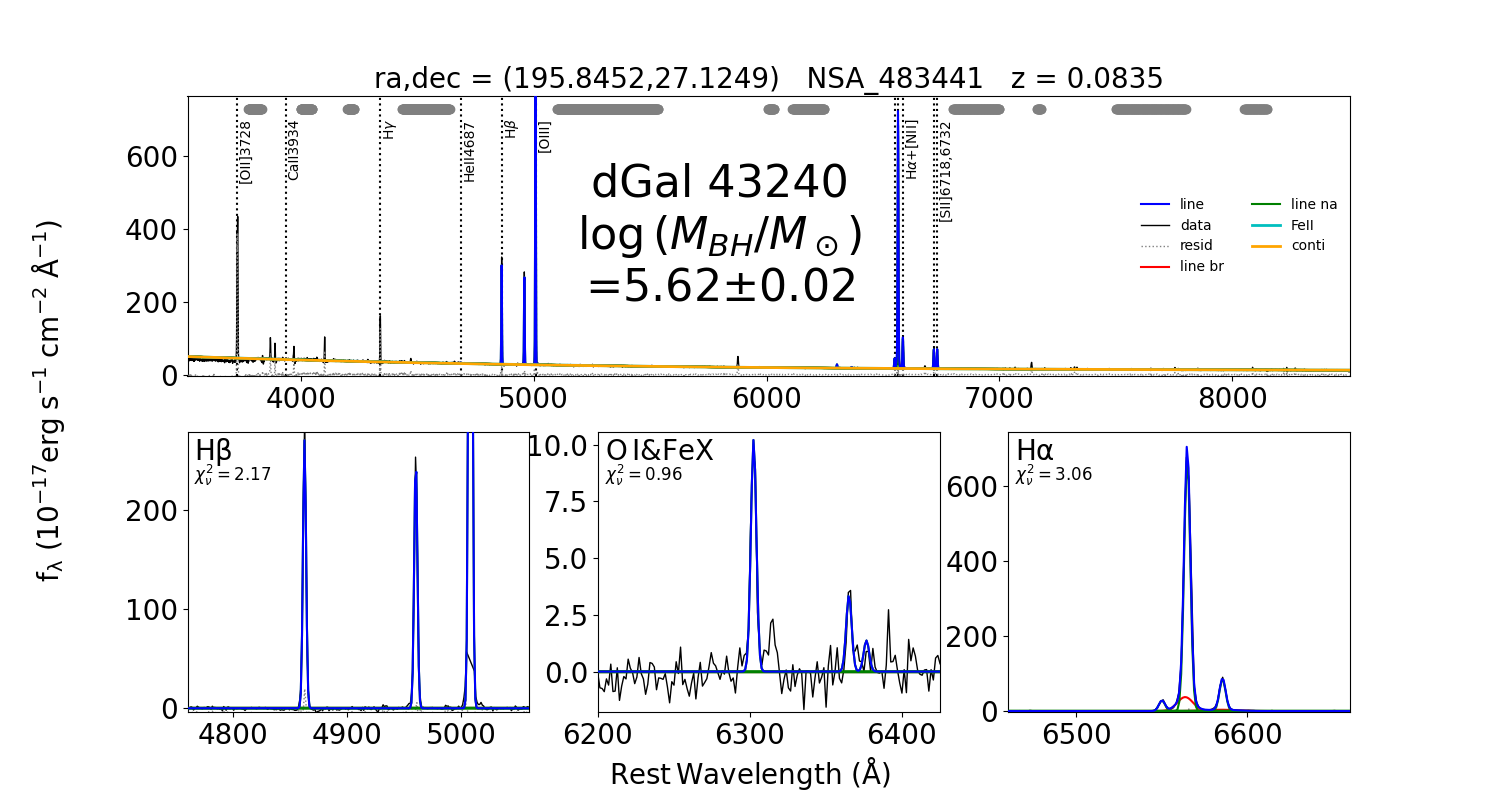}
        \includegraphics[width=\textwidth]{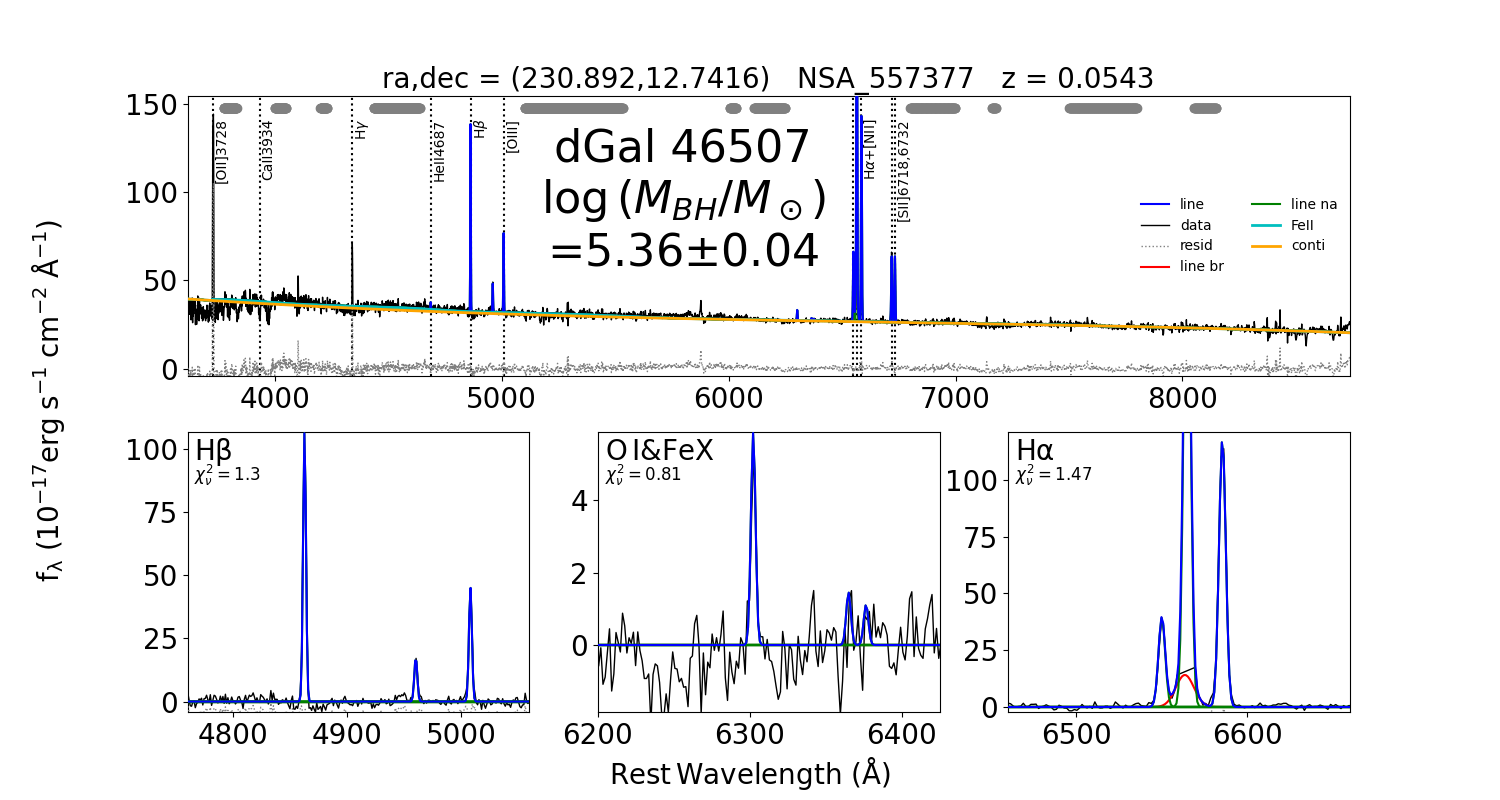}
        \caption{Spectra for two of the nine IMBH candidates, galaxies number 43240 and 46507, respectively. (Continuation of Figure \ref{fig:BH_spectra1}).}
        \label{fig:BH_spectra4}
\end{figure*}

\begin{figure*}
    \centering
        \includegraphics[width=\textwidth]{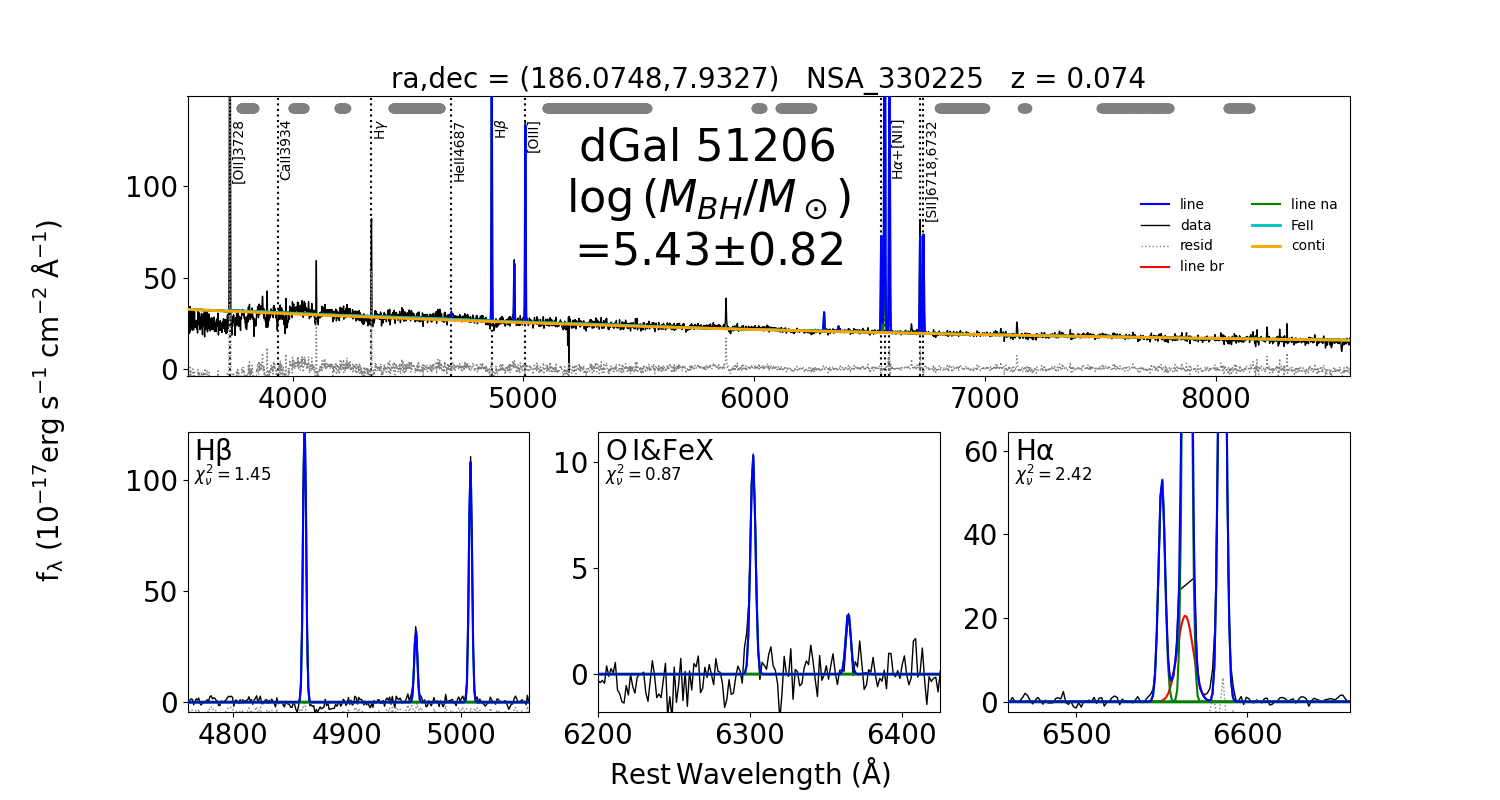}
        \caption{Spectra for one of the nine IMBH candidates, galaxy number 51206. (Continuation of Figure \ref{fig:BH_spectra1}).}
        \label{fig:BH_spectra5}
\end{figure*}

For the nine IMBH candidates ($M_{BH}\leq10^6M_\odot$), we also show results of the spectral fitting by \textit{PyQSOFit} in Figures \ref{fig:BH_spectra1}-\ref{fig:BH_spectra5}.
Each plot consists of the fitted spectrum and H$\alpha$, H$\beta$, and \ion{O}{1} \& \ion{Fe}{10} complexes alongside the BH mass derived from the fitting.

\bibliography{YSE_Search}{}

\begin{thebibliography}{}
\expandafter\ifx\csname natexlab\endcsname\relax\def\natexlab#1{#1}\fi
\providecommand{\url}[1]{\href{#1}{#1}}
\providecommand{\dodoi}[1]{doi:~\href{http://doi.org/#1}{\nolinkurl{#1}}}
\providecommand{\doeprint}[1]{\href{http://ascl.net/#1}{\nolinkurl{http://ascl.net/#1}}}
\providecommand{\doarXiv}[1]{\href{https://arxiv.org/abs/#1}{\nolinkurl{https://arxiv.org/abs/#1}}}

\bibitem[{{Abdurro'uf} {et~al.}(2022){Abdurro'uf}, {Accetta}, {Aerts}, {Silva Aguirre}, {Ahumada}, {Ajgaonkar}, {Filiz Ak}, {Alam}, {Allende Prieto}, {Almeida}, {Anders}, {Anderson}, {Andrews}, {Anguiano}, {Aquino-Ort{\'\i}z}, {Arag{\'o}n-Salamanca}, {Argudo-Fern{\'a}ndez}, {Ata}, {Aubert}, {Avila-Reese}, {Badenes}, {Barb{\'a}}, {Barger}, {Barrera-Ballesteros}, {Beaton}, {Beers}, {Belfiore}, {Bender}, {Bernardi}, {Bershady}, {Beutler}, {Bidin}, {Bird}, {Bizyaev}, {Blanc}, {Blanton}, {Boardman}, {Bolton}, {Boquien}, {Borissova}, {Bovy}, {Brandt}, {Brown}, {Brownstein}, {Brusa}, {Buchner}, {Bundy}, {Burchett}, {Bureau}, {Burgasser}, {Cabang}, {Campbell}, {Cappellari}, {Carlberg}, {Wanderley}, {Carrera}, {Cash}, {Chen}, {Chen}, {Cherinka}, {Chiappini}, {Choi}, {Chojnowski}, {Chung}, {Clerc}, {Cohen}, {Comerford}, {Comparat}, {da Costa}, {Covey}, {Crane}, {Cruz-Gonzalez}, {Culhane}, {Cunha}, {Dai}, {Damke}, {Darling}, {Davidson}, {Davies}, {Dawson}, {De Lee}, {Diamond-Stanic}, {Cano-D{\'\i}az}, {S{\'a}nchez},
  {Donor}, {Duckworth}, {Dwelly}, {Eisenstein}, {Elsworth}, {Emsellem}, {Eracleous}, {Escoffier}, {Fan}, {Farr}, {Feng}, {Fern{\'a}ndez-Trincado}, {Feuillet}, {Filipp}, {Fillingham}, {Frinchaboy}, {Fromenteau}, {Galbany}, {Garc{\'\i}a}, {Garc{\'\i}a-Hern{\'a}ndez}, {Ge}, {Geisler}, {Gelfand}, {G{\'e}ron}, {Gibson}, {Goddy}, {Godoy-Rivera}, {Grabowski}, {Green}, {Greener}, {Grier}, {Griffith}, {Guo}, {Guy}, {Hadjara}, {Harding}, {Hasselquist}, {Hayes}, {Hearty}, {Hern{\'a}ndez}, {Hill}, {Hogg}, {Holtzman}, {Horta}, {Hsieh}, {Hsu}, {Hsu}, {Huber}, {Huertas-Company}, {Hutchinson}, {Hwang}, {Ibarra-Medel}, {Chitham}, {Ilha}, {Imig}, {Jaekle}, {Jayasinghe}, {Ji}, {Johnson}, {Jones}, {J{\"o}nsson}, {Katkov}, {Khalatyan}, {Kinemuchi}, {Kisku}, {Knapen}, {Kneib}, {Kollmeier}, {Kong}, {Kounkel}, {Kreckel}, {Krishnarao}, {Lacerna}, {Lane}, {Langgin}, {Lavender}, {Law}, {Lazarz}, {Leung}, {Leung}, {Lewis}, {Li}, {Li}, {Lian}, {Liang}, {Lin}, {Lin}, {Lin}, {Lintott}, {Long}, {Longa-Pe{\~n}a}, {L{\'o}pez-Cob{\'a}}, {Lu},
  {Lundgren}, {Luo}, {Mackereth}, {de la Macorra}, {Mahadevan}, {Majewski}, {Manchado}, {Mandeville}, {Maraston}, {Margalef-Bentabol}, {Masseron}, {Masters}, {Mathur}, {McDermid}, {Mckay}, {Merloni}, {Merrifield}, {Meszaros}, {Miglio}, {Di Mille}, {Minniti}, {Minsley}, {Monachesi}, {Moon}, {Mosser}, {Mulchaey}, {Muna}, {Mu{\~n}oz}, {Myers}, {Myers}, {Nadathur}, {Nair}, {Nandra}, {Neumann}, {Newman}, {Nidever}, {Nikakhtar}, {Nitschelm}, {O'Connell}, {Garma-Oehmichen}, {Luan Souza de Oliveira}, {Olney}, {Oravetz}, {Ortigoza-Urdaneta}, {Osorio}, {Otter}, {Pace}, {Padilla}, {Pan}, {Pan}, {Parikh}, {Parker}, {Peirani}, {Pe{\~n}a Ram{\'\i}rez}, {Penny}, {Percival}, {Perez-Fournon}, {Pinsonneault}, {Poidevin}, {Poovelil}, {Price-Whelan}, {B{\'a}rbara de Andrade Queiroz}, {Raddick}, {Ray}, {Rembold}, {Riddle}, {Riffel}, {Riffel}, {Rix}, {Robin}, {Rodr{\'\i}guez-Puebla}, {Roman-Lopes}, {Rom{\'a}n-Z{\'u}{\~n}iga}, {Rose}, {Ross}, {Rossi}, {Rubin}, {Salvato}, {S{\'a}nchez}, {S{\'a}nchez-Gallego}, {Sanderson}, {Santana
  Rojas}, {Sarceno}, {Sarmiento}, {Sayres}, {Sazonova}, {Schaefer}, {Schiavon}, {Schlegel}, {Schneider}, {Schultheis}, {Schwope}, {Serenelli}, {Serna}, {Shao}, {Shapiro}, {Sharma}, {Shen}, {Shetrone}, {Shu}, {Simon}, {Skrutskie}, {Smethurst}, {Smith}, {Sobeck}, {Spoo}, {Sprague}, {Stark}, {Stassun}, {Steinmetz}, {Stello}, {Stone-Martinez}, {Storchi-Bergmann}, {Stringfellow}, {Stutz}, {Su}, {Taghizadeh-Popp}, {Talbot}, {Tayar}, {Telles}, {Teske}, {Thakar}, {Theissen}, {Tkachenko}, {Thomas}, {Tojeiro}, {Hernandez Toledo}, {Troup}, {Trump}, {Trussler}, {Turner}, {Tuttle}, {Unda-Sanzana}, {V{\'a}zquez-Mata}, {Valentini}, {Valenzuela}, {Vargas-Gonz{\'a}lez}, {Vargas-Maga{\~n}a}, {Alfaro}, {Villanova}, {Vincenzo}, {Wake}, {Warfield}, {Washington}, {Weaver}, {Weijmans}, {Weinberg}, {Weiss}, {Westfall}, {Wild}, {Wilde}, {Wilson}, {Wilson}, {Wilson}, {Wolf}, {Wood-Vasey}, {Yan}, {Zamora}, {Zasowski}, {Zhang}, {Zhao}, {Zheng}, {Zheng}, \& {Zhu}}]{2022ApJS..259...35A}
{Abdurro'uf}, {Accetta}, K., {Aerts}, C., {et~al.} 2022, \apjs, 259, 35, \dodoi{10.3847/1538-4365/ac4414}

\bibitem[{{Aihara} {et~al.}(2011){Aihara}, {Allende Prieto}, {An}, {Anderson}, {Aubourg}, {Balbinot}, {Beers}, {Berlind}, {Bickerton}, {Bizyaev}, {Blanton}, {Bochanski}, {Bolton}, {Bovy}, {Brandt}, {Brinkmann}, {Brown}, {Brownstein}, {Busca}, {Campbell}, {Carr}, {Chen}, {Chiappini}, {Comparat}, {Connolly}, {Cortes}, {Croft}, {Cuesta}, {da Costa}, {Davenport}, {Dawson}, {Dhital}, {Ealet}, {Ebelke}, {Edmondson}, {Eisenstein}, {Escoffier}, {Esposito}, {Evans}, {Fan}, {Femen{\'\i}a Castell{\'a}}, {Font-Ribera}, {Frinchaboy}, {Ge}, {Gillespie}, {Gilmore}, {Gonz{\'a}lez Hern{\'a}ndez}, {Gott}, {Gould}, {Grebel}, {Gunn}, {Hamilton}, {Harding}, {Harris}, {Hawley}, {Hearty}, {Ho}, {Hogg}, {Holtzman}, {Honscheid}, {Inada}, {Ivans}, {Jiang}, {Johnson}, {Jordan}, {Jordan}, {Kazin}, {Kirkby}, {Klaene}, {Knapp}, {Kneib}, {Kochanek}, {Koesterke}, {Kollmeier}, {Kron}, {Lampeitl}, {Lang}, {Le Goff}, {Lee}, {Lin}, {Long}, {Loomis}, {Lucatello}, {Lundgren}, {Lupton}, {Ma}, {MacDonald}, {Mahadevan}, {Maia}, {Makler},
  {Malanushenko}, {Malanushenko}, {Mandelbaum}, {Maraston}, {Margala}, {Masters}, {McBride}, {McGehee}, {McGreer}, {M{\'e}nard}, {Miralda-Escud{\'e}}, {Morrison}, {Mullally}, {Muna}, {Munn}, {Murayama}, {Myers}, {Naugle}, {Neto}, {Nguyen}, {Nichol}, {O'Connell}, {Ogando}, {Olmstead}, {Oravetz}, {Padmanabhan}, {Palanque-Delabrouille}, {Pan}, {Pandey}, {P{\^a}ris}, {Percival}, {Petitjean}, {Pfaffenberger}, {Pforr}, {Phleps}, {Pichon}, {Pieri}, {Prada}, {Price-Whelan}, {Raddick}, {Ramos}, {Reyl{\'e}}, {Rich}, {Richards}, {Rix}, {Robin}, {Rocha-Pinto}, {Rockosi}, {Roe}, {Rollinde}, {Ross}, {Ross}, {Rossetto}, {S{\'a}nchez}, {Sayres}, {Schlegel}, {Schlesinger}, {Schmidt}, {Schneider}, {Sheldon}, {Shu}, {Simmerer}, {Simmons}, {Sivarani}, {Snedden}, {Sobeck}, {Steinmetz}, {Strauss}, {Szalay}, {Tanaka}, {Thakar}, {Thomas}, {Tinker}, {Tofflemire}, {Tojeiro}, {Tremonti}, {Vandenberg}, {Vargas Maga{\~n}a}, {Verde}, {Vogt}, {Wake}, {Wang}, {Weaver}, {Weinberg}, {White}, {White}, {Yanny}, {Yasuda}, {Yeche}, \&
  {Zehavi}}]{2011ApJS..193...29A}
{Aihara}, H., {Allende Prieto}, C., {An}, D., {et~al.} 2011, \apjs, 193, 29, \dodoi{10.1088/0067-0049/193/2/29}

\bibitem[{{Albareti} {et~al.}(2017){Albareti}, {Allende Prieto}, {Almeida}, {Anders}, {Anderson}, {Andrews}, {Arag{\'o}n-Salamanca}, {Argudo-Fern{\'a}ndez}, {Armengaud}, {Aubourg}, {Avila-Reese}, {Badenes}, {Bailey}, {Barbuy}, {Barger}, {Barrera-Ballesteros}, {Bartosz}, {Basu}, {Bates}, {Battaglia}, {Baumgarten}, {Baur}, {Bautista}, {Beers}, {Belfiore}, {Bershady}, {Bertran de Lis}, {Bird}, {Bizyaev}, {Blanc}, {Blanton}, {Blomqvist}, {Bolton}, {Borissova}, {Bovy}, {Brandt}, {Brinkmann}, {Brownstein}, {Bundy}, {Burtin}, {Busca}, {Camacho Chavez}, {Cano D{\'\i}az}, {Cappellari}, {Carrera}, {Chen}, {Cherinka}, {Cheung}, {Chiappini}, {Chojnowski}, {Chuang}, {Chung}, {Cirolini}, {Clerc}, {Cohen}, {Comerford}, {Comparat}, {Correa do Nascimento}, {Cousinou}, {Covey}, {Crane}, {Croft}, {Cunha}, {Darling}, {Davidson}, {Dawson}, {Da Costa}, {Da Silva Ilha}, {Deconto Machado}, {Delubac}, {De Lee}, {De la Macorra}, {De la Torre}, {Diamond-Stanic}, {Donor}, {Downes}, {Drory}, {Du}, {Du Mas des Bourboux}, {Dwelly},
  {Ebelke}, {Eigenbrot}, {Eisenstein}, {Elsworth}, {Emsellem}, {Eracleous}, {Escoffier}, {Evans}, {Falc{\'o}n-Barroso}, {Fan}, {Favole}, {Fernandez-Alvar}, {Fernandez-Trincado}, {Feuillet}, {Fleming}, {Font-Ribera}, {Freischlad}, {Frinchaboy}, {Fu}, {Gao}, {Garcia}, {Garcia-Dias}, {Garcia-Hern{\'a}ndez}, {Garcia P{\'e}rez}, {Gaulme}, {Ge}, {Geisler}, {Gillespie}, {Gil Marin}, {Girardi}, {Goddard}, {Gomez Maqueo Chew}, {Gonzalez-Perez}, {Grabowski}, {Green}, {Grier}, {Grier}, {Guo}, {Guy}, {Hagen}, {Hall}, {Harding}, {Harley}, {Hasselquist}, {Hawley}, {Hayes}, {Hearty}, {Hekker}, {Hernandez Toledo}, {Ho}, {Hogg}, {Holley-Bockelmann}, {Holtzman}, {Holzer}, {Hu}, {Huber}, {Hutchinson}, {Hwang}, {Ibarra-Medel}, {Ivans}, {Ivory}, {Jaehnig}, {Jensen}, {Johnson}, {Jones}, {Jullo}, {Kallinger}, {Kinemuchi}, {Kirkby}, {Klaene}, {Kneib}, {Kollmeier}, {Lacerna}, {Lane}, {Lang}, {Laurent}, {Law}, {Leauthaud}, {Le Goff}, {Li}, {Li}, {Li}, {Li}, {Liang}, {Liang}, {Lima}, {Lin}, {Lin}, {Lin}, {Liu}, {Long}, {Lucatello},
  {MacDonald}, {MacLeod}, {Mackereth}, {Mahadevan}, {Maia}, {Maiolino}, {Majewski}, {Malanushenko}, {Malanushenko}, {Mallmann}, {Manchado}, {Maraston}, {Marques-Chaves}, {Martinez Valpuesta}, {Masters}, {Mathur}, {McGreer}, {Merloni}, {Merrifield}, {M{\'e}sz{\'a}ros}, {Meza}, {Miglio}, {Minchev}, {Molaverdikhani}, {Montero-Dorta}, {Mosser}, {Muna}, {Myers}, {Nair}, {Nandra}, {Ness}, {Newman}, {Nichol}, {Nidever}, {Nitschelm}, {O'Connell}, {Oravetz}, {Oravetz}, {Pace}, {Padilla}, {Palanque-Delabrouille}, {Pan}, {Parejko}, {Paris}, {Park}, {Peacock}, {Peirani}, {Pellejero-Ibanez}, {Penny}, {Percival}, {Percival}, {Perez-Fournon}, {Petitjean}, {Pieri}, {Pinsonneault}, {Pisani}, {Prada}, {Prakash}, {Price-Jones}, {Raddick}, {Rahman}, {Raichoor}, {Barboza Rembold}, {Reyna}, {Rich}, {Richstein}, {Ridl}, {Riffel}, {Riffel}, {Rix}, {Robin}, {Rockosi}, {Rodr{\'\i}guez-Torres}, {Rodrigues}, {Roe}, {Roman Lopes}, {Rom{\'a}n-Z{\'u}{\~n}iga}, {Ross}, {Rossi}, {Ruan}, {Ruggeri}, {Runnoe}, {Salazar-Albornoz}, {Salvato},
  {Sanchez}, {Sanchez}, {Sanchez-Gallego}, {Santiago}, {Schiavon}, {Schimoia}, {Schlafly}, {Schlegel}, {Schneider}, {Sch{\"o}nrich}, {Schultheis}, {Schwope}, {Seo}, {Serenelli}, {Sesar}, {Shao}, {Shetrone}, {Shull}, {Silva Aguirre}, {Skrutskie}, {Slosar}, {Smith}, {Smith}, {Sobeck}, {Somers}, {Souto}, {Stark}, {Stassun}, {Steinmetz}, {Stello}, {Storchi Bergmann}, {Strauss}, {Streblyanska}, {Stringfellow}, {Suarez}, {Sun}, {Taghizadeh-Popp}, {Tang}, {Tao}, {Tayar}, {Tembe}, {Thomas}, {Tinker}, {Tojeiro}, {Tremonti}, {Troup}, {Trump}, {Unda-Sanzana}, {Valenzuela}, {Van den Bosch}, {Vargas-Maga{\~n}a}, {Vazquez}, {Villanova}, {Vivek}, {Vogt}, {Wake}, {Walterbos}, {Wang}, {Wang}, {Weaver}, {Weijmans}, {Weinberg}, {Westfall}, {Whelan}, {Wilcots}, {Wild}, {Williams}, {Wilson}, {Wood-Vasey}, {Wylezalek}, {Xiao}, {Yan}, {Yang}, {Ybarra}, {Yeche}, {Yuan}, {Zakamska}, {Zamora}, {Zasowski}, {Zhang}, {Zhao}, {Zhao}, {Zheng}, {Zheng}, {Zhou}, {Zhu}, {Zinn}, \& {Zou}}]{2017ApJS..233...25A}
{Albareti}, F.~D., {Allende Prieto}, C., {Almeida}, A., {et~al.} 2017, \apjs, 233, 25, \dodoi{10.3847/1538-4365/aa8992}

\bibitem[{{Antonucci}(1993)}]{1993ARA&A..31..473A}
{Antonucci}, R. 1993, \araa, 31, 473, \dodoi{10.1146/annurev.aa.31.090193.002353}

\bibitem[{{Bahcall} \& {Ostriker}(1975)}]{1975Natur.256...23B}
{Bahcall}, J.~N., \& {Ostriker}, J.~P. 1975, \nat, 256, 23, \dodoi{10.1038/256023a0}

\bibitem[{{Baldassare} {et~al.}(2018){Baldassare}, {Geha}, \& {Greene}}]{2018ApJ...868..152B}
{Baldassare}, V.~F., {Geha}, M., \& {Greene}, J. 2018, \apj, 868, 152, \dodoi{10.3847/1538-4357/aae6cf}

\bibitem[{{Baldassare} {et~al.}(2020){Baldassare}, {Geha}, \& {Greene}}]{2020ApJ...896...10B}
---. 2020, \apj, 896, 10, \dodoi{10.3847/1538-4357/ab8936}

\bibitem[{{Baldry} {et~al.}(2010){Baldry}, {Robotham}, {Hill}, {Driver}, {Liske}, {Norberg}, {Bamford}, {Hopkins}, {Loveday}, {Peacock}, {Cameron}, {Croom}, {Cross}, {Doyle}, {Dye}, {Frenk}, {Jones}, {van Kampen}, {Kelvin}, {Nichol}, {Parkinson}, {Popescu}, {Prescott}, {Sharp}, {Sutherland}, {Thomas}, \& {Tuffs}}]{2010MNRAS.404...86B}
{Baldry}, I.~K., {Robotham}, A.~S.~G., {Hill}, D.~T., {et~al.} 2010, \mnras, 404, 86, \dodoi{10.1111/j.1365-2966.2010.16282.x}

\bibitem[{Baldwin {et~al.}(1981)Baldwin, Phillips, \& Terlevich}]{Baldwin_1981}
Baldwin, J.~A., Phillips, M.~M., \& Terlevich, R. 1981, Publications of the Astronomical Society of the Pacific, 93, 5, \dodoi{10.1086/130766}

\bibitem[{{Begelman} \& {Rees}(1978)}]{1978MNRAS.185..847B}
{Begelman}, M.~C., \& {Rees}, M.~J. 1978, \mnras, 185, 847, \dodoi{10.1093/mnras/185.4.847}

\bibitem[{{Blanton} {et~al.}(2011){Blanton}, {Kazin}, {Muna}, {Weaver}, \& {Price-Whelan}}]{2011AJ....142...31B}
{Blanton}, M.~R., {Kazin}, E., {Muna}, D., {Weaver}, B.~A., \& {Price-Whelan}, A. 2011, \aj, 142, 31, \dodoi{10.1088/0004-6256/142/1/31}

\bibitem[{{Bond} {et~al.}(1984){Bond}, {Arnett}, \& {Carr}}]{1984ApJ...280..825B}
{Bond}, J.~R., {Arnett}, W.~D., \& {Carr}, B.~J. 1984, \apj, 280, 825, \dodoi{10.1086/162057}

\bibitem[{{Brinchmann} {et~al.}(2004){Brinchmann}, {Charlot}, {White}, {Tremonti}, {Kauffmann}, {Heckman}, \& {Brinkmann}}]{2004MNRAS.351.1151B}
{Brinchmann}, J., {Charlot}, S., {White}, S.~D.~M., {et~al.} 2004, \mnras, 351, 1151, \dodoi{10.1111/j.1365-2966.2004.07881.x}

\bibitem[{{Burke} {et~al.}(2021{\natexlab{a}}){Burke}, {Liu}, {Chen}, {Shen}, \& {Guo}}]{2021MNRAS.504..543B}
{Burke}, C.~J., {Liu}, X., {Chen}, Y.-C., {Shen}, Y., \& {Guo}, H. 2021{\natexlab{a}}, \mnras, 504, 543, \dodoi{10.1093/mnras/stab912}

\bibitem[{{Burke} {et~al.}(2021{\natexlab{b}}){Burke}, {Shen}, {Blaes}, {Gammie}, {Horne}, {Jiang}, {Liu}, {McHardy}, {Morgan}, {Scaringi}, \& {Yang}}]{2021Sci...373..789B}
{Burke}, C.~J., {Shen}, Y., {Blaes}, O., {et~al.} 2021{\natexlab{b}}, Science, 373, 789, \dodoi{10.1126/science.abg9933}

\bibitem[{{Butler} \& {Bloom}(2011)}]{2011AJ....141...93B}
{Butler}, N.~R., \& {Bloom}, J.~S. 2011, \aj, 141, 93, \dodoi{10.1088/0004-6256/141/3/93}

\bibitem[{Cann {et~al.}(2019)Cann, Satyapal, Abel, Blecha, Mushotzky, Reynolds, \& Secrest}]{Cann_2019}
Cann, J.~M., Satyapal, S., Abel, N.~P., {et~al.} 2019, The Astrophysical Journal, 870, L2, \dodoi{10.3847/2041-8213/aaf88d}

\bibitem[{{Chen} {et~al.}(2012){Chen}, {Kauffmann}, {Tremonti}, {White}, {Heckman}, {Kova{\v{c}}}, {Bundy}, {Chisholm}, {Maraston}, {Schneider}, {Bolton}, {Weaver}, \& {Brinkmann}}]{2012MNRAS.421..314C}
{Chen}, Y.-M., {Kauffmann}, G., {Tremonti}, C.~A., {et~al.} 2012, \mnras, 421, 314, \dodoi{10.1111/j.1365-2966.2011.20306.x}

\bibitem[{{Conroy} {et~al.}(2009){Conroy}, {Gunn}, \& {White}}]{2009ApJ...699..486C}
{Conroy}, C., {Gunn}, J.~E., \& {White}, M. 2009, \apj, 699, 486, \dodoi{10.1088/0004-637X/699/1/486}

\bibitem[{{Coulter} {et~al.}(2022){Coulter}, {Jones}, {McGill}, {Foley}, {Aleo}, {Bustamante-Rosell}, {Chatterjee}, {Davis}, {Engel}, {Gagliano}, {Jacobson-Gal{\'a}n}, {Kilpatrick}, {Pan}, {Rojas-Bravo}, {Siebert}, {Taggart}, {Tinyanont}, \& {Wang}}]{2022zndo...7278430C}
{Coulter}, D.~A., {Jones}, D.~O., {McGill}, P., {et~al.} 2022, {YSE-PZ: An Open-source Target and Observation Management System}, v0.3.0,  Zenodo, \dodoi{10.5281/zenodo.7278430}

\bibitem[{{Coulter} {et~al.}(2023){Coulter}, {Jones}, {McGill}, {Foley}, {Aleo}, {Bustamante-Rosell}, {Chatterjee}, {Davis}, {Dickinson}, {Engel}, {Gagliano}, {Jacobson-Gal{\'a}n}, {Kilpatrick}, {Kutcka}, {Le Saux}, {Malanchev}, {Pan}, {Qui{\~n}onez}, {Rojas-Bravo}, {Siebert}, {Taggart}, {Tinyanont}, \& {Wang}}]{2023PASP..135f4501C}
---. 2023, \pasp, 135, 064501, \dodoi{10.1088/1538-3873/acd662}

\bibitem[{{Davis} {et~al.}(2019){Davis}, {Graham}, \& {Cameron}}]{2019ApJ...873...85D}
{Davis}, B.~L., {Graham}, A.~W., \& {Cameron}, E. 2019, \apj, 873, 85, \dodoi{10.3847/1538-4357/aaf3b8}

\bibitem[{{Gal-Yam}(2021)}]{2021AAS...23742305G}
{Gal-Yam}, A. 2021, in American Astronomical Society Meeting Abstracts, Vol. 237, American Astronomical Society Meeting Abstracts, 423.05

\bibitem[{{Geha} {et~al.}(2003){Geha}, {Alcock}, {Allsman}, {Alves}, {Axelrod}, {Becker}, {Bennett}, {Cook}, {Drake}, {Freeman}, {Griest}, {Keller}, {Lehner}, {Marshall}, {Minniti}, {Nelson}, {Peterson}, {Popowski}, {Pratt}, {Quinn}, {Stubbs}, {Sutherland}, {Tomaney}, {Vandehei}, \& {Welch}}]{2003AJ....125....1G}
{Geha}, M., {Alcock}, C., {Allsman}, R.~A., {et~al.} 2003, \aj, 125, 1, \dodoi{10.1086/344947}

\bibitem[{{Greene} {et~al.}(2020){Greene}, {Strader}, \& {Ho}}]{2020ARA&A..58..257G}
{Greene}, J.~E., {Strader}, J., \& {Ho}, L.~C. 2020, \araa, 58, 257, \dodoi{10.1146/annurev-astro-032620-021835}

\bibitem[{{Greene} {et~al.}(2016){Greene}, {Seth}, {Kim}, {L{\"a}sker}, {Goulding}, {Gao}, {Braatz}, {Henkel}, {Condon}, {Lo}, \& {Zhao}}]{2016ApJ...826L..32G}
{Greene}, J.~E., {Seth}, A., {Kim}, M., {et~al.} 2016, \apjl, 826, L32, \dodoi{10.3847/2041-8205/826/2/L32}

\bibitem[{Groves {et~al.}(2006)Groves, Heckman, \& Kauffmann}]{10.1111/j.1365-2966.2006.10812.x}
Groves, B.~A., Heckman, T.~M., \& Kauffmann, G. 2006, Monthly Notices of the Royal Astronomical Society, 371, 1559, \dodoi{10.1111/j.1365-2966.2006.10812.x}

\bibitem[{{Guo} {et~al.}(2018){Guo}, {Shen}, \& {Wang}}]{2018ascl.soft09008G}
{Guo}, H., {Shen}, Y., \& {Wang}, S. 2018, {PyQSOFit: Python code to fit the spectrum of quasars}, Astrophysics Source Code Library.
\newblock \doeprint{1809.008}

\bibitem[{{Haehnelt} \& {Rees}(1993)}]{1993MNRAS.263..168H}
{Haehnelt}, M.~G., \& {Rees}, M.~J. 1993, \mnras, 263, 168, \dodoi{10.1093/mnras/263.1.168}

\bibitem[{{Jones} {et~al.}(2021){Jones}, {Foley}, {Narayan}, {Hjorth}, {Huber}, {Aleo}, {Alexander}, {Angus}, {Auchettl}, {Baldassare}, {Bruun}, {Chambers}, {Chatterjee}, {Coppejans}, {Coulter}, {DeMarchi}, {Dimitriadis}, {Drout}, {Engel}, {French}, {Gagliano}, {Gall}, {Hung}, {Izzo}, {Jacobson-Gal{\'a}n}, {Kilpatrick}, {Korhonen}, {Margutti}, {Raimundo}, {Ramirez-Ruiz}, {Rest}, {Rojas-Bravo}, {Siebert}, {Smartt}, {Smith}, {Terreran}, {Wang}, {Wojtak}, {Agnello}, {Ansari}, {Arendse}, {Baldeschi}, {Blanchard}, {Brethauer}, {Bright}, {Brown}, {de Boer}, {Dodd}, {Fairlamb}, {Grillo}, {Hajela}, {Hede}, {Kolborg}, {Law-Smith}, {Lin}, {Magnier}, {Malanchev}, {Matthews}, {Mockler}, {Muthukrishna}, {Pan}, {Pfister}, {Ramanah}, {Rest}, {Sarangi}, {Schr{\o}der}, {Stauffer}, {Stroh}, {Taggart}, {Tinyanont}, {Wainscoat}, \& {Young Supernova Experiment}}]{2021ApJ...908..143J}
{Jones}, D.~O., {Foley}, R.~J., {Narayan}, G., {et~al.} 2021, \apj, 908, 143, \dodoi{10.3847/1538-4357/abd7f5}

\bibitem[{{Kauffmann} {et~al.}(2003{\natexlab{a}}){Kauffmann}, {Heckman}, {White}, {Charlot}, {Tremonti}, {Brinchmann}, {Bruzual}, {Peng}, {Seibert}, {Bernardi}, {Blanton}, {Brinkmann}, {Castander}, {Cs{\'a}bai}, {Fukugita}, {Ivezic}, {Munn}, {Nichol}, {Padmanabhan}, {Thakar}, {Weinberg}, \& {York}}]{2003MNRAS.341...33K}
{Kauffmann}, G., {Heckman}, T.~M., {White}, S. D.~M., {et~al.} 2003{\natexlab{a}}, \mnras, 341, 33, \dodoi{10.1046/j.1365-8711.2003.06291.x}

\bibitem[{{Kauffmann} {et~al.}(2003{\natexlab{b}}){Kauffmann}, {Heckman}, {Tremonti}, {Brinchmann}, {Charlot}, {White}, {Ridgway}, {Brinkmann}, {Fukugita}, {Hall}, {Ivezi{\'c}}, {Richards}, \& {Schneider}}]{2003MNRAS.346.1055K}
{Kauffmann}, G., {Heckman}, T.~M., {Tremonti}, C., {et~al.} 2003{\natexlab{b}}, \mnras, 346, 1055, \dodoi{10.1111/j.1365-2966.2003.07154.x}

\bibitem[{{Kewley} {et~al.}(2006){Kewley}, {Groves}, {Kauffmann}, \& {Heckman}}]{2006MNRAS.372..961K}
{Kewley}, L.~J., {Groves}, B., {Kauffmann}, G., \& {Heckman}, T. 2006, \mnras, 372, 961, \dodoi{10.1111/j.1365-2966.2006.10859.x}

\bibitem[{Kimura {et~al.}(2020)Kimura, Yamada, Kokubo, Yasuda, Morokuma, Nagao, \& Matsuoka}]{Kimura_2020}
Kimura, Y., Yamada, T., Kokubo, M., {et~al.} 2020, The Astrophysical Journal, 894, 24, \dodoi{10.3847/1538-4357/ab83f3}

\bibitem[{{Koushiappas} {et~al.}(2004){Koushiappas}, {Bullock}, \& {Dekel}}]{2004MNRAS.354..292K}
{Koushiappas}, S.~M., {Bullock}, J.~S., \& {Dekel}, A. 2004, \mnras, 354, 292, \dodoi{10.1111/j.1365-2966.2004.08190.x}

\bibitem[{{Koz{\l}owski}(2017)}]{2017A&A...597A.128K}
{Koz{\l}owski}, S. 2017, \aap, 597, A128, \dodoi{10.1051/0004-6361/201629890}

\bibitem[{{Lee}(1993)}]{1993ApJ...418..147L}
{Lee}, M.~H. 1993, \apj, 418, 147, \dodoi{10.1086/173378}

\bibitem[{{Loeb} \& {Rasio}(1994)}]{1994ApJ...432...52L}
{Loeb}, A., \& {Rasio}, F.~A. 1994, \apj, 432, 52, \dodoi{10.1086/174548}

\bibitem[{MacLeod {et~al.}(2010)MacLeod, Ivezić, Kochanek, Kozłowski, Kelly, Bullock, Kimball, Sesar, Westman, Brooks, Gibson, Becker, \& de~Vries}]{MacLeod_2010}
MacLeod, C.~L., Ivezić, Z., Kochanek, C.~S., {et~al.} 2010, The Astrophysical Journal, 721, 1014, \dodoi{10.1088/0004-637X/721/2/1014}

\bibitem[{{Madau} \& {Rees}(2001)}]{2001ApJ...551L..27M}
{Madau}, P., \& {Rees}, M.~J. 2001, \apjl, 551, L27, \dodoi{10.1086/319848}

\bibitem[{{Magnier} {et~al.}(2020){Magnier}, {Sweeney}, {Chambers}, {Flewelling}, {Huber}, {Price}, {Waters}, {Denneau}, {Draper}, {Farrow}, {Jedicke}, {Hodapp}, {Kaiser}, {Kudritzki}, {Metcalfe}, {Stubbs}, \& {Wainscoat}}]{2020ApJS..251....5M}
{Magnier}, E.~A., {Sweeney}, W.~E., {Chambers}, K.~C., {et~al.} 2020, Astrophysical Journal, Supplement, 251, 5, \dodoi{10.3847/1538-4365/abb82c}

\bibitem[{{Maraston} \& {Str{\"o}mb{\"a}ck}(2011)}]{2011MNRAS.418.2785M}
{Maraston}, C., \& {Str{\"o}mb{\"a}ck}, G. 2011, \mnras, 418, 2785, \dodoi{10.1111/j.1365-2966.2011.19738.x}

\bibitem[{{Maraston} {et~al.}(2013){Maraston}, {Pforr}, {Henriques}, {Thomas}, {Wake}, {Brownstein}, {Capozzi}, {Tinker}, {Bundy}, {Skibba}, {Beifiori}, {Nichol}, {Edmondson}, {Schneider}, {Chen}, {Masters}, {Steele}, {Bolton}, {York}, {Weaver}, {Higgs}, {Bizyaev}, {Brewington}, {Malanushenko}, {Malanushenko}, {Snedden}, {Oravetz}, {Pan}, {Shelden}, \& {Simmons}}]{2013MNRAS.435.2764M}
{Maraston}, C., {Pforr}, J., {Henriques}, B.~M., {et~al.} 2013, \mnras, 435, 2764, \dodoi{10.1093/mnras/stt1424}

\bibitem[{{Pacucci} {et~al.}(2021){Pacucci}, {Mezcua}, \& {Regan}}]{2021ApJ...920..134P}
{Pacucci}, F., {Mezcua}, M., \& {Regan}, J.~A. 2021, \apj, 920, 134, \dodoi{10.3847/1538-4357/ac1595}

\bibitem[{{Quinlan} \& {Shapiro}(1990)}]{1990ApJ...356..483Q}
{Quinlan}, G.~D., \& {Shapiro}, S.~L. 1990, \apj, 356, 483, \dodoi{10.1086/168856}

\bibitem[{{Reines} {et~al.}(2013){Reines}, {Greene}, \& {Geha}}]{2013ApJ...775..116R}
{Reines}, A.~E., {Greene}, J.~E., \& {Geha}, M. 2013, \apj, 775, 116, \dodoi{10.1088/0004-637X/775/2/116}

\bibitem[{{Schleicher} {et~al.}(2019){Schleicher}, {Arbet-Engels}, {Baack}, {Balbo}, {Biland}, {Blank}, {Bretz}, {Bruegge}, {Bulinski}, {Buss}, {Doerr}, {Dorner}, {Elsaesser}, {Grischagin}, {Hildebrand}, {Linhoff}, {Mannheim}, {Mueller}, {Neise}, {Neronov}, {Noethe}, {Paravac}, {Rhode}, {Schulz}, {Sedlaczek}, {Shukla}, {Sliusar}, {Willert}, \& {Walter}}]{2019Galax...7...62S}
{Schleicher}, B., {Arbet-Engels}, A., {Baack}, D., {et~al.} 2019, Galaxies, 7, 62, \dodoi{10.3390/galaxies7020062}

\bibitem[{Schmidt {et~al.}(2010)Schmidt, Marshall, Rix, Jester, Hennawi, \& Dobler}]{Schmidt_2010}
Schmidt, K.~B., Marshall, P.~J., Rix, H.-W., {et~al.} 2010, The Astrophysical Journal, 714, 1194, \dodoi{10.1088/0004-637x/714/2/1194}

\bibitem[{{Shen} {et~al.}(2019){Shen}, {Hall}, {Horne}, {Zhu}, {McGreer}, {Simm}, {Trump}, {Kinemuchi}, {Brandt}, {Green}, {Grier}, {Guo}, {Ho}, {Homayouni}, {Jiang}, {I-Hsiu Li}, {Morganson}, {Petitjean}, {Richards}, {Schneider}, {Starkey}, {Wang}, {Chambers}, {Kaiser}, {Kudritzki}, {Magnier}, \& {Waters}}]{2019ApJS..241...34S}
{Shen}, Y., {Hall}, P.~B., {Horne}, K., {et~al.} 2019, \apjs, 241, 34, \dodoi{10.3847/1538-4365/ab074f}

\bibitem[{{Taylor} {et~al.}(2011){Taylor}, {Hopkins}, {Baldry}, {Brown}, {Driver}, {Kelvin}, {Hill}, {Robotham}, {Bland-Hawthorn}, {Jones}, {Sharp}, {Thomas}, {Liske}, {Loveday}, {Norberg}, {Peacock}, {Bamford}, {Brough}, {Colless}, {Cameron}, {Conselice}, {Croom}, {Frenk}, {Gunawardhana}, {Kuijken}, {Nichol}, {Parkinson}, {Phillipps}, {Pimbblet}, {Popescu}, {Prescott}, {Sutherland}, {Tuffs}, {van Kampen}, \& {Wijesinghe}}]{2011MNRAS.418.1587T}
{Taylor}, E.~N., {Hopkins}, A.~M., {Baldry}, I.~K., {et~al.} 2011, \mnras, 418, 1587, \dodoi{10.1111/j.1365-2966.2011.19536.x}

\bibitem[{{Tremonti} {et~al.}(2004){Tremonti}, {Heckman}, {Kauffmann}, {Brinchmann}, {Charlot}, {White}, {Seibert}, {Peng}, {Schlegel}, {Uomoto}, {Fukugita}, \& {Brinkmann}}]{2004ApJ...613..898T}
{Tremonti}, C.~A., {Heckman}, T.~M., {Kauffmann}, G., {et~al.} 2004, \apj, 613, 898, \dodoi{10.1086/423264}

\bibitem[{Ulrich {et~al.}(1997)Ulrich, Maraschi, \& Urry}]{doi:10.1146/annurev.astro.35.1.445}
Ulrich, M.-H., Maraschi, L., \& Urry, C.~M. 1997, Annual Review of Astronomy and Astrophysics, 35, 445, \dodoi{10.1146/annurev.astro.35.1.445}

\bibitem[{{van Wassenhove} {et~al.}(2010){van Wassenhove}, {Volonteri}, {Walker}, \& {Gair}}]{2010MNRAS.408.1139V}
{van Wassenhove}, S., {Volonteri}, M., {Walker}, M.~G., \& {Gair}, J.~R. 2010, \mnras, 408, 1139, \dodoi{10.1111/j.1365-2966.2010.17189.x}

\bibitem[{{Vanden Berk} {et~al.}(2004){Vanden Berk}, {Wilhite}, {Kron}, {Anderson}, {Brunner}, {Hall}, {Ivezi{\'c}}, {Richards}, {Schneider}, {York}, {Brinkmann}, {Lamb}, {Nichol}, \& {Schlegel}}]{2004ApJ...601..692V}
{Vanden Berk}, D.~E., {Wilhite}, B.~C., {Kron}, R.~G., {et~al.} 2004, \apj, 601, 692, \dodoi{10.1086/380563}

\bibitem[{{Volonteri} {et~al.}(2008){Volonteri}, {Lodato}, \& {Natarajan}}]{2008MNRAS.383.1079V}
{Volonteri}, M., {Lodato}, G., \& {Natarajan}, P. 2008, \mnras, 383, 1079, \dodoi{10.1111/j.1365-2966.2007.12589.x}

\bibitem[{{Volonteri} \& {Natarajan}(2009)}]{2009MNRAS.400.1911V}
{Volonteri}, M., \& {Natarajan}, P. 2009, \mnras, 400, 1911, \dodoi{10.1111/j.1365-2966.2009.15577.x}

\bibitem[{{Wasleske} \& {Baldassare}(2024)}]{2024ApJ...971...68W}
{Wasleske}, E.~J., \& {Baldassare}, V.~F. 2024, \apj, 971, 68, \dodoi{10.3847/1538-4357/ad5442}

\bibitem[{{Wasleske} {et~al.}(2022){Wasleske}, {Baldassare}, \& {Carroll}}]{2022ApJ...933...37W}
{Wasleske}, E.~J., {Baldassare}, V.~F., \& {Carroll}, C.~M. 2022, \apj, 933, 37, \dodoi{10.3847/1538-4357/ac715b}

\bibitem[{Waters {et~al.}(2020)Waters, Magnier, Price, Chambers, Burgett, Draper, Flewelling, Hodapp, Huber, Jedicke, Kaiser, Kudritzki, Lupton, Metcalfe, Rest, Sweeney, Tonry, Wainscoat, \& Wood-Vasey}]{Waters_2020}
Waters, C.~Z., Magnier, E.~A., Price, P.~A., {et~al.} 2020, The Astrophysical Journal Supplement Series, 251, 4, \dodoi{10.3847/1538-4365/abb82b}

\end{thebibliography}
\bibliographystyle{aasjournal}

\end{document}